\documentclass[%
a4paper,
superscriptaddress,
notitlepage,
bibnotes,
 amsmath,amssymb,
 aps,
10pt,
twocolumn,
prx,
floatfix,
longbibliography
]{revtex4-2}

\usepackage{xcolor}
\colorlet{mylinkcolor}{blue!66!black!80}
\colorlet{mynewcolor}{black}
\colorlet{mynewcolor2}{black}

\usepackage{mathtools}

\usepackage[colorlinks=true,linkcolor=mylinkcolor,citecolor=mylinkcolor,filecolor=cyan,urlcolor=mylinkcolor,breaklinks=true]{hyperref}
\newcounter{SIequation}
\newcounter{SIfigure}
\makeatletter
\@addtoreset{equation}{SIequation}
\@addtoreset{figure}{SIfigure}
\makeatother

\newcommand{\hiddenref}[1]{\hyperref[#1]{\color{black}\ref*{#1}}}

\usepackage{xcolor}

\definecolor{blind1}{HTML}{0173b2}
\definecolor{blind2}{HTML}{de8f05}
\definecolor{blind3}{HTML}{029e73}
\definecolor{blind4}{HTML}{d55e00}
\definecolor{blind5}{HTML}{cc78bc}
\definecolor{blind6}{HTML}{ca9161}
\definecolor{blind7}{HTML}{fbafe4}
\definecolor{blind8}{HTML}{949494}
\definecolor{blind9}{HTML}{ece133}
\definecolor{blind10}{HTML}{56b4e9}

\usepackage[utf8]{inputenc}

\newcommand{\mat}[1]{\textbf{#1}}
\newcommand{\avg}[1]{\langle#1\rangle}
\newcommand{\avgc}[1]{\langle#1\rangle^\mathcal{A}}
\newcommand{\avgcStar}[1]{\langle#1\rangle^\text{loc}}
\newcommand{\avgloc}[1]{\langle#1\rangle^\text{loc}}
\newcommand{\avgdwell}[1]{\langle#1\rangle^\text{dwell}}
\newcommand{\avgtrans}[1]{\langle#1\rangle^\text{tr}}
\newcommand{\avgexit}[1]{\langle#1\rangle^\text{exit}}
\newcommand{\avgs}[1]{\langle#1\rangle^\text{single}}

\newcommand{\del}{\partial}
\newcommand{\ddel}[1]{\partial_{#1}}
\newcommand{\dd}{\mathrm{d}}
\newcommand{\bF}{\boldsymbol{F}}

\newcommand{\bnu}{\boldsymbol{\nu}}

\newcommand{\bu}{\boldsymbol{u}}
\newcommand{\bv}{\boldsymbol{v }}
\newcommand{\bx}{\boldsymbol{x}}

\newcommand{\by}{\boldsymbol{y}}

\newcommand{\e}{{\rm e}}

\newcommand{\mD}{\boldsymbol{\mathcal{D}}}
\newcommand{\mDc}{\mathcal{D}}
\newcommand{\mM}{\mathbf{M}}

\newcommand{\T}{\top}

\newcommand{\bra}[1]{\langle#1|}
\newcommand{\ket}[1]{|#1\rangle}

\newcommand{\id}{\mathbf{1}}

\newcommand{\wps}{p}
\newcommand{\twps}{\tilde{p}}
\newcommand{\phic}{\phi^\mathcal{A}}
\newcommand{\wpc}{\wp^\mathcal{A}}
\newcommand{\twpc}{\tilde{\wp}^\mathcal{A}}
\newcommand{\wpcStar}{\wp^\text{loc}}
\newcommand{\wpctrans}{\wp^\text{tr}}
\newcommand{\wpcexit}{\wp^\text{exit}}
\newcommand{\twpctrans}{\tilde{\wp}^\text{tr}}
\newcommand{\wpcdwell}{\wp^\text{dwell}}
\newcommand{\twpcdwell}{\tilde{\wp}^\text{dwell}}

\newcommand{\twpcStar}{\tilde{\wp}^\text{loc}}

\newcommand{\cA}{\boldsymbol{\mathcal{A}}}

\newcommand{\phiStar}{\phi^{\rm loc}}
\newcommand{\philoc}{\boldsymbol{\Phi}}
\newcommand{\tloc}{\boldsymbol{\mathcal{T}}}

\newcommand{\ploc}{\tilde{\boldsymbol{\mathcal{P}}}(s)}
\newcommand{\ploct}{\boldsymbol{\mathcal{P}}(t)}

\newcommand{\Ploc}{P_i}
\newcommand{\Jloc}{J_i}
\newcommand{\Sloc}{S_i}
\newcommand{\tPloc}{\tilde{P}_i}

\newcommand{\LF}{\hat{\mathcal{L}}^{\rm F}}
\newcommand{\dLF}{\hat{\mathcal{J}}^{\rm F}}
\newcommand{\LB}{\hat{\mathcal{L}}^{\rm B}}

\newcommand{\gOuter}{\psi^{\rm out}}
\newcommand{\gInner}{\psi^{\rm in}}
\newcommand{\JOuter}{J^{\rm out}}
\newcommand{\JInner}{J^{\rm in}}

\newcommand{\PA}{\id_\mathcal{A}}
\newcommand{\PAc}{\id_{\mathcal{A}^{\rm c}}}

\newcount\minute
\newcount\hour
\def\currenttime{%
    \minute\time
    \hour\minute
    \divide\hour60
    \the\hour:\multiply\hour60\advance\minute-\hour\the\minute}
\begin{document}
 \title{Emergent memory and kinetic hysteresis in strongly driven networks}
\author{David Hartich}
\email{david.hartich@mpibpc.mpg.de}
\affiliation{%
Mathematical bioPhysics Group, Max Planck Institute for Biophysical Chemistry, 37077 Göttingen, Germany}
 \author{Aljaž Godec}%
 \email{agodec@mpibpc.mpg.de}
\affiliation{%
Mathematical bioPhysics Group, Max Planck Institute for Biophysical Chemistry, 37077 Göttingen, Germany}



%
%

\begin{abstract}

Stochastic network-dynamics are typically assumed to be memory-less. Involving prolonged dwells interrupted by instantaneous transitions between nodes such Markov networks stand as a coarse-graining paradigm for chemical reactions, gene expression, molecular machines, spreading of diseases, protein dynamics, diffusion in energy landscapes, epigenetics and many others. However, as soon as transitions cease to be negligibly short, as often observed in experiments, the dynamics develops a memory. That is, state-changes depend not only on the present state but also on the past. Here, we establish the first thermodynamically consistent -- dissipation-preserving --  mapping of continuous dynamics onto a network, which reveals ingrained dynamical symmetries and an unforeseen kinetic hysteresis. These symmetries impose three independent sources of fluctuations in state-to state kinetics that determine the `flavor of memory'. The hysteresis between the forward/backward in time coarse-graining of continuous trajectories implies a new paradigm for the thermodynamics of active molecular processes in the presence of memory, that is, beyond the assumption of local detailed balance. Our results provide a new understanding of fluctuations in the operation of molecular machines as well as catch-bonds involved in cellular adhesion.

\end{abstract}

\maketitle


\begin{figure*}
\centering
\includegraphics[width=\textwidth]{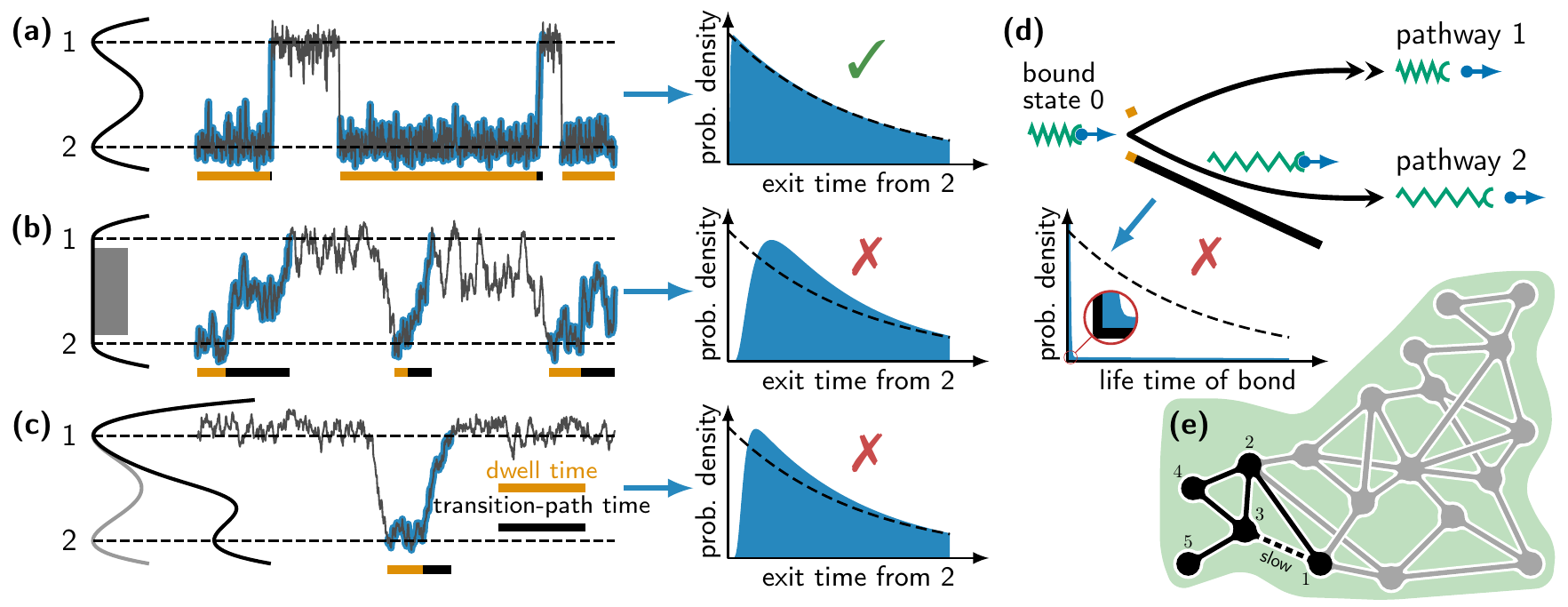}
  \caption{Breakdown of Markovian and  emergence of
      non-Markovian kinetics. 
      (a, left)~Diffusion in a double-well
    potential as a reduced model of the dynamics of a protein molecule
    transitioning between an unfolded (state 1) and a folded (state 2)
    conformation. Each exit-event from state 2 to state 1 is
    highlighted in blue. (a, right)~ The histogram (shaded region) of the exit
    time from state 2 is well approximated by a memory-less single exponential decay (dashed line). 
    Orange bars below the trajectory highlight dwell periods in
    the reduced state 2, and black bars
    the duration of transitions from state 2 to state
    1. (b, left)~Diffusion in a potential with a diffusive
    barrier. (b, right)~ The histogram (shaded region) of the exit
    time from state 2 alongside a single exponential decay with the same
    mean exit time (dashed line) that, however, only poorly approximates the statistics of exit.    
    (c)~Double-well potential from panel (a) ``tilted'' by an
    additional pulling-force that destabilizes the (folded) state 2.  
    (d)~Schematic of rupture-pathways
of a ``catch-bond'' under force. The bond can rupture along two
possible pathways: a fast pathway 1 (double arrow), and a slow pathway 2 that
involves an intermediate conformational change. As before the orange
and black bars denote dwell and transition periods, respectively. The
probability density of the life-time of the bond is shown below,
whereby the probability densities depicted by the histogram (shaded region) and dashed line have
the same mean. The stark disagreement between the two reflects that the
rupture is not memory-less. 
(e)~Schematics of a general network with a sub-network with 5
states highlighted in black. Transitions between states 1
and 3 (dashed line) are assumed to be slow.
}
   \label{fig:illustrantion_all}
\end{figure*}


\section{Introduction}\label{sec:intro}
In the presence of a time-scale separation the coarse-graining of continuous-space dynamics to
transitions on a network yields memory-less, Markovian
kinetics.
Such Markov networks are routinely used for the description of
chemical reactions \cite{gill77,gill07,elbe20}, gene expression \cite{mcad97,paul05},
molecular machines \cite{seif12}, spreading of diseases \cite{past15}, protein dynamics \cite{bowm10,wood14,chod14,bowm14,husi18},
diffusion in energy landscapes \cite{wale98}, epigenetics \cite{mori16} and many others.
Markov networks with only a few discrete states
are useful for the modeling of a physical systems
at large times in, for example,  molecular machines \cite{seif12} and  proteins \cite{bowm10,wood14,chod14,bowm14,husi18}.
One inherent feature of memory-less dynamics is that the waiting time between consecutive state changes is exponentially distributed \footnote{A stochastic waiting time $t$ is said to be memory-less if $\text{Prob}[t\ge t_1]=\text{Prob}[t\ge t_1+t_2]/\text{Prob}[t\ge t_2]$ for all $t_1,t_2\ge0$, which is satisfied if and only if $t$ is exponentially distributed \cite{fell71}.
} as captured e.g. by the Gillespie algorithm \cite{gill77,gill07}.

To highlight how memory-less state-to-state transitions arise
microscopically we depict in
Fig.~\ref{fig:illustrantion_all}a a realization of a
continuous-space diffusion in a double-well potential as function of time,
which may represent, e.g. the extension of a protein molecule
inter-converting between two conformational states \cite{neup16}. As
soon as the barrier between the two wells
is high enough the system locally equilibrates within each well
before transiting to the other, which renders the probability density
of the exit time from either well (exits from well 2 are highlighted
in Fig.~\ref{fig:illustrantion_all}a) to a good approximation
exponentially distributed (see right panel in
Fig.~\ref{fig:illustrantion_all}a). More generally, two conditions must be satisfied 
for memory-less kinetics between meta-stable states to emerge \cite{schu11}. To provide an intuition about these two conditions it is useful
to dissect each exit time into a transition period
\cite{humm04,bere18,maka21} (see black bars) and the rest that we will call
\emph{dwell time} (see orange bars). The \emph{first condition} requires that the
system, once it leaves any of the meta-stable states (e.g., state 2),
quickly transits to the next state (i.e., the transition-path time is
negligibly short) or rapidly returns to the initial state. In
Fig.~\ref{fig:illustrantion_all}a the latter is visible as short excursions
within the long ``dwell time'' periods highlighted in orange.  The
\emph{second condition} requires dwell periods to be long enough for the system to
reach a local equilibrium in the initial well, which guarantees that any potentially hidden degree of freedom has also reached equilibrium. Memory-less kinetics thus involves the interplay of long dwells  and short ``instantaneous'' transitions.

A two state Markov jump process -- representing the minimum-to-minimum hopping in Fig.~\ref{fig:illustrantion_all}a --
inherently neglects a finite duration of transitions that can nowadays
be probed in single-molecule
fluorescence  \cite{chun12,chun13} or force
\cite{neup12,ritc15,neup16a} spectroscopy experiments.
Even when they are short random transition times
encode important information about the topological shape of the
free energy barriers \cite{sati20}. This implies that a non-Markovian
network theory that explicitly incorporates transition-path times -- which
is the main aim of this work -- is desirable even in the presence of time-scale separation.

More importantly, \emph{prolonged transition-path times} \cite{chun12,neup12,chun13,ritc15,neup16a,glad19,kim20,zijl20}
resulting, e.g. from spatial transport of molecules in chemical reactions under
imperfect mixing
\cite{aqui17,zhan19}, the presence of a rugged energy landscape \cite{zwan88}
as shown in Fig.~\ref{fig:illustrantion_all}b, or external forces that
destabilize 
local minima as in Fig.~\ref{fig:illustrantion_all}c, are
bound to cause ``mild'' violations of Markovianity. Moreover,
dynamics in higher dimensions allows for the
coexistence of \emph{parallel transitions paths}
\cite{kim20,sati20}. 
Parallel transition paths as depicted in Fig.~\ref{fig:illustrantion_all}d
allow for the coexistence  of fast and slow time scales that can cause ``strong'' violations of Markovianity, manifested, e.g. as
so-called catch bonds in cellular adhesion \cite{thom06,thom08a,buck14} which we will discuss below in more detail. 

The idea to account for non-exponential waiting time distributions
is not new and is in fact at the heart of the \emph{generalized master
equation} \cite{land77} (see also Refs.~\cite{mont65,klaf87,haus87,fara04,shal06}
with numerous applications that go beyond the scope of this article).
While these models were constructed and applied phenomenologically
to unravel interesting phenomena such as anomalous diffusion
\cite{metz98,bark00,soko05}, their microscopic physical underpinning
remains elusive. Moreover, the phenomenological construction of the generalized master
equation \cite{land77,mont65,klaf87,haus87,fara04} only assures that it is
kinetically consistent, whereas it remains unclear under which
conditions the resulting renewal dynamics is thermodynamically consistent.
{\color{mynewcolor}
The latter turns out to be essential -- we show that \emph{coarse-graining
and time-reversal in fact do not commute} giving rise to a phenomenon we coin \emph{kinetic hysteresis}. This has important consequences for the quantification of dissipation.
}



\begin{figure}
\includegraphics{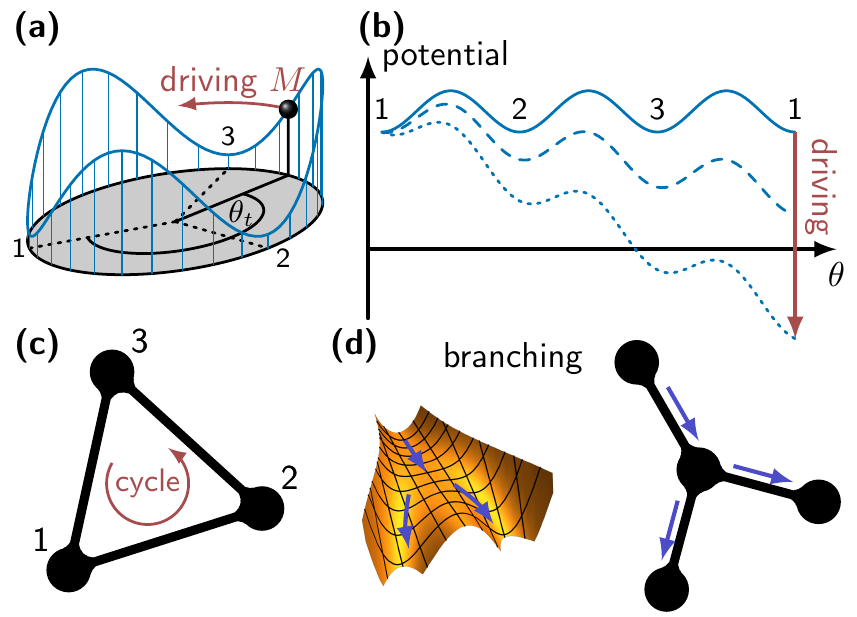}
 \caption{\color{mynewcolor}Rationale and building blocks of coarse-graining. (a-c) Unicyclic (driven) systems and (d)
   branching point.
 (a)~Rotation angle $\theta_t$ describing a system/observable
 driven by a non-equilibrium torque $M$ along a periodic potential
 with three metastable states.  (b)~The driving tilts the potential
 (dashed lines)  which leads to work exchange along a cycle $1\to 2
 \to 3\to 1$. (c)~Network representation of (a) as in
 Fig.~\ref{fig:illustrantion_all}e (for corresponding trajectories see
 Fig.~\ref{fig:S_rotator_3dtraj}). (d)~Multiple cycles require
 branching points (blue arrows), for example, generated by a
 multidimensional force field with a local pitchfork bifurcation
 \cite{mori16} (left panel depicts the corresponding local potential).}
 \label{fig:network_meaning}
\end{figure}

\begin{figure*}
 \centering
 \includegraphics{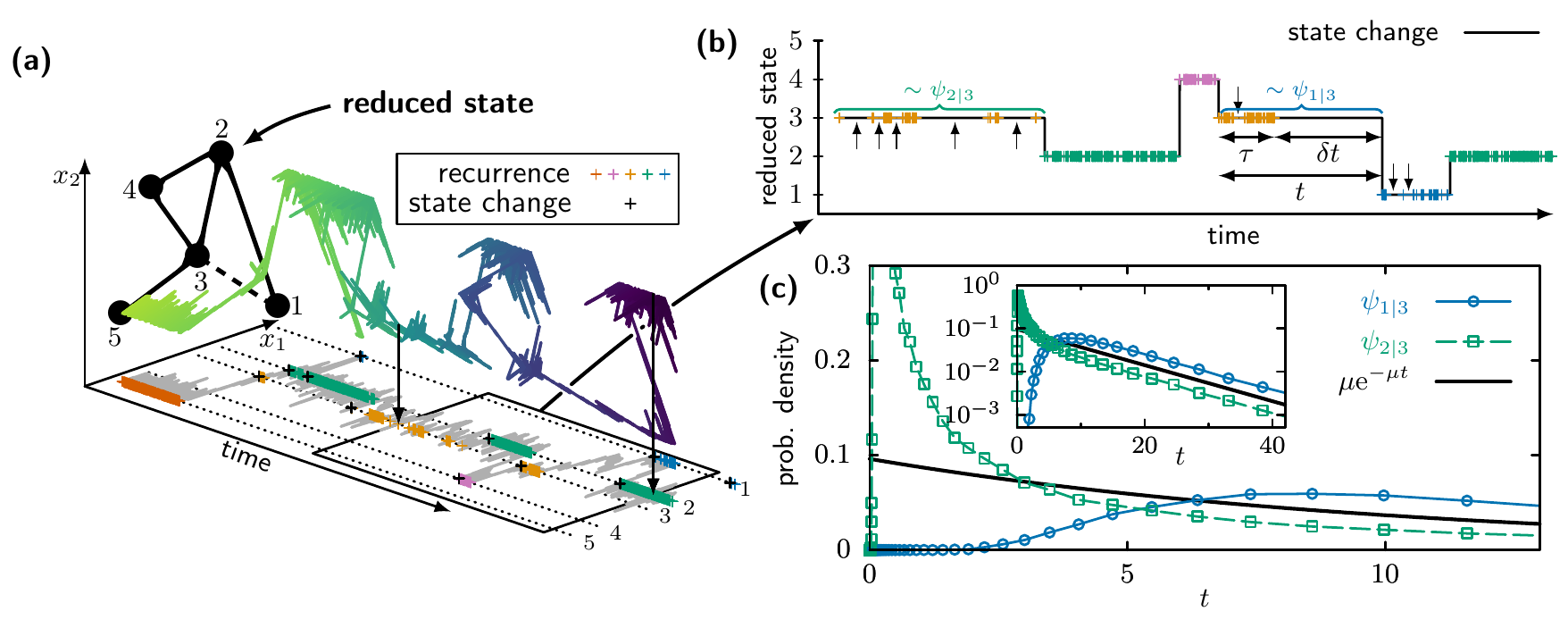}
 \caption{Coarse-grained dynamics.
   (a)~Projection of a trajectory of the full dynamics on the
   sub-graph in Fig.~\ref{fig:illustrantion_all}e onto a plane
   (spanned by $x_1x_2$). Time runs from bright to dark. The
   network is represented by black lines with each reduced state $1,\dots,5$. A second projection onto the
   $x_2=0$ plane (gray line) reveals recurrences (colored crosses) and
   state-changes (black crosses).     
 (b)~Coarse-graining yields a time-series of state-changes
 (solid line); the time interval corresponds to the box in (a).
 One dwell interval ``$\tau$'' and one transition-path time interval ``$\delta t$''
 are highlighted; their sum $t=\delta t+\tau$ is the local first passage time from state $i$ (here $i=3$) to a neighboring state $j$ (here $j=1$).
  Long recurrence times are
 highlighted by vertical arrows. 
 (c)~Normalized probability density of local first passage time $\psi_{j|i}\equiv\wp^{\rm
 loc}_{j|i}(t)/\phi^{\rm loc}_{j|i}$ (see also (b)) from $i=3$ to $j=1,2$. Details to the model are given in Appendix \ref{subsec:synthetic_network}.
 }
 \label{fig:blinking}
 \end{figure*}

To account for transitions with a finite duration in complex networks
as shown in Fig.~\ref{fig:illustrantion_all}e, we here develop a
theory embodying an \emph{exact projection} of continuous dynamics on a
graph onto a 
network with discrete states. Diffusion on a graph arises quite generally from the averaging of
fast degrees of freedom in Hamiltonian dynamics weakly coupled to a
heat bath \cite{frei93} (see also \cite{frei04,*frei07}), and includes both, a
position dependent force
and a position dependent diffusion coefficient \cite{hinc10,bere11}.
 The
coarse-grained dynamics evolve as jumps between the nodes. 
A state-change occurs once the trajectory enters
a new node for the first time {\color{mynewcolor}(see also Ref.~\cite{fara04}, where the
process is referred to as ``milestoning'')}.

{\color{mynewcolor}
Let us highlight two elementary ``building blocks'' of a network,
which we call \emph{cycles} and \emph{branching points} (see
Fig.~\ref{fig:network_meaning}), which are in fact not considered in Ref.~\cite{fara04}. 
The presence of cycles is required to enable a strongly driven system
to constantly exchange (free) energy with the environment, thereby
facilitating e.g.\ a biased transport.  
Consider an ATPase which has a 3-fold rotational symmetry
(see Fig.~\ref{fig:network_meaning}a). An external mechanical torque
applied to the system causes a rotation
\cite{toya11}, which in turn effectively gives rise to a non-conservative force field
as illustrated in Fig.~\ref{fig:network_meaning}b. In other words,
upon closing a cycle $1\to2 \to 3\to 1$ (see
Fig.~\ref{fig:network_meaning}c) the system has made a $360^\circ$ turn while dissipating $M\times 360^\circ$ of free energy. 
We say that a network is strongly driven if the energy exchange substantially exceeds the thermal energy $k_\text{B}T$, which prevents the emergence of an equilibrium Boltzmann distribution.
Note that the continuous dynamics on the graph shown in Fig.~\ref{fig:network_meaning}c
inter alia takes into account possible intermediate meta-stable states
observed e.g.\ in Ref.~\cite{yasu01}. 

Networks may have a genuinely multidimensional underlying topology
that in turn allows for branching points as illustrated in Fig.~\ref{fig:network_meaning}d. 
Branching points allow for the existence of multiple cycles,
i.e.\ account for multiple dissipative mechanisms. 
We consider the dynamics along the blue arrows in Fig.~\ref{fig:network_meaning}d to be
effectively one dimensional which
assumes that the degrees perpendicular to the arrows are quickly
relaxing \cite{fara04,bere11}.
More precisely, by considering graph networks as in Fig.~\ref{fig:illustrantion_all}e
we assume that the dynamics is effectively concentrated along
``tubes'' connecting metastable states and/or branches. For example,
complex topological free energy landscapes can display such tube-like
structures \cite{trib12}. Later we will relax the
assumption of paths concentrating along tubes and consider more general types of microscopic dynamics.  
In the following we first outline how one can utilize dynamics on a graph to
understand the emergence of non-Markovian dynamics on networks. 
}



%


%


\paragraph*{Coarse-graining.}
We first consider a sub-graph with 5 states
highlighted in Fig.~\ref{fig:illustrantion_all}e. A continuous trajectory
on the graph is depicted in Fig.~\ref{fig:blinking}a, where the
time runs from bright to dark.
Consider a \emph{gedanken experiment} in which we record a `blinking' whenever the continuous
trajectory enters a node that changes color upon
each state-change (Fig.~\ref{fig:blinking}a and b).
The time-series of state-changes arising from such a
forward-in-time coarse-graining  
is shown in Fig.~\ref{fig:blinking}b alongside recurrences, i.e. re-visitations of nodes (see solid line and crosses, respectively). 
We measure the (local) joint probability density to exit state $i$ after a time $t$ and
enter state $j$, $\wpcStar_{j|i}(t)$.
Its marginal over time -- the so-called
\emph{splitting probability} -- 
is defined as
\begin{equation}
 \phiStar_{j|i}=\int_{0}^\infty\wpcStar_{j|i}(t)\dd t
 \label{eq:phi_marginalization}
\end{equation}
and is
normalized according to $\sum_j\phiStar_{j|i}=1$. Whenever $\wpcStar_{j|i}(t)$
deviates appreciably from an exponential function as in
Fig.~\ref{fig:blinking}c, \emph{the continuous trajectory does not locally
equilibrate in $i$ before changing state to $j$, giving rise to
memory in the coarse-grained dynamics.}
{\color{mynewcolor}The reduced state change depicted in 
 Fig.~\ref{fig:blinking}c forms a semi-Markov process \cite{wang07,mart19a} (see also \cite{land77,mont65,klaf87,haus87}).}

 In the following we explain
the salient features of memory, and the constraints that it
imposes on the construction of thermodynamically consistent network-dynamics.

\subsection{Summary of the main results}\label{subsec:summary_main_results}


As our \emph{first main result} we prove that the splitting probability obeys a reflection identity -- the
\emph{generalization of local detailed balance} (see Sec.~\ref{sec:hysteresis}):
\begin{equation}
 k_{\rm B}T\ln(\phiStar_{j|i}/\phiStar_{i|j})=g(i)-g(j)+\int_i^j \bF(\bx)\cdot\dd \bx,
 \label{eq:local_detailed_balance_like}
\end{equation}
where $k_{\rm B}T$ is the thermal energy, the quantity $
g(i)-g(j)$ (defined in Eq.~\eqref{delG})
is strictly conservative, and the last term denotes the force integrated along the link from node $i$ to node $j$.
Eq.~\eqref{eq:local_detailed_balance_like} connects the
mesoscopic dynamics in the presence of the memory embodied in $\phiStar_{j|i}$ 
 to the microscopic dissipative
force $\bF$ on the underlying graph.
The force $\bF$ may have a globally
non-conservative contribution, and alone
encodes any violation of microscopic reversibility (i.e. detailed balance). {\color{mynewcolor}The last term in Eq.~\eqref{eq:local_detailed_balance_like} allows for an exchange of mechanical or  (electro) chemical energy.}
{\color{mynewcolor2} Eq.~\eqref{eq:local_detailed_balance_like} breaks down if we do hide a way cycles \cite{pugl10,teza20}, which we address in the dedicated Sec.~\ref{subsec:marginal}.}
Crucially, in \textcolor{mynewcolor2}{either of both cases} the presence of memory the coarse-graining into a
discrete-state dynamics must \emph{not} commute with the time reversal,
which gives rise to a phenomenon we refer to as \emph{kinetic hysteresis} that is explained in Sec.~\ref{sec:hysteresis} {\color{mynewcolor}and illustrated in Fig.~\ref{fig:cg_traj2}.
In addition, we explain how the kinetic hysteresis resolves a puzzling
conflict between two mutually contradicting views on irreversibility \cite{wang07} and \cite{mart19a}.}

By means of the \emph{gedanken experiment}
in  Fig.~\ref{fig:blinking}a 
we dissect each waiting time $t$ between two consecutive state-changes
as depicted in Fig.~\ref{fig:blinking}b
into a dwell period $\tau$, spanning the time between the last state-change
until the last recurrence before the next state-change, and the transition-path time $\delta t$, which is the time between the last recurrence and the next state-change. The waiting time becomes the sum $t=\tau+\delta t$.
This decomposition is in fact the \emph{key step towards
  understanding the emergence and manifestations of memory} in network dynamics. 
  As our \emph{second main result} we
  prove the statistical independence of local dwell and transition
  times emerging from an exact coarse-graining of the underlying continuous dynamics (proof shown in Appendix \ref{sec:greens}), i.e. 
\begin{equation}
 \psi_{j|i}(t)\equiv\frac{\wpcStar_{j|i}(t)}{\phiStar_{j|i}}=
 \int_0^t\wpctrans_{j|i}(t-\tau)\wpcdwell_{i}(\tau)\dd\tau,
 \label{eq:local_decomposition}
\end{equation}
where $\wpctrans_{j|i}$ and $\wpcdwell_{i}$ are the
probability densities of transition-path and dwell time,
respectively.\ {\color{mynewcolor}Using Eq.~\eqref{eq:phi_marginalization} one finds that
the independence holds if the new state $j$ is already known.}
Eq.~\eqref{eq:local_decomposition} embodies the following symmetries:
(i) the dwell time is a \emph{state variable} -- it does not
depend on the final state $j$ -- and  (ii) the transition-path time is
reflection-symmetric,
$\wpctrans_{i|j}(t)=\wpctrans_{j|i}(t)$ (see also \cite{bere06}).
We prove both symmetries in
Appendix~\ref{sec:greens}
and illustrate symmetry~(i)
in Fig~\ref{fig:S_catch_hist2}
while symmetry (ii) is demonstrated in Fig.~\ref{fig:S_ATPase_FB_trans}c as well as Table~\ref{tab:ATPase_forward/backward_symmetry}. 
{\color{mynewcolor}Eq.~\eqref{eq:local_decomposition} is somewhat surprising since we find that slow
transition kinetics, i.e.\ $\wpctrans_{j|i}(t)\neq \delta (t)$,  
in fact (seemingly paradoxically) affect the \emph{statistics} of dwell time $\wpcdwell_{i}$.}

As our \emph{third and main practical result}  we derive explicit
formulas for the moments of transition-path- and dwell-time, which are
given in Eqs.~\eqref{eq:trans_methods}-\eqref{eq:binomial}. 
While moments of transition-path times are found to obey recursion integral formulas
\cite{zhan07} we identify redundant integrals in the first two moments
of the transition-path time that can be omitted and, interestingly, lead to an
independent proof of the main finding in Ref.~\cite{sati20}. Moreover,
we derive, for the first time, analytical formulas for the moments of
the dwell-time. The main consequence of
this result is that transitions dictate the amplitude of fluctuations of the waiting time between any consecutive state-change (see Sec.~\ref{subesec:transitions_dictate}).
We apply our main finding to two opposing scenarios. First, we show that
large ``super-Markovian'' fluctuations in the   waiting time (life time)
observed in experiments with catch-bonds
\cite{thom06,buck14} are a unique signature parallel transitions
between states that are \emph{unequally fast}. Second, we show that symmetric stopping-times of the entropy
production (here called waiting time) in stationary driven systems
\cite{neri17,neri19,neri20} automatically imply \emph{equally fast} transition
times, which in turn yields ``sub-Markovian'' dynamics, i.e., 
suppressed waiting time fluctuations. Our theoretical results are
directly applicable to the analysis of experimental time-series.

{\color{mynewcolor}
\subsection{Relation to previous works and nomenclature}
%
%
%

Before we derive our results let us briefly explain how the three main
quantities, transition-path time $\delta t$, dwell time $\tau$ and
waiting time $t$ (see Fig.~\ref{fig:blinking}b), relate to, and
appeared in, previous works. The following paragraph summarizes the different
terminologies used across the disciplines, which in turn helps us to identify
and clarify the core of the conflict between Refs.~\cite{wang07} and \cite{mart19a}.

First, the term waiting time \cite{wang07} frequently appears 
 under the terms (conditional) first passage time \cite{beni09,bere19}, life time \cite{bere19}, stopping time \cite{neri17}, residence time, folding time \cite{chun13}, and even dwell time \cite{kolo05,kolo07,zijl20} or cycle time \cite{kolo07}.
 Note that the waiting  and dwell time coincide
once the transtion-paths become instantaneous as in
\cite{kolo05,kolo07}, and the inverse of the mean waiting time is also
called Kramer's reaction velocity (rate) \cite{kram40}. As illustrated in Fig.~\ref{fig:illustrantion_all}d  the life-time of a (catch) bond \cite{bart02,mars03,thom06,thom08a,buck14} represents a waiting time in the bound state.
Second, the transition-path time
\cite{chun12,neup16a,coss18,glad19,kim20} is sometimes also referred
to as transition-event duration \cite{zhan07}, translocation time \cite{bere06}, and occasionally transition time \cite{neup12,kim20} or transit time \cite{neup16a}.
Third, the dwell time is also referred as residence time \cite{kim15} or ``loops'' \cite{bere18,maka21}.
The distinction between these three quantities is important for understanding the following puzzling conflict \textcolor{mynewcolor2}{between the fundamental notion ``irreversibility'' \cite{wang07,mart19a}}.

{\color{mynewcolor2} A trajectory satisfying Hamilton equation of motion is physically reversible but mathematically irreversible. That is, if we na\"ively mathematically revert in time a phase space trajectory  satisfying Hamilton equation, the resulting trajectory will violate the equation of motion, unless one ``physically''  takes into account the well-known fact that momenta change sign under time reversal. In thermodynamics this reversibility translates into the concept of detailed balance, which implies that
 probability of any path
is identically to the probability of the \emph{physically}  time-reverted path.

A network with clear time scale separation as in Fig.~\ref{fig:illustrantion_all}a violating detailed balance
 may still ``locally equilibrate'' with all connected reservoirs prior to changing to the next state in the network. In this case local detailed balance relation  connects the forward and backward transition probability (or rate) to the entropy flux \cite{seif11,maes21}. If links between states connected to different reservoirs the network can maintain fluxes between them in a non-equilibrium steady state that breaks time-reversal symmetry \cite{seif12}. If  a local equilibration ceases to exist this connection becomes more subtle. In particular
}
the deep connection between breaking of detailed balance and the
breaking of \textcolor{mynewcolor2}{(mathematical)} time-reversal symmetry in semi-Markov processes (as
depicted in Fig.~\ref{fig:blinking}b) \cite{mart19a} has been put into
question in Ref.~\cite{wang07}.  In fact the example in Fig.~\ref{fig:blinking} turns out to invalidate
the main conclusion of Ref.~\cite{mart19a}. In order to restore the
view put forward in Ref.~\cite{mart19a} we find that 
\emph{transition paths must be ``odd'' under time reversal}, which
gives rise to a phenomenon we call \emph{kinetic hysteresis}. \textcolor{mynewcolor2}{It turns out the kinetic hysteresis restores the connection between the  breaking of time-reversal symmetry and the braking of detailed balance.}}

\subsection{Structure of the article}\label{subsec:structure}
The remainder of this article is structured as follows. In section
\ref{sec:model} we define diffusion on a graph depicted in
Fig.~\ref{fig:blinking} along with the precise coarse-graining into digitized
states. We discuss the limitations of the coarse-graining, define
transition-path time and dwell-time functionals, and explain their independence
and symmetries that follow from 
Eq.~\eqref{eq:local_decomposition}. In
Sec.~\ref{subsec:main_practical} we present the main practical result.
In Sec.~\ref{sec:hysteresis} we derive
Eq.~\eqref{eq:local_detailed_balance_like} 
and prove the thermodynamic consistency of the coarse-graining.
Surprisingly we
find in Sec.~\ref{subsec:hysteresis} that the coarse-graining must not commute with time-reversal,
which gives rise to a \emph{kinetic hysteresis}. {\color{mynewcolor}The kinetic
hysteresis turns out to reconcile two contradicting views on the
thermodynamics of irreversibility, namely those between \cite{wang07}
and \cite{mart19a}.}
The central  implications of
Eqs.~\eqref{eq:phi_explicit}-\eqref{eq:binomial} are discussed in
Sec.~\ref{sec:three_noises}, where we identify three fundamental
sources of noise in the waiting time 
and explain the practical implications of deviations from
Markovianity, in particular the emergence of sub-Markovian fluctuations in driven periodic systems in
Sec.~\ref{subsec:sub_Markov} and super-Markovian fluctuations in the
presence of parallel transition paths in
Sec.~\ref{subsec:super_Markov}. 
{\color{mynewcolor}Sec.~\ref{sec:cg_general} provides a broader perspective on our
results including the relation between the coarse-graining and
milestoning \cite{schu11,bere19} (see Ref.~\cite{elbe20,elbe20a} for a
more elaborate expose).}
We conclude in Sec.~\ref{sec:conclusion}.

The derivations are rather involved and therefore relegated to a
series of Appendices.
Details about stochastic differential equations on a graph and their numerical implementation are given in Appendix \ref{sec:S_stoch_dyn}. 
The proof of Eq.~\eqref{eq:local_decomposition} along with the
entailed symmetries is shown in Appendix~\ref{sec:greens}.
{\color{mynewcolor}Our results are derived on the basis of a novel decomposition of paths 
shown in Appendix~\ref{sec:path_decomposition}, which represents a generalization of the renewal theorem \cite{sieg51}.}
The quite lengthy and tedious derivation of
Eqs.~\eqref{eq:phi_explicit}-\eqref{eq:binomial} is explained in 
the Supplemental Material (SM) \footnote{See Supplementary Material for the derivation of the main practical result}. 
Eq.~\eqref{eq:local_detailed_balance_like} is proven in
Appendix~\ref{sec:SI_thermodynamic_consistency} and the symmetries
tested
in Appendix~\ref{sec:SI_examples}.

\section{Model} \label{sec:model}
\subsection{Diffusion on a graph}\label{subsec:model_graph}

\begin{figure}
\includegraphics[width=\columnwidth]{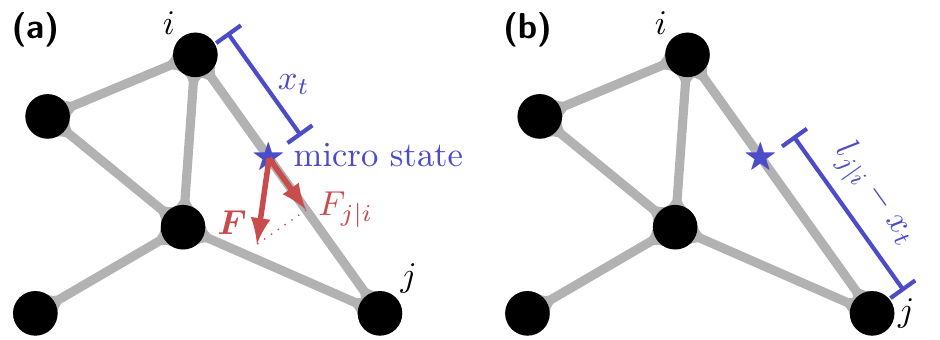}
\caption{Microstate versus network state on the path between nodes $i$ and $j$ separated by a distance $l_{j|i}$. (a)~The microstate $\bx\equiv(x_t,j,i)$
at distance $x_t$ from node $i$ in the direction towards node $j$ is driven by a force $\bF$, where  $F_{j|i}$ denotes the parallel component of the force. (b)~The
microstate measured by the distance $\tilde x_t=l_{j|i}-x_t$ from node
$j$ is equivalent to the one from (a), i.e.,  $\bx=(x_t,j,i)$ and $\tilde\bx=(\tilde x_t,i,j)$ are equivalent.}
\label{fig:network_microstate_redundancy}
\end{figure}

The full system's dynamics is assumed to
evolve as piece-wise continuous space-time Markovian diffusion on a
graph as shown in Fig.~\ref{fig:blinking}a with potential (weak) discontinuities at the set of all nodes $i\in\Omega$. We denote all neighbor-nodes of
$i$ by $\mathcal{N}_i\subset \Omega$. For example, in Fig.~\ref{fig:blinking}a the set of neighboring states of state 2 are $\mathcal{N}_2=\{1,3,4\}$.
At any time $t$ between the last passage by $i$ in the direction of
$j$ at time $t_{\rm ini}$ and the next visit
of a node $j\in\mathcal{N}_i$ or the return to $i$ at time $t_{\rm fin}$, i.e. $t_{\rm ini}<t<t_{\rm fin}$, the system is assumed to
satisfy the anti-It\^o (or H\"anggi-Klimontovich \cite{haen82,klim90}) Langevin  equation 
\begin{align}
 \dot x_{t}=\beta D_{j|i}(x_t)F_{j|i}(x_t)+\sqrt{2D_{j|i}(x_{t})}\circledast\xi_t,
  \label{eq:Langevin_def}
\end{align}
where $x_t$ (see Fig.~\ref{fig:network_microstate_redundancy}a) denotes the instantaneous distance from node $i$ in the
leg $i-j$ with $0<x_t<l_{j|i}$, $D_{j|i}(x)$ and $F_{j|i}(x)$ are the diffusion landscape and
force field along the leg directed from $i$ to $j$, respectively,
$\beta\equiv 1/(k_{\rm B}T)$, $\xi_t$ is standard Gaussian
white noise with zero mean, i.e. $\avg{\xi_t}=0$ and
$\avg{\xi_t\xi_{t'}}=\delta (t-t')$. The symbol ``$\circledast$'' denotes the
anti-It\^o product (see Appendix \ref{subsec:S_stoch_dyn_traj}) and $l_{j|i}=l_{i|j}$ denotes the length of the path connecting nodes $i$ and $j$ (see Fig.~\ref{fig:network_microstate_redundancy}).


Once a node $i$ is reached from within a leg 
the
consecutive leg is chosen, without loss of generality, randomly from the set of all neighbors
$j'\in\mathcal{N}_i$ with equal probability, i.e., the microstate  $\bx=(x_t,j,i)$ in Fig.~\ref{fig:network_microstate_redundancy}a changes
$(0,j,i)\to (0,j',i)$.
 Thereupon the dynamics
again evolves according to Eq.~\eqref{eq:Langevin_def} until the next visit of
a node.
Similarly, as soon as the node $j$ is reached, the microstate
 changes to $(l_{j|i},j,i)\to (0,k,j)$ with $k$ randomly chosen among the neighbors of node $j$  ($k\in\mathcal{N}_j$) with equal probability.
 This  fully specifies the full system's dynamics.
In Appendix~\ref{subsec:FPE} we translate the Langevin equation \eqref{eq:Langevin_def} into a Fokker-Planck equation and in 
%
Appendix~\ref{subsec:SI_discontinuous_landscapes}
we explain in detail how  one 
can account for discontinuities in the diffusion landscape and force-field.

Three remarks are in order.
First, in what follows we assume the Langevin equation \eqref{eq:Langevin_def}
to determine the time evolution of the microstate
along any link depicted by the gray lines in
Fig.~\ref{fig:network_microstate_redundancy}.  It is shown that such a
dynamics naturally emerges when possibly hidden degrees of freedom
(perpendicular to the gray lines) are quickly relaxing \cite{bere11},
that is, the full system's trajectories concenrate along ``tubes''. 
Strikingly, ignoring a possibly
higher dimensional embedding renders the diffusion coefficient spatially dependent \cite{bere11} due to entropic effects. Diffusion on a graph 
also emerges from
Hamiltonian dynamics weakly coupled to a heat bath \cite{frei93}.

Second, the Langevin equation \eqref{eq:Langevin_def} can globally
violate detailed balance. Nevertheless, for any $i\in\Omega$ and
$j\in\mathcal{N}_i$ the force translates along any link into the \emph{local} potential, 
\begin{equation}
 U_{j|i}(x)\equiv-\int_0^xF_{j|i}(y)dy
 \label{eq:local_potential}
\end{equation}
for  $0<x<l_{i|j}$. The representation of the microstate by design entails a redundancy, meaning that  $\bx=(x_t,j,i)$ and $\tilde \bx=(l_{j|i}-x_t,i,j)$
are the \emph{same} microstate (see Fig.~\ref{fig:network_microstate_redundancy}).
This imposes the following reflection symmetries:
$D_{i|j}(x)=D_{j|i}(l_{j|i}-x)$ as well as
$F_{i|j}(x)=-F_{j|i}(l_{j|i}-x)$, and hence
$U_{i|j}(x)=U_{j|i}(l_{j|i}-x)-U_{j|i}(l_{j|i})$. If a \emph{global}
potential $\mathcal U$ exists, that is $U_{j|i}(l_{j|i})=\mathcal{U}_j-\mathcal{U}_i$,
$\forall
i,j\in \Omega$ with $j\in\mathcal{N}_i$ the dynamics is said to obey \emph{detailed
balance}.
Conversely, if no such global potential exist microscopic
reversibility is said to be broken
(see also Appendix \ref{subsec:SI_cycles}).

Third,  we
propagate Eq.~\eqref{eq:Langevin_def} numerically
using the
Milstein scheme provided in Appendix~\ref{subsubsec:Milstein}
 whenever the diffusion coefficient is non-constant
 ``$D_{j|i}(x)\neq\text{const}$''. Otherwise we use the stochastic
 Runge-Kutta scheme \cite{chan87a} described in Appendix~\ref{subsubsec:Runge_Kutta}.


%
%
\subsection{Coarse-graining to state-changes on a network}
\label{subsec:coarse-graining}

According to the \emph{gedanken experiment} outlined in
Fig.~\ref{fig:blinking}
the continuous trajectory is coarse-grained into a time-series of
recurrences and state-changes on a
network. Consecutive visits of the continuous trajectory of the same node correspond to
recurrences (see colored crosses in Fig.~\ref{fig:blinking}a and b), whereas
transitions between distinct nodes yield state-changes (see black
crosses in Fig.~\ref{fig:blinking}a and line in Fig.~\ref{fig:blinking}b).  
In-between two consecutive state-changes the reduced network-state
remains in the initial state (see Fig.~\ref{fig:blinking}b). This exactly specifies the coarse-grained
trajectory on the network.

The dwell time $\tau$ corresponds to the sum of all consecutive
recurrence times $t_{r}$ since the last state-change. One dwell period
is highlighted in Fig.~\ref{fig:blinking}b. The transition
time $\delta t$ corresponds to the time between the last recurrence and the
instance of the state-change. The local \emph{waiting time} $t$ for a transition
$i\to j$ is formally the sum of the dwell time and transition-path time, $t=\delta t+\tau$, and
corresponds to the time-interval between two consecutive first
entrances of nodes. Since the complete dynamics is stochastic these
quantities correspond to random variables. The joint probability
density of a waiting time at $i$ and consecutive transition to $j$ corresponds to $\wpcStar_{j|i}(t)$, and the
dwell and transition-path time are distributed according to  $\wpcdwell_{i}(\tau)$ and
$\wpctrans_{j|i}(\delta t)$, respectively. 
Precise formal definitions of the waiting, dwell
and transition-path time functionals entering
Eq.~\eqref{eq:local_decomposition} are given in the following subsection. 
In Appendix \ref{subsubsec:functionals} we explain the numerical evaluation of both, dwell and transition-path time.

Let us briefly discuss the \emph{strengths and the limitations of the coarse-graining}.
First, as we show Sec.~\ref{sec:hysteresis}  the coarse graining
\emph{preserves the thermodynamic entropy production}. Second, the
coarse-graining \emph{preserves node-to-node currents}
\cite{bara15,horo20} and its fluctuations which are discussed in
Sec.~\ref{subsec:sub_Markov}. Third,  \emph{first passage functionals
of the full, microscopic model are preserved} (see Appendix~\ref{subsec:from_starlike}).
These are, for instance, crucial for catch-bond rupture-experiments carried out in \cite{bart02,mars03,thom06,thom08a,buck14} (see Sec.~\ref{subsec:super_Markov}).
Fourth, the coarse-graining \emph{retains vital information} \cite{sati20} \emph{encoded in transition-path times},
which are the key to understanding the emergence of memory in the network-kinetics that we discuss in Sec.~\ref{sec:three_noises}.

However, the coarse-graining has one \emph{shortcoming}. Due to the
aforementioned  redundancy (see
Fig.~\ref{fig:network_microstate_redundancy}) the \emph{statistics of occupation time} \cite{lapo20} (also known as ``local time'' or ``empirical density'' \cite{bara15c})
\emph{is not preserved}.  Occupation-time statistics on graphs  were studied, for example, in \cite{beni09}. We note that in the presence of a time scale separation
transitions become effectively instantaneous 
(see Fig.~\ref{fig:illustrantion_all}a),
and in this limit occupation times within the meta-stable regions are
preserved, while concurrently the kinetics becomes memory-less (see Sec.~\ref{subsec:limit:LDB}).

\subsection{Definition of transition-path-time and dwell-time functionals}\label{subsec:def_trans}
Using the gedanken experiment depicted in Fig.~\ref{fig:blinking}b
we define the \emph{dwell time} $\tau$ as the time between the first ``blinking'' and the last ``blinking'' (last recurrence) of the same color (state), while the \emph{transition-path time} denotes the time-span between
the last recurrence and the first following change of color. While the gedanken experiment allows for an intuitive definition of the dwell and transition-path time, we now provide precise formal definitions that allow us to relate 
the gedanken experiment to existing definitions of transition-path times  \cite{humm04}.

The transition path from node $i$ to node $j$ starts with the last recurrence to node $i$ and ends with the first visit of  another node $j$ conditioned that $i$ has not been visited in between.
Suppose that $x_t$ denotes the distance from node $i$ towards node $j$
satisfying the Langevin equation \eqref{eq:Langevin_def} between said nodes.
Then the transition-path time is defined as the random variable \cite{humm04}
\begin{equation}
 \delta t= \lim_{y\to 0}\inf_t\{t|x_t=l_{j|i}\wedge x_0=y \wedge (0<x_\tau\;\forall\; 0\le \tau\le t)\}
 \label{eq:trans-time_def}
\end{equation}
whose probability density function is denoted by $\wpctrans_{j|i}(\delta t)$.
Note that an unsuccessful transition attempt terminates as soon as
$x_t=0$, whereas a transition is successfully completed once
$x_t=l_{j|i}$. Since transitions correspond to successful attempts only, we need to discard all unsuccessful attempts by introducing the transition Green's function, $G_{j|i}^{\rm tr}$, defined as follows.

The probability density starting from $x_0=y$ to be found 
after time $t$ at distance $x$ from node $i$ in direction to node $j$,
while never having either returned to state $i$ or reached state $j$,
will be denoted by $G_{j|i}^{\rm tr}(x,t|y)$.
The probability density satisfies  the
initial condition $G_{j|i}^{\rm tr}(x,0|y)=\delta(x-y)$.
We translate the Langevin equation \eqref{eq:Langevin_def}
into a Fokker-Planck equation \cite{gard04} (see also Appendix~\ref{subsec:FPE})
$\del_tG_{j|i}^{\rm tr}(x,t|y)=-\del_x\dLF_{j|i}(x) G_{j|i}^{\rm tr}(x,t|y)$, where $\dLF_{j|i}(x)\equiv \beta D_{j|i}(x)F_{j|i}(x)-D_{j|i}(x)\del_x$ is the current operator
and the boundary conditions are absorbing $G_{j|i}^{\rm tr}(0,t|y)=G_{j|i}^{\rm tr}(l_{j|i},t|y)=0$.
The absorbing boundaries effectively terminate the process once either of the nodes $i$ or $j$ is reached. The transition-path time statistics are determined by
taking
the limit of
successful trajectories in Eq.~\eqref{eq:trans-time_def}, that is,
$y\to 0$ (starting from node $i$) and $x\to l_{j|i}$ (ending in node $j$). The corresponding probability density of transition-path time reads
\begin{equation}
 \wpctrans_{j|i}(\delta t)=\lim_{y\to0}\lim_{x\to l_{j|i}}\frac{\dLF_{j|i}(x) G_{j|i}^{\rm tr}(x,\delta t|y)}{\int_0^\infty \dLF_{j|i}(x) G_{j|i}^{\rm tr}(x,t|y)\dd t}.
 \label{eq:ptrans_Gtr_limit}
\end{equation}

The dwell time is defined as follows.
First, we define in node $i$ the
state $j$-dependent conditional first passage time $t_j=\inf\{t|x_0=0\wedge x_t=l_{j_t|i}\}$,
where
 the index $j=j_t$ denotes the randomly chosen state
following state $i$, which in turn can be used to
define the dwell time as
\begin{equation}
 \tau=\sup_t\{t|x_0=0\wedge x_t=0\wedge t\le t_j\}.
  \label{eq:dwell-time_def}
\end{equation}
The probability density of dwell time is denoted by $\wpcdwell_{j|i}(\tau)$.
We prove in Appendix~\ref{sec:greens} that the dwell time $\tau$
 has in fact the \emph{same} distribution for all final states $j$, which
 is manifested in the property
 $\wpcdwell_{j|i}(\tau)=\wpcdwell_{i}(\tau)$ -- symmetry (i)
 in Eq.~\eqref{eq:local_decomposition}. Moreover, we prove in
 Appendix~\ref{sec:greens}  the independence of dwell and
 transition-path times, which allows us to represent the probability density of the conditional waiting time as a convolution as in Eq.~\eqref{eq:local_decomposition}.

\subsection{Main practical result} \label{subsec:main_practical}
A straightforward translation of Eq.~\eqref{eq:Langevin_def} into a
Fokker-Planck equation with appropriate boundary and internal
continuity conditions allows us to obtain explicit results for the
splitting probability and the statistics of dwell and transition-path time,
which follow from some quite tedious algebra
(see SM \cite{Note2}).

More precisely, in
the Supplemental Section 1 in \cite{Note1}
we show that the path decomposition from Appendix~\ref{sec:path_decomposition} can be ``inverted''
to conveniently derive the statistics of waiting time $\wpcStar_{j|i}(t)$,
which after insertion of the results derived in
the Supplemental Section 2 in \cite{Note2}
-- so-called unconditioned first passage times --  finally yields the main practical result, Eqs.~\eqref{eq:phi_explicit}-\eqref{eq:binomial}, as shown in 
the Supplemental Section 3 in \cite{Note2}

For convenience we introduce the following essential auxiliary integrals
\begin{equation}
 I_{j|i}^{(k)}
 \equiv\int_0^{l_{j|i}}\dd y_1\int_0^{y_1}\dd y_2\ldots\int_0^{y_{k-1}}\dd y_kg_{j|i}^{(k)},
\label{eq:Idef}
\end{equation}
where $g_{j|i}^{(k)}$ are depicted in Tab.~\ref{tab:g_def}, with the local potential $U_{j|i}$ defined in Eq.~\eqref{eq:local_potential}. In the following we require only the first five integrals  $I_{j|i}^{(k)}$ ($k=1,\ldots,5$).
\begin{table}
\caption{Integrands entering Eq.~\eqref{eq:Idef} at a glance.}
 \label{tab:g_def}
 \centering
\begin{tabular}{rc}
 $k$&$g^{(k)}_{j|i}\phantom{\big|^{M}}$\\[0.6mm]\hline \\[-3mm]
 $1$&$\dfrac{\e^{\beta U_{j|i}(y_1)}}{D_{j|i}(y_1)}$ \\[2.5mm]
 $2$&$\dfrac{\e^{\beta U_{j|i}(y_1)-\beta U_{j|i}(y_2)}}{D_{j|i}(y_1)}$
 \\[2.5mm]
 $3$&$\dfrac{\e^{\beta U_{j|i}(y_1)-\beta U_{j|i}(y_2)+\beta U_{j|i}(y_3)}}{D_{j|i}(y_1)D_{j|i}(y_3)}$
 \\[2.5mm]
 $4$&$\dfrac{\e^{\beta U_{j|i}(y_1)-\beta U_{j|i}(y_2)+\beta U_{j|i}(y_3)-\beta U_{j|i}(y_4)}}{D_{j|i}(y_1)D_{j|i}(y_3)}$
 \\[2.5mm]
 $5$&$\dfrac{\e^{\beta U_{j|i}(y_1)-\beta U_{j|i}(y_2)+\beta U_{j|i}(y_3)-\beta U_{j|i}(y_4)+\beta U_{j|i}(y_5)}}{D_{j|i}(y_1)D_{j|i}(y_3)D_{j|i}(y_5)}$\\[2.5mm]\hline
\end{tabular}

\end{table}
Using the auxiliary integrals in Eq.~\eqref{eq:Idef} the splitting probabilities read
\begin{equation}
 \phiStar_{j|i}=
\bigg( \sum_{k\in \mathcal{N}_i}I_{j|i}^{(1)}/I_{k|i}^{(1)}\bigg)^{-1}
 \label{eq:phi_explicit}
\end{equation}
and the first two moments of the transition-path time become
\begin{equation}
\avgtrans{\delta t}_{j|i}=\frac{I_{j|i}^{(3)}}{I_{j|i}^{(1)}} \quad\text{and} \quad
\avgtrans{\delta t^2}_{j|i}=2(\avgtrans{\delta t}_{j|i})^2-2\frac{I_{j|i}^{(5)}}{I_{j|i}^{(1)}},
\label{eq:trans_methods}
\end{equation}
where the second moment is generally sub-Markovian, i.e. $\avgtrans{\delta t^2}_{j|i}\le 2(\avgtrans{\delta t}_{j|i})^2$. See also Ref.~\cite{sati20} for an alternative proof, where $\avgtrans{\delta t^2}_{j|i}\le 2(\avgtrans{\delta t}_{j|i})^2$ corresponds to a coefficient of variation being smaller than one.
Some further extended algebra yields the first two moments of the average local dwell time
\begin{align}
  \avgdwell{\tau}_i&=\sum_{k\in \mathcal{N}_i}\phiStar_{k|i}\Big[I_{k|i}^{(2)}-\avgtrans{\delta t}_{k|i}\Big],
  \nonumber\\
  \avgdwell{\tau^2}_i&=2(\avgdwell{\tau}_i)^2
  \label{eq:dwell_methods}
  \\&+\sum_{k\in \mathcal{N}_i}\phiStar_{k|i}\Big[2I_{k|i}^{(2)}\avgtrans{\delta t}_{k|i}
  -2I_{k|i}^{(4)}-\avgtrans{\delta t^2}_{k|i}\Big],
  \nonumber
\end{align}
wherefrom follows the variance of the dwell time $\sigma^2_{{\rm
    dwell}, i}=\avgdwell{\tau^2}_i-(\avgdwell{\tau}_i)^2$.
The independence of dwell and transition-path times in Eq.~\eqref{eq:local_decomposition}
immediately yields the binomial sum for the $n$-th moment of the local first passage time
\begin{equation}
\avgloc{t^n}_{j|i}=\sum_{l=0}^n\binom{n}{l}\avgloc{\delta t^l}_{j|i}\avgloc{\tau^{n-l}}_{i},
\label{eq:binomial}
\end{equation}
where the forward/backward symmetry implies $\avgloc{\delta
  t^l}_{j|i}=\avgloc{\delta t^l}_{i|j}$. The $n$-th moment of the exit
time is then simply given by $\avg{t^n}^{\rm
  exit}_i=\sum_k\phiStar_{k|i}\avgloc{t^n}_{k|i}$, yielding the variance
$\sigma_{{\rm exit},i}^2=\avgexit{t^2}_i-(\avgexit{t}_i)^2$. 
Note that $\avgexit{t}_i$ is given in
Eq.~(S47)
 and 
$\avgexit{t^2}_i$ can be found in 
Eq.~(S50)
in the SM \cite{Note2}.
According to
Eq.~\eqref{eq:binomial} the latter can be decomposed into three noise contributions
$\sigma_{{\rm exit},i}^2=\sigma_{{\rm dwell},i}^2+\sigma_{{\rm tr,
    int},i}^2+\sigma_{{\rm tr,ext},i}^2$, where $\sigma_{{\rm
    dwell},i}^2=\avgdwell{\tau^2}_i-(\avgdwell{\tau}_i)^2$, the
intrinsic noise due to transition-path time is given $\sigma_{{\rm tr, int},i}^2=\sum_k\phiStar_{k|i}[\avgtrans{\delta t^2}_{k|i}-(\avgtrans{\delta t}_{k|i})^2]$,
and the extrinsic noise among different transition paths is given by
$\sigma_{{\rm tr, ext},i}^2=\sum_k \phiStar_{k|i}(\avgtrans{\delta t}_{k|i})^2-(\sum_k \phiStar_{k|i}\avgtrans{\delta t}_{k|i})^2$.

Eqs.~\eqref{eq:phi_explicit}-\eqref{eq:binomial} are the main practical result of this paper. Notably, in Eq.~\eqref{eq:dwell_methods} we determine, for the first time, the moments of dwell time.  We emphasize that the results Eqs.~\eqref{eq:phi_explicit}-\eqref{eq:binomial} \emph{contain no redundant integrals} that were eliminated in a quite tedious calculation shown in the SM \cite{Note2} (See Supplementary Section 2.D and 3) 
This final step is crucial for the derivation of the main result in
Sec.~\ref{sec:three_noises}. Moreover, due to the positivity of the surviving auxiliary  integrals \eqref{eq:Idef}, Eq.~\eqref{eq:trans_methods} provides an independent proof of the main finding of Ref.~\cite{sati20}.
\textcolor{mynewcolor2}{If the network contains infinitely long legs ($l_{j­|i}\to\infty$) the auxiliary integrals diverge leading to divergent moments of dwell time, which in turn may trigger interesting phenomena such as anomalous diffusion \cite{metz98,bark00,soko05}. In this case the independence between transition time and dwell time, Eq.~\eqref{eq:local_decomposition}, is expected to still hold. }
We now address  the thermodynamic consistency of the coarse-graining.


%


%
%
%

\section{Thermodynamic consistency of the coarse-graining}
\label{sec:hysteresis}
\subsection{Splitting probability  encodes thermodynamics}%
\label{subsec:splitting_thermodynamics}%
In this section we derive our first main result, Eq.~\eqref{eq:local_detailed_balance_like}, and  explain its implications. In particular, we show that the coarse-graining into the reduced-state dynamics
preserves the dissipation (entropy production) of the underlying microscopic continuous dynamics in the presence of memory. The emergence of a \emph{kinetic hysteresis} is discussed in the following section.
Using Eq.~\eqref{eq:phi_explicit}
one  obtains
\begin{equation}
 \ln\frac{\phiStar_{j|i}}{\phiStar_{i|j}}=\ln\Big[\sum_{k\in\mathcal{N}_j}\frac{1}{I_{k|j}^{(1)}}\Big]
 -\ln\Big[\sum_{k\in\mathcal{N}_i}\frac{1}{I_{k|i}^{(1)}}\Big]
 +\ln\frac{I^{(1)}_{i|j}}{I^{(1)}_{j|i}},
 \label{eq:local_detailed_balance_like_pre}
\end{equation}
where $I_{j|i}^{(1)}=\int_0^{l_{j|i}}D_{j|i}(x)^{-1}\e^{\beta U_{j|i}(x)}\dd x$ [cf. Eq.~\eqref{eq:Idef} and Tab.~\ref{tab:g_def}].
To derive Eq.~\eqref{eq:local_detailed_balance_like} we multiply Eq.~\eqref{eq:local_detailed_balance_like_pre} by the thermal energy
and define 
\begin{equation}
g(\alpha)\equiv-k_{\rm B} T\ln\Big[\sum_{k\in\mathcal{N}_\alpha}\frac{1}{I_{k|\alpha}^{(1)}}\Big]
\label{delG},
\end{equation}
with $\alpha=i,j$. It remains to be shown that the  last term in Eq.~\eqref{eq:local_detailed_balance_like_pre}
is in fact
the force integrated along the path starting from node $i$ and and ending in node $j$ as in Eq.~\eqref{eq:local_detailed_balance_like}, which we prove in the following paragraph.

Using the the auxiliary integrals from Eq.~\eqref{eq:Idef}
we find
\begin{multline}
 \ln\frac{I^{(1)}_{i|j}}{I^{(1)}_{j|i}}
 =\ln\left[\frac{\int_0^{l_{j|i}}\frac{\e^{\beta U_{i|j}(x)}}{D_{i|j}(x)}\dd x}{\int_0^{l_{j|i}}\frac{\e^{\beta U_{j|i}(x)}}{D_{j|i}(x)}\dd x}\right]
 \\
=
 \ln\left[\frac{\int_0^{l_{j|i}}\frac{\e^{\beta U_{j|i}(l_{j|i}-x)-\beta U_{j|i}(l_{j|i})}}{D_{j|i}(l_{j|i}-x)}\dd x}{\int_0^{l_{j|i}}\frac{\e^{\beta U_{j|i}(x)}}{D_{j|i}(x)}\dd x}\right]
 =-\beta U_{j|i}(l_{j|i})
 ,
 \label{eq:LDB_like_last_term1}
\end{multline}
where in the second line we used the symmetries $D_{i|j}(x)=D_{j|i}(l_{j|i}-x)$ and $U_{i|j}(x)=U_{j|i}(l_{j|i}-x)-U_{j|i}(l_{j|i})$, which are discussed in the paragraph following Eq.~\eqref{eq:local_potential} in Sec.~\ref{subsec:model_graph}, and in the last step we  use the fact that
the integrals are identical up to the constant $\e^{-\beta U_{j|i}(l_{j|i})}$. We  use $\beta=1/(k_{\rm B}T)$ and insert the definition of the local potential in Eq.~\eqref{eq:local_potential}, $U_{j|i}(l_{j|i})=-\int_0^{l_{j|i}} F_{j|i}(x)\dd x$, into 
 Eq.~\eqref{eq:LDB_like_last_term1}
to  obtain
\begin{equation}
k_{\rm B}T \ln\frac{I^{(1)}_{i|j}}{I^{(1)}_{j|i}}
=\int_0^{l_{j|i}} F_{j|i}(x)\dd x =\int_{i}^{j}\bF(\bx)\cdot\dd \bx,
 \label{eq:LDB_like_last_term2}
\end{equation}
where in the last step we identified $F_{j|i}$ as the component of the force $\bF$ along the link $i\to j$ (see Fig.~\ref{fig:network_microstate_redundancy}a). 
Inserting Eqs.~\eqref{delG} and \eqref{eq:LDB_like_last_term2} into
Eq.~\eqref{eq:local_detailed_balance_like_pre} finally yields
Eq.~\eqref{eq:local_detailed_balance_like}, which completes the proof of the first main result.

\subsection{Entropy production rate}\label{subsec:entropy}

It is important to understand why Eq.~\eqref{eq:local_detailed_balance_like} in fact encodes thermodynamic consistency, that is, why the coarse-graining preserves the total entropy production rate of the underlying system \emph{at long times}.
During a long time interval of length $t$ we will observe $n_{j|i}(t)$
transitions  from state $i$ to state $j$, which is a random number that in the limit of long times
displays a non-negative stationary probability flow
$\dot n_{j|i}\equiv\lim_{t\to\infty} n_{j|i}(t)/t\ge0$.
This stationary probability flow (see, e.g.,  Ref.~\cite{gira03,mart19a})
can be calculated from
$\dot n_{j|i}=\phiStar_{j|i}\pi_i/\sum_k{\avgexit{t}_k\pi_k}$,
where $\boldsymbol{\pi}$ is the unit eigenvector of the splitting
matrix, i.e. $\pi_j=\sum_i\phiStar_{j|i}\pi_i$.
While detailed balance implies $\dot n_{j|i}-\dot n_{i|j}=0$, the
violation of this equality, i.e.\
$\dot n_{j|i}\neq \dot n_{i|j}$, reflects  a genuine breaking of detailed balance. 
Using the force-field along the continuous graph
one can conveniently express the entropy production of the microscopic dynamics as
\begin{equation}
 \dot S^{\rm ss}\equiv\sum_{i,j}\dot n_{ij}\int_i^j\dd \bx\cdot \bF/T,
 \label{eq:entropy_detailed}
\end{equation}
where $\int_i^j\dd \bx\cdot \bF$ is the dissipated ``work'' along a transition $i\to j$ at temperature $T$.
Inserting Eq.~\eqref{eq:local_detailed_balance_like} into Eq.~\eqref{eq:entropy_detailed}
we obtain
\begin{align}
 \dot S^{\rm ss}&=k_{\rm B}\sum_{i,j}\dot n_{j|i}\bigg[\frac{g(j)}{T}-\frac{g(i)}{T}+\ln\frac{\phiStar_{j|i}}{\phiStar_{i|j}}\bigg]
 \nonumber\\
 &=k_{\rm B}\sum_{i,j}\dot n_{j|i}\ln\frac{\phiStar_{j|i}}{\phiStar_{i|j}}\label{eq:entropy_cg}
 ,
\end{align}
where in the final step we used Kirchhoff's law, stating that all incoming flows
and outgoing flows are conserved, i.e.  $\sum_j \dot n_{i|j} =\sum_j \dot n_{j|i} $. We emphasize that Eq.~\eqref{eq:entropy_cg}
allows us to express the entropy production
of the underlying microscopic dynamics in Eq.~\eqref{eq:entropy_detailed}
solely in terms of the coarse-grained network dynamics ($\phi_{j|i}$ and $\dot n_{j|i}$). This
renders the coarse-graining (see Fig.~\ref{fig:blinking}) thermodynamically consistent. 
Morever,
Eq.~\eqref{eq:entropy_cg} explicitly does \emph{not} require the underlying microscopic force field $\bF$ 
entering the right hand side of Eq.~\eqref{eq:entropy_detailed} to be known.

A few additional remarks are in order. 
First, the coarse-graining preservers the stationary entropy production rate since it does not hide cycles \cite{pugl10}.
The preservation of cycles by the coarse-grainig is
explicitly explained in Appendix~\ref{subsec:SI_cycles}.
{\color{mynewcolor}Possible extensions to the theory including hidden cycles will be discussed in Sec.~\ref{sec:cg_general}.}
\textcolor{mynewcolor2}{At the presence of hidden cycles Eq.~\eqref{eq:entropy_cg} will expected to  underestimate the entropy production rate.}
Second, in the limit
of a time scale separation Eq.~\eqref{eq:entropy_cg} coincides with the 
entropy production in Markov networks \cite{schn76}, whereby
Eq.~\eqref{eq:entropy_detailed} encapsulates the local detailed
balance relation \cite{seif12}. In the following subsection we briefly
address this limit, which arises in the presence of high local free
energy barriers that in turn yield memory-less kinetics.


\subsection{The peculiar limit of local detailed balance}
\label{subsec:limit:LDB}

\begin{figure}
 \centering
  \includegraphics{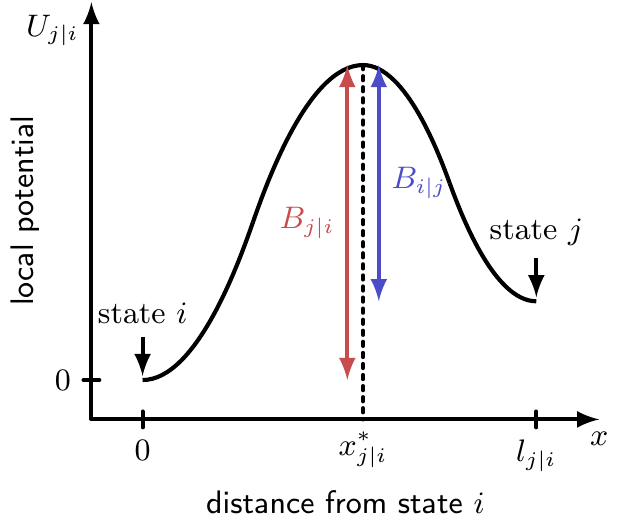}
 \caption{Local potential with local equilibration. Local potential $U_{j|i}(x)=-\int_0^x F_{j|i}(y)\dd y$ between \emph{one} pair of nodes $i$ and $j$. The two states are separated by a single maximum at $x^*_{j|i}$ of the local potential characterized by $F_{j|i}(x^*_{j|i})=0$, while satisfying $F_{j|i}(x)<0$ if $x<x^*_{j|i}$ and $F_{j|i}(x)>0$ if $x>x^*_{j|i}$. Local equilibration occurs if the local free energy barriers are high, meaning that $B_{j|i}\gg k_{\rm B}T$ and $B_{i|j}\gg k_{\rm B}T$.}
 \label{fig:loc_equilibration_methods}
\end{figure}

In the case of high local (free) energy barriers corresponding to
$B_{j|i}\to\infty$ and $B_{i|j}\to\infty$ in
Fig.~\ref{fig:loc_equilibration_methods} for each pair $i,j$, the full microscopic trajectory locally
equilibrates in each  well prior to any transition. In this limit the
transition-rate to jump from node $i$ to node $j$ becomes $w_{i\to
  j}\equiv \phiStar_{j|i}/\avgexit{t}_i$
and in turn Eq.~\eqref{eq:local_detailed_balance_like}
implies (see proof in Appendix \ref{subsec:S_local_detailed_balance})
\begin{equation}
 k_{\rm B}T\ln\frac{w_{i\to j}}{w_{j\to i}}\simeq \underbrace{\int_i^j\bF(\bx)\cdot\dd\bx+\mathcal{U}_j-\mathcal{U}_i}_{\text{``work'' along transition $i\to j$}}+\overbrace{\mathcal{F}_i-\mathcal{F}_j}^{\hidewidth\text{free energy difference}},
 \label{eq:methods_local_detailed_balance}
\end{equation}
where $\mathcal{U}_i$ denotes the potential energy of node $i$, the first term on
the right denotes the  external free energy input along the
transition (i.e. ``dissipated work''), and the free energy of state $i$ is defined by $\mathcal{F}_i=-k_{\rm B}T\ln\mathcal{Z}_i$ with partition function $\mathcal{Z}_i\equiv\sum_{k\in\mathcal{N}_i}\int_0^{x^*_{k|i}}\e^{-\beta [\mathcal{U}_i+U_{k|i}(x)]}\dd x$. 
The symbol ``$\simeq$'' denotes asymptotic equality ``$=$'', here taken in the limit of high local (free) energy barriers ($B_{j|i}\to\infty$ and $B_{i|j}\to\infty$).
Eq.~\eqref{eq:methods_local_detailed_balance} reflects the so-called local detailed balance \cite{seif12} {\color{mynewcolor}(or partial equilibrium \cite{elbe20})}. When there is no work performed along the transition local detailed balance also implies global detailed balance.
Local detailed balance is violated as soon as a \emph{single} barrier
$B_{j|i}$ ceases to be high. Interestingly, \emph{local detailed balance
(Eq.~\eqref{eq:methods_local_detailed_balance}) can be violated even
in systems obeying detailed balance globally}.  In this case the waiting time distribution becomes non-exponential as in Fig.~\ref{fig:blinking}c. 

{\color{mynewcolor}It is worth mentioning that the approximation \eqref{eq:methods_local_detailed_balance}
only affects the free energy difference. This approximation still
exactly satisfies
\begin{align}
 T\dot S^{\rm ss} &=k_{\rm B}T\sum_{ij}\dot{n}_{j|i}\ln\frac{w_{i\to j}}{w_{j\to i}}\nonumber\\
 &
 =\sum_{ij}\dot{n}_{j|i}\bigg[\int_i^j\bF(\bx)\cdot\dd\bx+\mathcal{U}_j-\mathcal{U}_i+\mathcal{F}_i-\mathcal{F}_j\bigg].
\end{align}
The equality follows from Kirchhoff's law stating that
incoming and outgoing currents balance each other
$\sum_{j}\dot{n}_{j|i}=\sum_{j}\dot{n}_{i|j}$, as well as  from Eq.~\eqref{eq:entropy_detailed} and \eqref{eq:entropy_cg}
with $w_{i\to j}=\phiStar_{j|i}/\avgexit{t}_i$.}



{\color{mynewcolor}
\section{Time reversal and kinetic hysteresis}\label{subsec:hysteresis}

The dissipation in a system was found to be closely linked  to the breaking of time reversal symmetry
(measured by the Kullback-Leibler divergence) in
Hamiltonian systems under time dependent driving \cite{jarz06,kawa07}, Markovian diffusion \cite{croo98,seif05,andr07}, and Markov jump dynamics \cite{lebo99,gasp04a} to name but a few.
These findings imply 
that a microscopic  trajectory
$\boldsymbol{\varGamma}_\tau=\bx(t)_{0\le t\le \tau}$ 
in a stationary ensemble of paths with measure $\mathcal{P}$
relates to the steady state dissipation rate
via
\begin{equation}
 \dot S_{\rm KL}=k_{\rm B}\lim_{\tau\to\infty} \frac{1}{\tau}\left\langle\ln\frac{\mathcal{P}[\boldsymbol{\varGamma}_\tau]}{\mathcal{P}[\boldsymbol{\varGamma}_\tau^{\rm R}]}\right\rangle,
 \label{eq:entropy_irr}
\end{equation}
where $\boldsymbol{\varGamma}_\tau^{\rm R}=\bx(\tau-t)_{0\le t\le \tau}$ is the time reversed trajectory  and $\avg{\cdots}$ is the average over the forward path measure $\mathcal{P}[\boldsymbol{\varGamma}_\tau]$. 
Note that we consider  overdamped dynamics, i.e., the micro-state
instantaneously ``loses'' momentum which is odd under time-reversal (e.g., see \cite{jarz06,kawa07}).
In Fig.~\ref{fig:blinking}a the time reversed trajectory $\boldsymbol{\varGamma}_t^{\rm R}$ corresponds to the color-gradient line 
with time running from dark to bright.
In fact by design  the entropy production rates in Eq.~\eqref{eq:entropy_irr}  coincide with the entropy production rate in Eq.~\eqref{eq:entropy_detailed}, and, therefore, also with the one deduced from the coarse-grained trajectory Eq.~\eqref{eq:entropy_cg}, i.e., $ \dot S^{\rm ss}_{\rm KL}=\dot S^{\rm ss}$ holds. Thus the entropy production rate, $\dot S^{\rm ss}$, measures both the breaking of time-reversal symmetry of the underlying diffusive dynamics and the breaking of detailed balance.

Two contradicting views have been put forward \cite{wang07,mart19a}
when addressing  coarse-grained dynamics depicted in
Fig.~\ref{fig:blinking}b that represents a semi-Markov chain. While
Ref.~\cite{wang07} showed that the breaking of time-reversibility does
not imply breaking of detailed balance, Ref.~\cite{mart19a} came to
the exactly opposite conclusion. We now show that this conflict in fact
unravels a counter-intuitive phenomenon.

Determining the breaking of time-reversal symmetry in a
coarse-grained process according to Eq.~\eqref{eq:entropy_irr} can in
general be challenging. However, for a  semi-Markov process one can
elegantly determine the relative entropy rate in 
Eq.~\eqref{eq:entropy_irr} from the waiting time density \cite{gira03,mart19a}
\begin{align}
\dot{S}^{\rm cg}_{\rm KL}
&=\dot S^{\rm ss}+k_{\rm B}\sum_{i,j,k}\phiStar_{k|j} \dot n_{j|i} D_{\rm KL}[\psi_{k|j}(t)\|\psi_{i|j}(t)],
\label{eq:entropy_inconsistent}
\end{align}
which follows immediately from the main result in Ref.~\cite{mart19a} [see Eqs. (2)-(4) therein]
along with the insertion of  Eq.~\eqref{eq:entropy_cg} and the definition of the Kullback-Leibler divergence $D_{\rm KL}[p(t)\|q(t)]\equiv \int_0^\infty p(t)\ln p(t)/q(t)\dd t\ge0$.  Eq.~\eqref{eq:entropy_inconsistent} quantifies the \textcolor{mynewcolor2}{mathematical} time-irreversibility of the coarse-grained process depicted in Fig.~\ref{fig:blinking}b.

In contradiction to Ref.~\cite{wang07}, the
last term in Eq.~\eqref{eq:entropy_inconsistent}
was believed to allow for the detection of 
 ``broken detailed balance in the absence of observable currents'' \cite{mart19a}.
Here we surprisingly find that the Kullback-Leibler divergence overestimates the entropy production, i.e. $\dot  S_{\rm KL}^{\rm cg}\ge \dot  S_{\rm KL}=\dot S^{\rm ss}$.
Notably, the process in Fig.~\ref{fig:blinking}a, which is a manifestly
equilibrium process with $\dot S^{\rm ss}=0$, would paradoxically
display a strictly positive rate $\dot S^{\rm cg}_{\rm KL}>0$. This
follows immediately from the fact that the waiting time densities in 
Fig.~\ref{fig:blinking}c are not equal.  Thus, (coupled) anisotropic waiting time distributions as in Fig.~\ref{fig:blinking}c 
are a signature of \emph{mathematical} irreversibility \cite{wang07}, 
whereas they are in general \emph{not} a signature of
the breaking of detailed balance as apparently erroneously concluded
in \cite{mart19a}. Interestingly, we could not find any
technical mistake in the calculation in Ref.~\cite{mart19a}, yet our
model provides a counter example. How can we reconcile this?

%

%

\begin{figure}
 \includegraphics[width=\columnwidth]{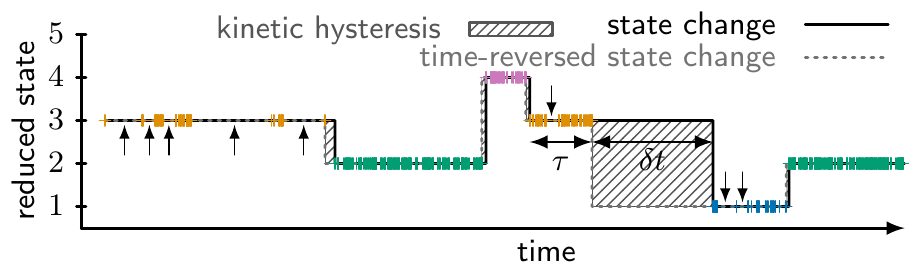}
 \caption{Kinetic hysteresis. The coarse-graining does not commute with time reversal. Transitions are ``odd'' under time reversal, while dwell periods
 are ``even'', i.e., dwell periods commute with time reversal. The state change is the same as in Fig.~\ref{fig:blinking}. }
 \label{fig:cg_traj2}
\end{figure}

It turns out that the coarse-grained trajectory depicted in Fig.~\ref{fig:blinking}b
displays the following counterintuitive phenomenon.
If we coarse-grain the same trajectory backward in time we
discover,  somewhat surprisingly, a \emph{kinetic hysteresis}.
That is, the time-reversed coarsened trajectory (see
dotted gray line Fig.~\ref{fig:cg_traj2}), where time is running from right to left, differs from the forward
one. This hysteresis allows for a \emph{unique decomposition} of each   waiting time $t$ in any given node into a \emph{dwell time}
  $\tau$ -- the interval in which the forward and time-reversed coarsened
  trajectory coincide -- and
  a \emph{transition-path time} $\delta t$ -- the interval in which they
  differ (see Fig.~\ref{fig:cg_traj2}).

  To physically revert time we  must also physically revert the gedanken experiment of the state visits (see colored crosses in Fig.~\ref{fig:cg_traj2} and Fig.~\ref{fig:blinking}b).   
  Thus, each state visit at the end of
a forward-in-time dwell period marks the first state change in the
time-reversed experiment. 
\textcolor{mynewcolor2}{In other words the time-reversal must be employed \emph{before} coarse-graining and \emph{not after} the coarse graining.}
This restores the connection between the
breaking of physical time-reversal symmetry and violations of detailed balance,
i.e. it reconciles the opposing views put forward in \cite{wang07} and \cite{mart19a}.

There is an analogy between transition paths and momenta, which
explains the problem in Ref.~\cite{mart19a}. If we were to reverse in
time a trajectory in an equilibrium system (without changing the sign
of momenta), we would obtain an unphysical time-reversed trajectory that can never
be observed, i.e. Eq.~\eqref{eq:entropy_irr}  would
yield a diverging entropy production rate at equilibrium. 
To avoid this unphysical result one must take into account that momenta in fact change sign under time reversal \cite{lebo99,jarz06,kawa07}.
Hence we find that
 the transition paths, similar to momenta in underdamped systems
 \cite{jarz06,kawa07}, in some sense are ``odd'' under time reversal,
 which gives rise to the kinetic hysteresis in Fig.~\ref{fig:cg_traj2}.
In other words, the``coarse-graining'' and ``time reversal'' must not commute,
which will lead to a paradigm shift in the understanding of time
reversal in the presence transition paths with a finite duration.
}







Some further remarks are in order.
The thermodynamically inconsistent second term
in 
Eq.~\eqref{eq:entropy_inconsistent} vanishes if the waiting time distribution is decoupled from the state change \cite{mart19a} as studied in 
\cite{espo08,maes09,andr08a}, which in fact follows from
Eq.~\eqref{eq:local_decomposition} 
in the limit of instantaneous transition-path times $\wpctrans_{j|i}(t)=\delta (t)$. That is, in this limit the waiting time distribution, $\psi_{j |i}(t)=\wpcdwell_i(t)$, does not depend on $j$. 
Second,  our finding $\dot S^{\rm cg}_{\rm KL}\ge \dot S^{\rm ss}$
does \emph{not} contradict Refs.~\cite{kawa07,gome08} since the path
weight of the coarse-grained process is \emph{not} a marginal path weight
of the full one (see Sec.~\ref{subsec:coarse-graining} and Fig.~\ref{fig:network_microstate_redundancy}).
Third, until now we considered the coarse-graining into individual
nodes (i.e., all cycles were preserved).
It has been found that for certain network topologies  
the lumping of nodes that hides cycles
may lead to what is called a ``second order semi-Markov process''
\cite{mart19a}. {\color{mynewcolor}The kinetics in the presence
of such ``lumped'' nodes is discussed
in Sec.~\ref{subsec:lumping}. In this case the connection between the
entropy production rate and coarse-grained dynamics embodied in Eqs.~\eqref{eq:entropy_detailed} and \eqref{eq:entropy_cg} is expected to disappear.}

\section{Three sources of fluctuations}
\label{sec:three_noises}
\subsection{Transtion noise dictates the amplitude of fluctuations}\label{subesec:transitions_dictate}

%

Memory in state-changes emerges \emph{locally}
as a result of long recurrence and transition-path times. Long recurrence times
arise whenever the continuous trajectory becomes trapped in the
legs of the subgraph without
changing state (see vertical arrows in Fig.~\ref{fig:blinking}b or Fig.~\ref{fig:cg_traj2}).
Long transition-path times are due to slow dynamics between a pair of adjacent nodes. 
Imagine that only one leg in
Fig.~\ref{fig:illustrantion_all}e, say $3\to1$, displays slow or recurrent dynamics,
e.g. because of slow diffusion and/or the absence of an energy barrier. Then
not only $\psi_{1|3}$ is clearly non-exponential (see blue line
in Fig.~\ref{fig:blinking}c) but strikingly also $\psi_{2|3}$ and all
others become non-exponential (see green line
in Fig.~\ref{fig:blinking}c)
-- the waiting time distribution becomes ``coupled''
\cite{klaf87,metz98,greb18a} to the state change. Note that this
problem cannot be solved within the framework of the generalized
master equation \cite{land77} because the coupling has to be ``put in by hand''.

\begin{figure*}
\includegraphics{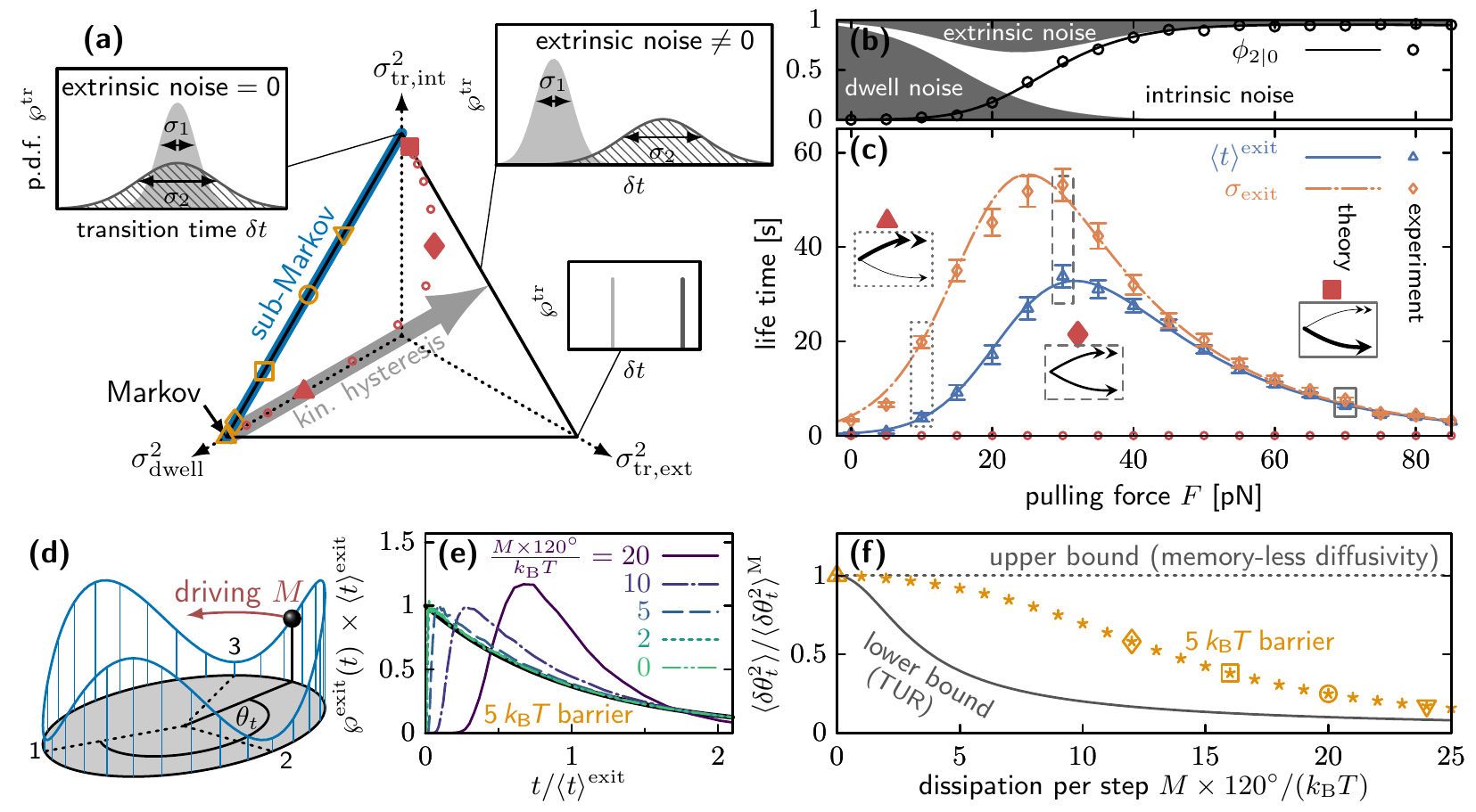}
 \caption{Manifestations of transition noise and ``flavor
     of memory''. (a)~The partitioning of noise-contributions mapped onto a triangle; the
   center of the triangle represents the equi-partitioning of
   noise-sources, $\sigma_{\rm dwell}^2=\sigma_{\rm tr,
     int}^2=\sigma_{\rm tr, ext}^2=\sigma_{\rm exit}^2/3$. The left
   corner corresponds to Markov kinetics,  $\sigma_{\rm tr, ext}^2=\sigma_{\rm tr, int}^2=0$, and the left edge (blue line) to ``sub-Markov'' kinetics,
   $\sigma_{\rm tr, ext}^2=0$.  The top corner
   corresponds to a vanishing dwell- and extrinsic transition-noise, $\sigma_{\rm dwell}^2=\sigma_{\rm
     tr,ext}^2=0$, whereas the right corner
   depicts the limit of vanishing dwell- and intrinsic 
 transition-noise, $\sigma_{\rm dwell}^2=\sigma_{\rm tr, int}^2=0$. The boxed histograms
 are shown for illustrative purposes. The kinetic hysteresis
 increases along the gray
 arrow. The red circles depicts results of the catch-bond example for
 different pulling forces $F$ shown in panel (b) and (c)
 (results depicted by larger symbols are additionally illustrated in
 (c)), and orange symbols show the results for the driven ATPase shown in
 panel (f).
(b,~c)~Reconstructed catch-bond experiment; (b) shows
$\phi_{2|0}$--the probability of taking the slow pathway 2 (black
line); the shaded areas depict the fraction of dwell ($\sigma_{\rm dwell}^2/\sigma_{\rm exit}^2$), extrinsic ($\sigma_{\rm tr,
  ext}^2/\sigma_{\rm exit}^2$), and intrinsic ($\sigma_{\rm tr,
  int}^2/\sigma_{\rm exit}^2$) noise, respectively. (c) depicts the
mean, $\avgexit{t}$, and standard deviation, $\sigma_{\rm exit}$, of
the bond's life-time. Lines correspond to exact results and symbols
are deduced from 500 rupture experiments. Red circles depict pulling
forces considered in panel (a). The density of the life-time of the
bond at $F=20$\,pN is shown in Fig.~\ref{fig:illustrantion_all}d.
(d-f)~Driven molecular motor displaying a vanishing extrinsic
transition-noise, $\sigma_{\rm tr, ext}=0$. (d) Free energy
landscape as function of the angle $\theta_t$ with a barrier-height of $5\,k_{\rm B}T$
(blue line; see Appendix~\ref{subsec:ATPase}
for details) that becomes
tilted due the action of the torque $M$ (red arrow); dotted lines
denote network-states, i.e. free energy minima. (e) Scaled probability density of exit-time from a state (i.e. first
passage-time to an angular displacement of $\pm 2\pi/3=\pm 120^\circ$) as
a function of the dissipation per step (i.e. torque $M$ multiplied by
the rotation-step $120^\circ$) of magnitude $0,2,5,10$ and $20\,_{\rm B}T$. (f)
Steady-state mean squared angular deviation
$\langle \delta \theta_t^2\rangle$ compared
with the a Markov-jump approximation $\langle \delta
\theta_t^2\rangle^{\rm M}$ as a function of dissipation; The full line depicts
the lower bound \eqref{eq:ineq_TUR} derived in Appendix~\ref{subsec:TUR_proof} using the Thermodynamic Uncertainty Relation
(TUR) \cite{bara15,horo20}; the
individual noise-contributions for selected points (open symbols) are shown in panel (a).
Parameters: (b,c) see Appendix~\ref{subsec:SI_catch}, (d-f) see Appendix~\ref{subsec:ATPase}. 
}
 \label{fig:allresults}
\end{figure*}

In order to  understand the emergence these anisotropic 
local `waiting times' we
dissect fluctuations 
of
time required to exit
state $i$ as, $\wp_i^{\rm
  exit}(t)\equiv\sum_{j}\wpcStar_{j|i}(t)$.
The independence of dwell and transition-path
times in Eq.~\eqref{eq:local_decomposition} implies three independent
contributions to fluctuations 
\begin{equation}
 \sigma_{\rm exit}^2=\sigma_{\rm dwell}^2+\sigma_{\rm tr, int}^2+\sigma_{\rm tr,ext}^2,
 \label{eq:noise_3parts}
\end{equation}
where $\sigma^2\equiv\avg{t^2}-\avg{t}^2$ denotes the variance, and we further decomposed fluctuations of
transition-path time into intrinsic fluctuations along the respective legs of the
subgraph, $\sigma_{\rm tr, int}^2=\sum_j\phiStar_{j|i}\sigma^2_{{\rm
    tr},j|i}$, and the extrinsic
scatter of
mean transition-path times among distinct legs, $\sigma_{\rm
  tr,ext}^2\equiv\sum_j\phiStar_{j|i}(\avg{\delta t}_{j|i}^{\rm
  tr})^2-(\sum_j\phiStar_{j|i}\avg{\delta t}_{j|i}^{\rm
  tr})^2$. The three contributions in Eq.~\eqref{eq:noise_3parts} are explained in
Fig.~\ref{fig:allresults}a and given explicitly in Sec.~\ref{subsec:main_practical}.

When  $\sigma_{\rm
  tr,ext}^2$ vanishes,
i.e. $\avg{\delta t}_{j|i}^{\rm tr}=\avg{\delta  t}_{k|i}^{\rm tr}$ for all $j,k$ (see Fig.~\ref{fig:allresults}a left),
the fluctuations of exit time are \emph{sub-Markovian} since $\sigma_{\rm exit}\le
\avg{t}^{\rm exit}_j$. In turn, \emph{super-Markovian} fluctuations,  that
is $\sigma_{\rm exit}\ge \avgexit{t}$, necessarily imply the
existence of multiple exit-pathways with distinct
transition-path times (see Fig.~\ref{fig:allresults}a,
right).
\emph{This proves that one can infer, in general, the existence of parallel transition pathways
  without actually resolving individual pathways,}
which is our third main result of this paper 
(for proof see  last subsection in SM \cite{Note2}).
 Below we illustrate this main finding by means of two opposing examples.

\subsection{Super-Markovian exit dynamics reflect parallel unequally fast transition paths}\label{subsec:super_Markov}

In a first demonstration of the practical implications of our results
we address the counter-intuitive catch-bond phenomenon
\cite{bart02,thom08a} depicted in
Fig.~\ref{fig:illustrantion_all}d (see also Refs.~\cite{thom06,buck14}).
A ligand bound to a receptor is pulled by a constant force $F$
until the bond ruptures (details about the model are given in Appendix~\ref{subsec:SI_catch}). The time of rupture
corresponds mathematically to the exit time from the bound state. A characteristic of
catch-bonds is that they rupture along two possible pathways. One pathway involves a
conformational change of the receptor that prolongs the
transition-path time. In turn this gives rise to a non-monotonic force-dependence
of the rupture-time (see
Fig.~\ref{fig:allresults}c). 
That is, within a certain
interval of $F$ -- the so-called catch-bond phase
\cite{mars03,thom06,thom08a,buck14}
-- the bond counter-intuitively survives longer if we pull stronger.
The mean life-time $\avgexit{t}$ and its standard
deviation $\sigma_{\rm exit}$ reconstructed according to Ref.~\cite{bart02,thom08a} are depicted in
Fig.~\ref{fig:allresults}c,  where the lines denote exact results
(see Sec.~\ref{subsec:main_practical}) and symbols were deduced from 500 simulated rupture
events. A larger pulling-force increases the likelihood of choosing
the slow path (see black line in Fig.~\ref{fig:allresults}b) and in
turn amplifies extrinsic noise (see shaded areas
reflecting relative noise contributions in Fig.~\ref{fig:allresults}b
as well as red symbols in Fig.~\ref{fig:allresults}a). 
The observed fluctuations are evidently \emph{super-Markovian},
i.e. $\sigma_{\rm exit}\ge \avgexit{t}$, and therefore immediately
imply the existence of at least two rupture pathways that are not equally fast,
$\sigma_{\text{tr,ext}}\neq 0$.
We show the decomposition of the life time  of the bond into dwell-
and transition-path time along the individual pathways in
Appendix~\ref{subsec:SI_catch} (see Fig.~\ref{fig:S_catch_hist2}). 
{\color{mynewcolor}If transition-path times can be measured explicitly, one can alternatively detect parallel paths  by means of the coefficient of variation as explained in \cite{sati20},
which for the sake of completeness is shown in Appendix~\ref{paragraph:sati20} (see 
Fig.~\ref{fig:S_CV}).}



\subsection{Symmetry in transtions causes sub-Markovian exit dynamics} \label{subsec:sub_Markov}
We now consider the scenario where extrinsic transition noise
vanishes, implying $\sigma_{\rm exit}\le \avgexit{t}$. Particularly important examples are the steady-state
operation of driven molecular machines and the more abstract ``stopping
times'' of the thermodynamic entropy production \cite{neri17,neri19,neri20}. We
consider an ATPase\- operating under
the influence of a non-equilibrium torque $M$,
were $M=0$ refers to the torque at which the ATPase\- stalls
\cite{toya11}. The rotation of the ATPase\- evolves as diffusion in a periodic
potential with period $2\pi/3$, reflecting the $120^\circ$ rotational
symmetry of the motor, and a barrier height of $5\,k_{\rm B}T$
separating the well-defined minima (see Fig.~\ref{fig:illustrantion_all}d). The torque is accounted for by tilting  the
potential (see Fig.~\ref{fig:network_meaning}a,b and Appendix~\ref{subsec:ATPase}). The continuous rotation is
coarse-grained into a uni-cyclic network with three rotational states,
whereby the statistics of rotational state-changes remain exact.
The statistics of exit time from either minimum are depicted
Fig.~\ref{fig:allresults}e.  


The probability densities to make a step in the forward ($+$)
and backward ($-$) direction after time $t$ are given by $\wpcStar_\pm(t)=\phi_\pm\wpcexit(t)$,
yielding a mean squared angular deviation \cite{fell71,land77}, $\langle \delta \theta_t^2\rangle\equiv\avg{\theta_t^2}-\avg{\theta_t}^2$ 
(see proof in Appendix~\ref{subsec:ATPase}, Eq.~\eqref{eq:S8_error_propagation_proof})
\begin{equation}
\frac{\langle \delta \theta_t^2\rangle}{(120^\circ)^2}\simeq 4\phi_+\phi_-\frac{t}{\avgexit{t}}+(\phi_+-\phi_-)^2\frac{t\sigma_{\rm exit}^2}{(\avgexit{t})^3},
\label{eq:error_propagation}
\end{equation}
where $\simeq$ denotes asymptotic equality in the
limit $t\to\infty$. Memory-less, Markovian kinetics would predict $\sigma_{\rm exit}^{\rm M}\equiv\avgexit{t}\ge \sigma_{\rm exit}$
 and thus systematically
overestimate fluctuations (dotted line in Fig.~ \ref{fig:allresults}f). 
\emph{Memory is
  particularly pronounced in the regime of strong driving}, i.e. $\phi_+\gg \phi_-$ or $\phi_-\gg \phi_+$. Notably, we find the so-called thermodynamic uncertainty relation \cite{bara15,horo20}
to bound fluctuations from below (Fig.~\ref{fig:allresults}f,
solid line; for a proof see Appendix~\ref{subsec:TUR_proof}). Our results thereby yield \emph{a ``sandwich''
bound on actual fluctuations in driven cycle-graph (i.e., ring-shaped)
networks.}

{\color{mynewcolor}
\section{Marginal observations and hidden cycles}\label{sec:cg_general}
So far we have discussed a non-Markovian network theory that accounts for transition paths without hidden cycles. 
We now show how one can generalize our results to systems with hidden
cycles. Strikingly, we show that an evaluation of the mean waiting
time from a partial observation alone can reveal non-Markovian
fingerprints of the full network dynamics. Moreover, we discuss future
perspectives and show what can be learned from established coarse-graining schemes known as \emph{milestoning}.

\subsection{Marginal observations reveal fingerprints of non-Markovianity}
\label{subsec:marginal}
Detecting even a single cycle in an experiment can be challenging
because it requires us to resolve two parallel transition paths as
achieved e.g.\ in Ref.~\cite{kim20}. It thus may not always be feasible to monitor more than one existing cycle.

\begin{figure}
 \centering
 \includegraphics[width=\columnwidth]{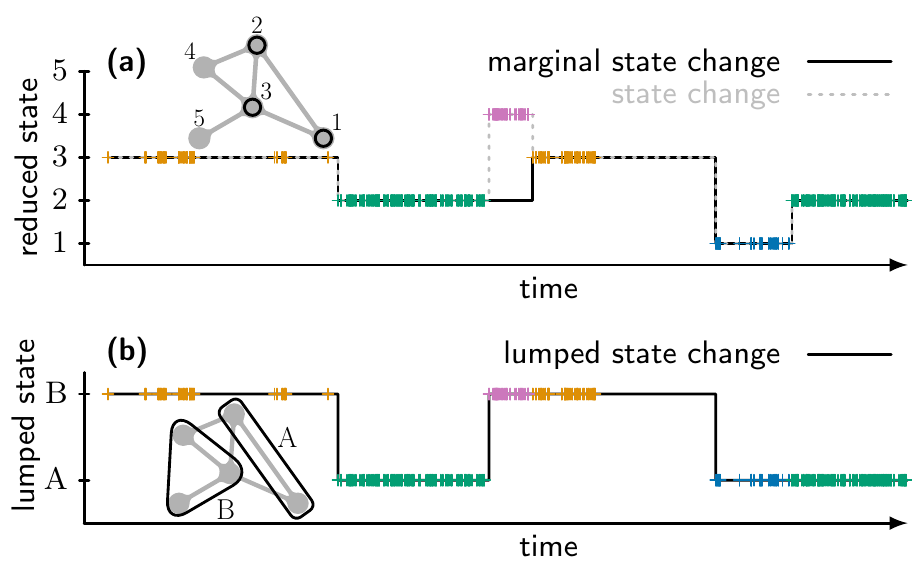}
 \caption{\color{mynewcolor}Marginal observations and lumping of states. (a)~Marginal
   state-change trajectory (black line) if states 4 and 5 are
   hidden. (b)~Lumping of states into $\text{A}=\{1,2\}$ and
   $\text{B}=\{3,4,5\}$. The colored crosses correspond to state
   visits of 
   all five states as in Fig.~\ref{fig:blinking}. }
 \label{fig:cg_traj_lumping}
\end{figure}

To describe such scenarios we need to account for possible hidden nodes that cannot be detected as shown in Fig.~\ref{fig:cg_traj_lumping}a.
For example, among the five states listed Fig.~\ref{fig:blinking}
it may be that we can only monitor three states, say states 1, 2, and
3. In other words, recurrences and state changes to states 4 and 5 are
assumed not to be monitored.  In this case we can directly observe the
nontrivial cycle $1\to2\to 3$, whereas the second elementary cycle $3\to 2\to 4\to 3$
shown in Fig.~\ref{fig:blinking} is not directly visible -- it corresponds
$3\to 2\to 3$ in the marginal observation (compare dotted and solid
line in Fig.~\ref{fig:cg_traj_lumping}a). 
Once a network has hidden cycles with a nonzero affinity (see \cite{schn76} and Appendix~\ref{subsec:SI_cycles}), we expect Eq.~\eqref{eq:local_detailed_balance_like} not to hold
in conjunction with the possible breakdown of the forward/backward symmetry of the transition-path time within the marginal observation as demonstrated in Ref.~\cite{glad19}.
It is worth mentioning  that both, the dotted and the solid line
depicted in Fig.~\ref{fig:cg_traj_lumping}a will generally represent a
semi-Markov process, and that
the independence between transition-path time and dwell time as in Eq.~\eqref{eq:local_decomposition} is expected to be preserved.

We explain  in Appendix~\ref{subsec:from_starlike}  how one can
apply the network theory we developed to study marginal
observations.   
More precisely, the waiting time distribution of a marginal state change
to the other states within a subset can be described by all the local waiting time distributions $\wpcStar_{j|i}(t)$ [see Eq.~\eqref{eq:S2_montroll}].
Thus the path-decomposition, which generalizes the classical renewal
theorem \cite{sieg51} that is fully explained in
Appendix~\ref{sec:path_decomposition} and used to derive our results,
can be used further to generalize our results in order to 
include marginal observations.

Appendix~\ref{subsec:fingerprints_ill} shows that the marginal
dynamics along $3\to 1$ is 
\emph{slower} and concurrently also
\emph{less likely} than the marginal dynamics along $3\to 2$.
Note that both, Markov-state kinetics \cite{gill77,gill07,mcad97,wale98,paul05,bowm10,seif12,wood14,chod14,bowm14,past15,husi18,mori16}
as well as isotropic (decoupled) renewal processes \cite{land77,bark00,espo08,andr08a,metz98} would invariably infer transition $3\to 1$ to be  erroneously faster on average
than $3\to 2$ (see Appendix~\ref{subsec:fingerprints_ill}).
%
This is because the direct transition
  $3\to 1$ takes longer than the paths  $3\to2$ that involve a
  detour through the intermediate state $4$. Markov models, for example,
  do not allow for this to happen because they assume transitions to occur
  instantaneously. \textcolor{mynewcolor2}{Therefore, non-Markovianity of the underlying network can be detected from just the mean waiting times of the marginal observation.} Similarly, symmetric waiting time distributions in
  renewal processes render the duration of all transitions from any
  state equal, such that any path involving a detour is bound to take longer on average.

  In other words, 
  memory in the marginal observation in
  Fig.~\ref{fig:cg_traj_lumping} \emph{cannot} emerge solely from 
  ignoring states 4 and 5. Notably, this has a mean waiting time along $3\to1$
  that is different from the mean waiting time along $3\to2$, which
  was concluded to be a sign of time-reversal symmetry breaking in Ref.~\cite{wang07}
  and 
  confused to be a signature of broken detailed balance in Ref.~\cite{mart19a}.  
  Since the model satisfies detailed balance these findings need to be
  revised for both, the marginal observation and the full observation,
  as soon as transition paths are not instantaneous. We thus expect
  the kinetic hysteresis, which accounts for transition paths being
  ``odd'' under time reversal as explained in Fig.~\ref{fig:cg_traj2},
  to extend to marginal observations. Note that if the marginal observation
  hides away cycles, we expect Eq.~\eqref{eq:entropy_cg} to be a lower
  bound to the entropy production rates, while the rate in
  Eq.~\eqref{eq:entropy_inconsistent} for the marginal observation in
  Fig.~\ref{fig:cg_traj_lumping}a becomes non-zero at equilibrium.





  

  A correct interpretation of marginal
  observations in the presence of memory therefore requires the
  thermodynamically consistent coarse-graining derived in our work. 
%
  More severe manifestations of memory are found in driven networks
(see catch-bond analysis in Fig.~\ref{fig:illustrantion_all}d or
Table~\ref{tab:asymmetry_catch}).

\subsection{Lumping nodes into meso states}\label{subsec:lumping}

In the following we discuss 
the coarse-graining 
that lumps multiple states 
into a few ``meso states'' as shown in
Fig.~\ref{fig:cg_traj_lumping}b. In this figure we lump 5 five states
into a pair of clustered meso-states $\mathrm{A}=\{1,2\}$ and $\mathrm{B}=\{3,4,5\}$.
The lumped state changes form a  \emph{hidden semi-Markov process}
\cite{yu10}, for which we expect the waiting and dwell time to \emph{not} depend only
on the current state, that is, Eq.~\eqref{eq:local_decomposition} is expected to be violated.

Lumped dynamics emerging from
hidden Markov models 
have been studied thoroughly from a thermodynamic point of view
\cite{espo12}. However, in contrast to our example
Fig.~\ref{fig:cg_traj_lumping}b, Ref.~\cite{espo12} presumed
 transition paths to be infinitely fast.  The trajectory depicted Fig.~\ref{fig:cg_traj_lumping}b displays three transitions between clusters A and B, which occur along three distinct \emph{continuous} transition pathways,
 while the last transition involves the prolonged transition $\delta t$ between states 1 and 3 which is highlighted in Figs.~\ref{fig:blinking}b and \ref{fig:cg_traj2}.
  
 As soon as states 
 are lumped the waiting time density in a lumped state,
 Eq.~\eqref{eq:local_decomposition},
 depends also on the states visited before said state. The dwell time
 statistics then depends not only on the current state. For example,
 the dwell time statistics in certain lumped networks was found to
 develop a ``transition memory'' \cite{hawk11}, which depends on the
 future $j$, current $i$, and the past $k$ state \cite{mart19a}, i.e.,
 the dwell time symmetry Eq.~\eqref{eq:local_decomposition} is
 violated $\wpcdwell_{i}\to \wpcdwell_{j|ik}$. In this case we
 furthermore expect  the independence between dwell time and
 transition-path time to be violated.
 
Notably, one of our examples shown in Fig.~\ref{fig:allresults}b,c
in fact involves lumped states. There we measured the life time
 of a bond that can rupture along two pathways. After rupture the bond
 remains in two distinct states that we effectively lumped into one
 ruptured state (see Fig.~\ref{fig:illustrantion_all}d). Furthermore, any exit time distribution from a state $i$ represents a passage from $i$
 to the lumped set $\mathcal{N}_i$ of all states adjacent to $i$.
 
 
 We demonstrate in Appendix~\ref{subsec:fingerprints_ill} (see also discussion in the previous subsection) how one can deduce
 the waiting time distribution from one state within B into either of
 both states in the lumped mesostate $\mathrm{A}=\{1,2\}$. Choosing
 the set of target states to be either of the two mesostates
 ($\mathcal{A}=\mathrm{A}$ or $\mathcal{A}=\mathrm{B}$), we can deduce
 the waiting time statistics to any cluster.

To obtain a physically consistent coarse-graining transition paths
 should be considered to be ``odd'' under time reversal. The
 kinetic hysteresis is thus also important to properly account for the
 thermodynamic entropy production rate in the presence of lumping. As
 soon as the lumping hides cycles,
 Eq.~\eqref{eq:entropy_cg} is expected bound the entropy production in
 the lumped system from below \cite{pugl10,espo12,mart19a}.  If all
 lumped states relax quickly \emph{and} the transitions between all lumped states are fast,
 the lumped dynamics becomes approximately Markovian \cite{schu11}.

Next we explain how one can conceptually extend the applicability of our results to
   more general types of underlying microscopic 
   dynamics, i.e., other than diffusion on a graph, via a 
 proxy --
the
 so-called ``milestoning''
 methodology (see, e.g., Refs.~\cite{fara04,shal06,schu11,hawk11,bere19a,elbe20,elbe20a}).

%


\subsection{Milestoning}
Milestoning is a (numerical) method for deriving discrete state
kinetics from an underlying continuous dynamics
\cite{elbe20,elbe20a}. We now highlight how milestoning allows us to
extend our results to systems with microscopic dynamics other than
diffusion on a graph. 
To this end, we identify/replace the nodes (or
states) in the network by ``milestones'' such that each recurrence in Fig.~\ref{fig:blinking} becomes a passage through a milestone \cite{bere19}. 

Milestones can represent hypersurfaces \cite{fara04} or
hypervolumes, i.e., cores \cite{schu11}.  
In Ref.~\cite{fara04} milestones were introduced as a
hypersurface-to-hypersurface hopping, while the underlying dynamics
perpendicular to the milestones was assumed to be the slowest degree
of freedom that is effectively one-dimensional.  
Our work now additionally introduces cycles and branch points (see
Fig.~\ref{fig:network_meaning}), which allow for a non-vanishing rate
of steady state dissipation. Eq.~\eqref{eq:local_detailed_balance_like} shows, for
 the first time, that a coarse-graining based on milestoning robustly
preserves the rate of dissipation both, arbitrarily far from
equilibrium \emph{and} in the presence of memory. In other words, local detailed
balance \cite{seif12} is allowed to be violated. 
 It will be interesting to further account for the thermodynamics of
 transition memory  \cite{hawk11}, which can be caused by fast
 inter-hypersurface dynamics or by lumping of 
 hypersurfaces that may
 lead to what is called a semi-Markov process of second order
 \cite{mart19a}. In this latter case a splitting probability of the transition
 $i\to j$ that depends also on the state $k$ directly preceding $i$.
 
 A slightly different milestoning approach presented in Ref.~\cite{schu11} is based on cores which are small volumes in phase space.
  In the scenario when transition-path times become short milestoning
  already emerged as a valuable tool to deduce a (memory-less) master
  equation in a kinetically consistent manner \cite{schu11} (cf. third
  paragraph in the Introduction therein). While
  Eq.~\eqref{eq:local_detailed_balance_like} may not hold anymore
  exactly, we expect that it will still provide a useful estimate for entropy production.  \textcolor{mynewcolor2}{Our result illustrates that milestoning leads to a robust and thermodynamically consistent coarse-graining without requiring requiring the dynamics to be Markovian.}

We therefore anticipate the milestoning based on cores to become a valuable tool
for extending our results also to those types of microscopic dynamics that
may not be directly described by a diffusion on a graph. 
To illustrate the rationale behind this idea we briefly sketch how a
generic 
discrete-state dynamics obtained by milestoning relates to the results
derived from our model of diffusion on a graph
 and the marginal observation depicted in
 Fig.~\ref{fig:cg_traj_lumping}a.
 
In a Markov jump processes each recurrence is in fact extended in time
and exponentially distributed. To capture this in our model we must
simply 
adopt finite cores \cite{schu11} that effectively ``smear out'' the
nodes (i.e.\ crosses
in Fig.~\ref{fig:cg_traj_lumping}) to continuous segments. The
transition time $(2\to 3)$ in the marginal observation becomes
the time span over which we would detect the hidden state 4. That is,
the full and marginal state change differ. We may now conceptually 
replace, to a very good approximation, all hidden parts of the network
with a sufficiently dense Markov network with the appropriate
topology, and adopt
Ref.~\cite{bere19a} which explains how to deduce transition-path time statistics 
between any pair of marginally observed states in Markov jump networks
in quite 
general context. 
 The entropy production of the reduced network dynamics was in turn evaluated
 in Ref.~\cite{wang07}. We may thus consider
 Refs.~\cite{wang07,schu11,bere19a} and the coarse-graining outlined
 in Fig.~\ref{fig:blinking} to provide a generic blueprint for constructing a thermodynamically consistent coarse-graining.
}
\section{Conclusion}\label{sec:conclusion}

Emerging from the mapping of continuous dynamics are three elementary, independent sources of
fluctuations in state-transitions on a network: dwell-time
fluctuations and the intrinsic and extrinsic noise arising from random
transition-path times. The balance of these noise channels, depicted in the \emph{noise triangle} in
Fig.~\ref{fig:allresults}a,
yields Markovian, sub- or super-Markovian fluctuations and thus sets the
`flavor of memory'. 
A vanishing extrinsic transition-noise causes sub-Markovian dynamics
as in the driven ATPase\- (Fig.~\ref{fig:allresults}a, orange symbols). Markovian dynamics is dominated by dwell-noise
(left corner).  Super-Markovian fluctuations (observed e.g. in
catch-bond dynamics) are dominated by
extrinsic transition-noise (right corner). The noise triangle allows for a conclusive inference of
underlying dominant, hidden continuous paths in general networks solely
from the observed fluctuations in
state-transitions. The kinetic hysteresis between forward- and
time-reversed state-trajectories that arises in the presence of transition-noise
(grey arrow) provides a new understanding of the breaking of time-reversal symmetry
in the presence of memory \cite{mart19a,espo08,andr08a,maes09}. The widely
adopted principle of local detailed balance is found to be a
peculiarity of the Markovian limit, not a general feature
time-reversal symmetry. 
Our results
pave the way towards a deeper understanding of network-dynamics far from equilibrium including 
current-fluctuations in active molecular systems \cite{bara15,horo20}.
\textcolor{mynewcolor2}{Even though we deduced the dissipation in the long time limit, the generalization of the local detailed balance relation, Eq.~\eqref{eq:local_detailed_balance_like}, holds at any time, including networks with infinitely many states \emph{and} transient dynamics. Our work thus enable further studies with transisent network dynamics.}

\begin{acknowledgments}
 The financial support from the German Research Foundation (DFG) through the Emmy Noether Program GO 2762/1-1 to A. G. is gratefully acknowledged
\end{acknowledgments}
%
%
%



\appendix

\section{From diffusion on a graph to  state-changes on a network: rigorous coarse-graining}\label{sec:S_stoch_dyn}


In this Appendix we first describe the stochastic dynamics of individual trajectories and its numerical implementation. The translation of the Langevin equation for the time evolution of individual trajectories into a Fokker-Planck equation for the time evolution of probability densities on a graph is explained in the last subsection. 

\subsection{Stochastic differential equation on a graph}\label{subsec:S_stoch_dyn_traj}

We parametrize $\bx_t$ (the micro-state at time $t$) in such a way that the reduced state $i_t$ represents the last visited network-state $i_t\in\Omega=\{1,\ldots,N\}$.
The micro-state is assumed to be fully
characterized by $\bx_t=(x_t,j_t,i_t)$, where $x_t$ denotes  the distance from the last visited network-state $i_t$ along the the link to network-state $j_t$
as shown in Fig.~\ref{fig:network_microstate_redundancy}a, where $i_t=i$ and $j_t=j$.
The variable $\bx_t=(x_t,j_t,i_t)$ fully determines the micro-state configuration on the graph at time $t$.
Denoting the distance between two nodes $i$ and $j$ by $l_{i|j}=l_{j|i}$, the distance function $x_t$ must lie within the interval $0\le x_t\le l_{j_t|i_t}$.
The ``instantaneously'' targeted state $j_t$ and the last visited state $i_t$
do not change until $x_t$ reaches either the ``inner boundary'' $x_t=0$
or the ``outer boundary'' $x_t=l_{j_t|i_t}$. After this the variables
change according to the rule described in Tab.~\ref{tab:update_boundary}.
\begin{table*}
\caption{Detailed description of the update rules of the micro-state during time step ``$t\to t+\dd t$''.
Each variable $\beta_t$
is randomly chosen amongst the states adjacent to $k$,  $\mathcal{N}_k$,
with a uniform probability $1/|\mathcal{N}_k|$, where
$|\mathcal{N}_k|$ denotes the number of elements in the set $\mathcal{N}_k$ and $k=i_t,j_t$. Note that the rules listed in the table do \emph{not} change the micro-state (see
Fig.~\ref{fig:network_microstate_redundancy}).
}
\label{tab:update_boundary}
 \begin{tabular}{c|c||c|c}
\hline\hline
\multicolumn{2}{c||}{}&\multicolumn{2}{c}{update conditions at the border}\\\hline
\bfseries varaiable& \bfseries meaning&\textbf{outer boundary} ($x_t\ge l_{i_t|j_t}$)&\textbf{inner boundary} ($x_t\le0$)\\\hline
  $i_t$&latest visited state& $i_{t+\dd t}= j_t$&$i_{t+\dd t}= i_t$\\\hline
  $x_t$ &current distance from latest visited state&$x_{t+\dd t}= (l_{i_t|j_t}-x_t)$&$x_{t+\dd t}=-x_t$\\\hline
 $j_t$ &currently targeted new network state&
 \begin{tabular}{c}generate randomly $\beta_t\in\mathcal{N}_{j_t}$\\
 and set $j_{t+\dd t}= \beta_t$  with \end{tabular}& \begin{tabular}{c}generate randomly $\beta_t\in\mathcal{N}_{i_t}$\\
 and set $j_{t+\dd t}= \beta_t$  with \end{tabular}\\\hline\hline
 \end{tabular}
\end{table*}
%
%
%
%
In Fig.~\ref{fig:blinking}a each cross corresponds to the visit of a state, where each revisit of the state (recurrence) with the same color corresponds to all incidents $x_t=0$
in which the ``inner boundary is hit''.
Conversely, each state-change corresponds to hitting the ``outer boundary''
$x_t=l_{j_t|i_t}$ (see thick black crosses in Fig.~\ref{fig:blinking}a)
 after which the latest visited state becomes $i_{t+0}=j_t$, and the instantaneously targeted state is chosen, without loss of generality, with equal probability among the neighbors $j_{t+0}\in\mathcal{N}_{j_t}$.
The micro-state description deliberately contains a
redundancy since micro-states $\bx_t=(x_t,j_t,i_t)$
and  $\smash{\tilde{\bx}_t=(l_{j_t|i_t}-x_t,i_t,j_t)}$
correspond to exactly the same micro-state configuration, even though the last component  of $\smash{\tilde{\bx}_t}$ does not
represent the last visited state (see Fig.~\ref{fig:network_microstate_redundancy}).

The micro-state $\bx_t=(x_t,j_t,i_t)$ evolves such that both the last visited state $i_t$ and the instantaneously targeted state $j_t$
remain unchanged during each interval, in which the distance $x_t$ lies within
the interval $0<x_t<l_{j_t|i_t}$, which corresponds to $x_t\neq0$ and $x_t\neq l_{j_t|i_t}$.
During the time when both $j_t$ and $i_t$
are constant the distance $x_t$ between two connected nodes
evolves according to the anti-It\^o  Langevin
equation \eqref{eq:Langevin_def},
which can also translated into the following equivalent  It\^o-Langevin equation 
\begin{equation}
 \dot x_t=\beta D_{j_t|i_t}(x_t) F_{j_t|i_t}(x_t)+D_{j_t|i_t}'(x_t)+\sqrt{2D_{j_t|i_t}(x_t)}\xi_t,
 \label{eq:S3_Langevin_Ito}
\end{equation}
where  $D_{j|i}'(x)=\del_xD_{j|i}(x)$, $i=i_t$ and $j=j_t$.
Note that anti-It\^o differential equation is also referred to as H\"anggi-Klimontovich \cite{haen82,klim90} (see also more recently Ref.~\cite{pigo17,bo19}), while Ref.~\cite{klim90}
derives
the Stratonovich variant of Eq.~\eqref{eq:S3_Langevin_Ito} called ``kinetic form''.
Equations~\eqref{eq:Langevin_def} and \eqref{eq:S3_Langevin_Ito}
follow from the assumption that the inverse friction coefficient (mobility)
satisfies the Einstein relation, $\mu_{j|i}(x)=\beta D_{j|i}(x)$,  readily inserted in the first term of Eq.~\eqref{eq:S3_Langevin_Ito}.

Equations~\eqref{eq:Langevin_def} and \eqref{eq:S3_Langevin_Ito} describe the time evolution
of the first component of the micro-state $\bx_t=(x_t,j_t,i_t)$.
Numerical schemes to propagate Eqs.~\eqref{eq:Langevin_def} and \eqref{eq:S3_Langevin_Ito} are presented below,
where Appendix~\ref{subsubsec:naive} shows a na\"ive simple Euler method,
and Appendix~\ref{subsubsec:Milstein} the celebrated Milstein scheme \cite{kloe94}.
For systems with multiplicative noise (i.e. with a micro-state dependent noise amplitude) we generate trajectories according to the Milstein scheme described in Appendix~\ref{subsubsec:Milstein}.
For systems with additive noise (constant noise amplitude, that is $D_{j|i}(x)=\text{const.}$), we use the scheme shown in
Appendix~\ref{subsubsec:Runge_Kutta} (adopted from Ref.~\cite{chan87a}). Functionals of trajectories, such as the dwell and transition-path time periods, are evaluated irrespective of the chosen numerical integration scheme of the Langevin equation as will be explained in Appendix~\ref{subsubsec:functionals}.


\subsection{Na\"ive anti-It\^o Euler scheme (strong order 0.5)}\label{subsubsec:naive}
The simplest way to numerically integrate the  anti-It\^o Langevin equation \eqref{eq:Langevin_def}
 from $t$ to time $ t+\varDelta t$ is the following
anti-It\^o Euler scheme
\begin{align}
 \tilde x_t
 &
 =x_t+\beta D_{j_t|i_t}(x_t) F_{j_t|i_t}(x_t)\varDelta t+\sigma_{j_t|i_t}(x_t) \hat{Z}_t,\nonumber\\
  x_{t+\varDelta t}&= \tilde x_t+[\sigma_{j_t|i_t}(\tilde x_t) -\sigma_{j_t|i_t}(x_t)]\hat{Z}_t,
  \label{eq:S3_euler_antiIto}
\end{align}
where $\sigma_{j|i}(x)\equiv\sqrt{2D_{j|i}(x)\varDelta t}$ and $\hat Z_{t}$ is a standard normally distributed random number, i.e., $\hat Z_{t}\sim\mathcal N(0,1)$.
The first line of Eq.~\eqref{eq:S3_euler_antiIto} estimates the updated position
with the value $\tilde{x}_t$ after which the second line effectively
``replaces'' the last term of the first line by $\sigma_{j_t|i_t}(\tilde x_t) \hat{Z}_t$.
Eq.~\eqref{eq:S3_euler_antiIto} becomes the well-known Euler–Maruyama method
if $D_{i|j}(x)$ is constant, since the second line then simplifies to
$x_{t+\varDelta t}=\tilde x_t$. Once the position exceeds the outer boundary $x_{t+\varDelta t},\tilde x_t>l_{j_t|i_t}$ or the inner boundary $x_{t+\varDelta t},\tilde x_t<0$,
$j_t$ and $i_t$ are updated according to Tab.~\ref{tab:update_boundary}.
We note the pathwise error of the Euler scheme \eqref{eq:S3_euler_antiIto} (i.e. the strong error) \cite{kloe94} is expected to scale as $\propto\varDelta t^{0.5}$, i.e. the scheme is of the strong order 0.5.



\subsection{Milstein scheme (strong order 1.0)}\label{subsubsec:Milstein}
Since to our knowledge higher order stochastic Runge-Kutta schemes can only be found in the literature for It\^o or Stratonovich integrals \cite{kloe94} we will
now use the It\^o representation of the equation of motion \eqref{eq:S3_Langevin_Ito}.
In the case of multiplicative noise $D_{j|i}'(x)\neq0$ the Euler
scheme from Appendix~\ref{subsubsec:naive}  can be improved according to the 
Milstein scheme \cite{kloe94}, which is of strong order 1.0  (i.e., pathwise error scales as $\propto\varDelta t^{1.0}$). This scheme propagates the system from time $t$ to time $ t+\varDelta t$
according to
 \begin{align}
  x_{t+\varDelta t}={}&x_t+\Big[\beta D_{j_t|i_t}(x_t)F_{j_t|i_t}(x_t)+D_{j_t|i_t}'(x_t)\Big]\varDelta t\nonumber\\
  &+\sqrt{2D_{j_t|i_t}(x_{t})\varDelta t} Z_t+\frac{D_{j_t|i_t}'(x_{t})}{2}[ Z_t^2-1]\varDelta t\nonumber\\
  ={}&x_t+\beta D_{j_t|i_t}(x_t)F_{j_t|i_t}(x_t) \varDelta t
  \nonumber\\
  &+\sqrt{2D_{j_t|i_t}(x_{t})\varDelta t}Z_t+\frac{D_{j_t|i_t}'(x_{t})}{2}[ Z_t^2+ 1]\varDelta t, 
  \label{eq:S3_Milstein}
 \end{align}
where $Z_t\sim\mathcal{N}(0,1)$. The last term in the second line  of Eq.~\eqref{eq:S3_Milstein} reduces the
 pathwise error from ${\rm \epsilon}\propto\varDelta t^{0.5}$ to ${\rm \epsilon}\propto\varDelta t^{1.0}$. In the last step in Eq.~\eqref{eq:S3_Milstein} we solely combined the terms containing the derivative of the diffusion coefficient.
Once the position exceeds the outer boundary $x_{t+\varDelta t}>l_{j_t|i_t}$ or the inner boundary $x_{t+\varDelta t}<0$,
$j_t$ and $i_t$ are updated according to Tab.~\ref{tab:update_boundary}.

 \subsection{Stochastic Runge-Kutta with additive noise (strong order 1.5)}\label{subsubsec:Runge_Kutta}
For the simulation of stochastic trajectories with a constant diffusion
coefficient $D_{j|i}(x)=\text{const.}=D$ (i.e., additive noise),
we use an explicit stochastic Runge-Kutta scheme of strong order 1.5 from Ref.~\cite{kloe94} (see also Ref.~\cite{chan87a}), which involves the following steps assuming a time increment $\varDelta t$.
In order to update from $x_t$ to $x_{t+\varDelta t}$
we first generate two independent standard normally distributed random numbers,
$\hat Z_t\sim\mathcal{N}(0,1)$ and  $\zeta_t\sim\mathcal{N}(0,1)$, calculate
$\hat R_t=\hat Z_t/2+\zeta_t\cdot\sqrt{3}/6$
and then update the position according to \cite{chan87a}
 \begin{align}
  q_t&=x_t+\frac{1}{2}\beta D F_{j_t|i_t}(x_t)\varDelta t,\nonumber\\
  q_t^*&=q_t+\frac{3}{2}\sigma \hat R_t, \label{eq:S3_Runge_Kutta}\\
  x_{t+\varDelta t}&=x_t+\sigma \hat Z_t+\Big[\frac{\beta D F_{j_t|i_t}(q_t)+2\beta D F_{j_t|i_t}(q_t^*)}{3}\Big]\varDelta t,\nonumber
 \end{align}
%
where $\sigma=\sqrt{2D \varDelta t}$. We emphasize that this stochastic Runge-Kutta scheme is of strong order 1.5 and assumes the diffusion coefficient to be constant. Moreover, this scheme requires \emph{two} random numbers instead of \emph{one} in each iteration step.
A quite comprehensive collection of further higher order stochastic integration schemes can be found in Ref.~\cite{kloe94}, which in contrast to Eq.~\eqref{eq:S3_Runge_Kutta}
require generating non-Gaussian random numbers.

\subsection{Evaluation of dwell and transition-path time functionals}\label{subsubsec:functionals}
The waiting time in one reduced network state spans the time period between the first entrance into a network state on a graph until the first entrance to another state (see Fig.~\ref{fig:blinking}), i.e., the time between two state changes. The dwell and transition-path time dissect the residence time interval into two separate intervals, in which the last recurrence (revisit) of the same state before changing to another state
terminates the dwell time $\tau$ \emph{and} initiates the transition-path time period $\delta t$ that in turn spans the remaining time until the state changes. To numerically evaluate dwell and transition-path time functionals defined in Eqs.~\eqref{eq:trans-time_def} and \eqref{eq:dwell-time_def}
we perform the following computational steps.

\begin{table*}
\caption{Update of dwell and transition-path time functionals.
\normalfont The initial position is assumed to be $x_0=0$, time of the last
recurrence
is initially set to $T_{\rm rec}=0$ along with the time of the last state change to $T_{\rm last}=0$.
Each passage accross an outer boundary results in a transition-path time $\delta t$ and dwell time $\tau$ that correspond to one transition event along a single transition $\gamma$.}
\label{tab:update_functionals}
 \centering
 \begin{tabular}{r|c|c|c}\hline\hline
\multicolumn{2}{c|}{}&\multicolumn{2}{c}{update conditions at the boundary}\\\hline
 step&\textbf{functional}&\textbf{outer boundary} ($x_t\ge l_{j_t|i_t}$)&\textbf{inner boundary} ($x_t\le0$)\\\hline
 $1^\text{st}$& -- &&set $T_{\rm rec}=t$\\\hline
 $2^\text{nd}$ & splitting transition & store one transition $\gamma=(i_t\to j_t)$& \\\hline
 $3^\text{rd}$ & dwell time $\tau$& store $\tau=T_{\rm rec}-T_{\rm last}$  in transition $\gamma$& \\\hline
 $4^\text{th}$& transition-path time $\delta t$& store $\delta t=t-T_{\rm rec}$ in transition $\gamma$ \\\hline
 $5^\text{th}$& --& set $T_{\rm last}=T_{\rm rec}=t$\\\hline\hline
 \end{tabular}
\end{table*}

Whenever the position $x_t$ exceeds the ``outer boundary'' $x_{t}\ge
l_{j_t|i_t}$ (i.e. the state changes)  or the ``inner boundary''
$x_{t}\le 0$ (i.e. to a recurrence) which both represent a ``state visit'' the variables $j_t,i_t$ are updated according to Tab.~\ref{tab:update_boundary}. Any update of $j_t,i_t$ according to Tab.~\ref{tab:update_boundary} is accompanied with a change of dwell time $\tau$ and transition-path time $\delta t$ according to Tab.~\ref{tab:update_functionals}. Thereby, $T_{\rm rec}$ denotes the last recurrence of a network state and $T_{\rm last}$ the time of the last state-change. Each transition event is stored in a list  for all transitions $\gamma$ (see second step in Tab.~\ref{tab:update_functionals}).

\subsection{Fokker-Planck equation on local star-like graph}\label{subsec:FPE}
The preceding subsection dealt with single trajectories.
One can cast the Langevin equation \eqref{eq:S3_Langevin_Ito}
into a partial differential equation for the probability density
function -- the socalled Fokker-Planck equation -- as follows \cite{gard04}.
We pick without loss of generality a state of interest, $i$, and focus us
on a local star-like graph spanned by the $i$-th state.
For a pair of neighboring states $j,k\in \mathcal{N}_i$ at distances $x$ and $y$
within $0\le x\le l_{j|i}$ and $0\le y\le l_{k|i}$ 
the probability density to find the system in the state $(x,j,i)$
after time $t$ starting initially from $(y,k,i)$
denoted by
$\Ploc(x,j,t|y,k)$ satisfies the Fokker-Planck equation
\begin{multline}
 \ddel{t}\Ploc(x,j,t|y,k)=-\ddel{x}\Jloc(x,j,t|y,k)=\\
 =\ddel{x}\e^{-\beta U_{j|i}(x)}D_{j|i}(x)\ddel{x}\e^{\beta U_{j|i}(x)}\Ploc(x,j,t|y,k)\\
 \equiv\LF_{j|i}(x)\Ploc(x,j,t|y,k),
 \label{eq:S3_FPE_forward}
\end{multline}
where $U_{j|i}(x)=-\int_0^x F_{j|i}(x')\dd x'$ and without any loss of generality we assume the diffusion constant to be continuous $D_{j|i}(0)=D_{k|i}(0)$ for all $j,k\in\mathcal{N}_i$. Note that $\Jloc(x,j,t|y,k)\equiv D_{j|i}(x)[\beta F_{j|i}(x)-\del_x]\Ploc(x,j,t|y,k)$ denotes the probability flux away from $i$ and $\LF_{j|i}(x)$ denotes the (forward) Fokker-Planck operator.
The initial probability density is set to $\Ploc(x,j,t|y,k)=\delta(x-y)\delta_{j k}$, where $\delta(x-y)$ is the delta-function and $\delta_{j k}$ the Kronecker-delta.
The inner boundary conditions read
\begin{equation}
\label{eq:S3_forward_boundary}
 \begin{aligned}
 &\Ploc(0,j,t|y,k)=\Ploc(0,j',t|y,k), \quad\forall j'\in\mathcal{N}_i,\\
 &\sum_j \Jloc(0,j,t|y,k)=0,
\end{aligned}
\end{equation}
which reflect that trajectories are continuous and fluxes are conserved according to Kirchhoff's law.
Note that the generalization to both, diverging force-kicks and discontinuous diffusion landscapes, is explicitly discussed in
Appendix~\ref{subsec:SI_discontinuous_landscapes}. Hence, we may derive all results based on Eq.~\eqref{eq:S3_forward_boundary} in order to render the derivations less tedious.

There are two distinct boundary conditions at the outer end of the $j$th leg adjacent to node $i$ (i.e. $j\in\mathcal{N}_i$),
which correspond to
\begin{equation}
\label{eq:S3_forward_boundary2}
 \begin{aligned}
  \Ploc(l_{j|i},j,t|y,k)&=0&&
  \left[
  \begin{aligned}
  &\text{if the $j$-th outer}\\[-1mm]
  &\text{boundary is absorbing},
  \end{aligned}\right.\\
  \Jloc(l_{j|i},j,t|y,k)&=0&&
    \left[ \begin{aligned}
  &\text{if the $j$-th outer}\\[-1mm]
  &\text{boundary is reflecting}.
  \end{aligned}\right.
 \end{aligned}
\end{equation}
For all absorbing ends  $\Ploc(x,j,t|y,k)\dd x$ is the probability  that a trajectory starting from
distance $y$ from state $i$ in direction towards state $k$
will be at time $t$ within the interval $x$ and $x+\dd x$
having never reached any of the neighboring states $\neq i$. In this case
the survival probability,
$\Sloc(t|y,k)=\sum_j\int_0^{l_{j|i}}\Ploc(x,j,t|y,k)\dd x$,
decays monotonically in time from $\Sloc(0|y,k)=1$ to $\Sloc(\infty|y,k)=0$.
More precisely,
if $\Ploc(l_{j|i},j,t|y,k)=0,$ $\forall j\in\mathcal{N}_i$,
we obtain
\cite{redn01,gard04}
\begin{align}
  \wpcStar_{j|i}(t)
  &
=\Jloc(l_{j|i},j,t|0,k)
  \nonumber
  \\
  &
  =-D_{j|i}(l_{j|i})\ddel{x}\Ploc(x,j,t|y,k)\big|_{x=l_{j|i}},
\end{align}
where $\wpcStar_{j|i}$ is the local state-to-state kinetics with $-\del_t \Sloc(t|0,k)=\sum_j\wpcStar_{j|i}(t)=\wpcexit_i(t)$ for all $k$; the Laplace transform of $\wpcStar_{j|i}$ is given in Eq.~\eqref{eq:twpc}.

\section{Proof of symmetry and independence of dwell and transition-path time using Green's function theory}
\label{sec:greens}
In the following we prove that diffusive dynamics on a graph
\eqref{eq:S3_FPE_forward} renders dwell and transition-path times conditionally independent functionals. We first show that the aforementioned conditional independence follows directly from the definition of the coarse-graining (last visited state) based on the \emph{gedanken experiment} from
Fig.~\ref{fig:blinking}.
Using Green's function theory we then prove the following two symmetries entering Eq.~\eqref{eq:local_decomposition}: (i)~the dwell-time statistics solely depend on the initial state, and  (ii) the transition-path time obeys a forward/backward symmetry.

\subsection{Proof of conditional independence between transition-path and dwell time}\label{subsec:informal}
The independence of transition-path and dwell time follows immediately from the coarse-graining of the full trajectory once we realize that it effects an ``erasure of memory''.

The micro-state $\bx_t=(x_t,j_t,i_t)$ is characterized by the last visited state $i_t$,  and the distance $x_t$ from the last visited state in direction to the instantaneously targeted state $j_t$. Each recurrence in Fig.~\ref{fig:blinking}, highlighted by colored crosses,
represents a state-visit $x_t=0$, which in turn fully determines the micro-state via $\bx_t=(0,j_t,i_t)\mathrel{\widehat{=}}(0,k,i_t)\mathrel{\widehat{=}}(l_{j_t|i_t},i_t,j_t)$, where the symbol ``$\widehat{=}$'' refers to parameters corresponding to the \emph{same} instantaneous micro-state (see Fig.~\ref{fig:network_microstate_redundancy}). Since the micro-state $\bx_t$ is assumed 
to follow Markovian kinetics we find that the future state-visit depends only on the last state-visit
\emph{not on the state-visits before}, which triggers a renewal of the dynamics.
Since a transition spans the time \emph{after} the last revisit  of a state  (recurrence)
and the dwell time spans the time \emph{before} the last revisit of a state (see Fig.~\ref{fig:blinking}b), said revisit of a state causes their statistical independence.
This completes the proof of independence between transition-path and dwell time.
In the following we derive symmetries of transition-path and dwell times using the underlying Fokker-Planck equation on a graph.

%


\subsection{Laplace transform of the Fokker-Planck equation on a graph}
Let us first write the Fokker-Planck equation
in terms of the current operator $\dLF_{j|i}(x)=-D_{j|i}(x)\e^{-\beta U_{j|i}(x)}\del_x\e^{\beta U_{j|i}(x)}$ which allows to rewrite
Eq.~\eqref{eq:S3_FPE_forward} in the form
\begin{equation}
 \ddel{t}\Ploc(x,j,t|y,k)=-\ddel{x}\dLF_{j|i}(x)\Ploc(x,j,t|y,k).
 \label{eq:S4_FPEcurrent}
\end{equation}
The Laplace transform $\tPloc(x,j,s|y,k)=\int_0^\infty \e^{-st}\Ploc(x,j,t|y,k)\dd t$
transforms the Fokker-Planck equation \eqref{eq:S4_FPEcurrent} into
\begin{equation}
 \big[s+\ddel{x}\dLF_{j|i}(x)\big]\tPloc(x,j,s|y,k)=\delta(x-y)\delta_{jk}.
  \label{eq:S4_FPE_Laplace}
\end{equation}
From Eq.~\eqref{eq:S4_FPE_Laplace}
follows the continuity of probability $\tPloc(x,j,s|y,j)\big|_{x=y+0}=\tPloc(x,j,s|y,j)\big|_{x=y-0}$ and jump-discontinuity at $x=y$ of the current $\dLF_{j|i}(x)\tPloc(x,j,s|y,j)\big|_{x=y+0}-\dLF_{j|i}(x)\tPloc(x,j,s|y,j)\big|_{x=y-0}=1$.
Let us now express the solutions $\tPloc$ in terms of
the homogeneous solutions $\gInner_{j|i}(x),\gOuter_{j|i}(x)$ satisfying
\begin{equation}
  \big[s+\ddel{x}\dLF_{j|i}(x)\big]\gInner_{j|i}(x)=0,\quad
  \big[s+\ddel{x}\dLF_{j|i}(x)\big]\gOuter_{j|i}(x)=0,
  \label{eq:S4_sol_hom}
\end{equation}
with $\gInner_{j|i}(0)=0$ and $\gOuter_{j|i}(l_{j|i})=0$.
Introducing further the current functions
 $\JInner_{j|i}(x)\equiv\dLF_{j|i}(x)\gInner_{j|i}(x)$ and $\JOuter_{j|i}(x)\equiv\dLF_{j|i}(x)\gInner_{j|i}(x)$,
 one can easily check that the limit $y\to 0$ yields the solution 
\begin{equation}
  \tPloc(x,j|0,k)=\frac{
  \gOuter_{j|i}(x,s)
/
  \gOuter_{j|i}(0,s)
  }{\sum_{n\in\mathcal{N}_i}
  \JOuter_{n|i}(0,s)
/
  \gOuter_{n|i}(0,s)}
  ,
  \label{eq:S4_Pcenter}
\end{equation}
which is equal for all $k\in\mathcal{N}_i$ and continuous in $x=0$.
Conversely, the currents are discontinuous at $x=0$ according to
\eqref{eq:S4_FPE_Laplace}.
Eq.~\eqref{eq:S4_Pcenter} also solves $\tPloc(l_{j|i},j|0,k)=0$ for all $j,k\in\mathcal{N}_i$.
The Laplace transform of the first passage time density, $\twpcStar_{j|i}(s)$, is obtained from
 the outward current \cite{redn01,gard04} at position $x=l_{j|i}$, finally yielding
\begin{align}
   \twpcStar_{j|i}(s)& \equiv\dLF_{j|i}(x)\tPloc(x,j|0,k)\Big|_{x=l_{j|i}}
   \nonumber\\
   &=
   \frac{
   \JOuter_{j|i}(l_{j|i},s)
   /
   \gOuter_{j|i}(0,s)
   }{
   \sum\limits_{k\in\mathcal{N}_i}
   \JOuter_{k|i}(0,s)
   /
   \gOuter_{k|i}(0,s)
   }
  \label{eq:S4_twpcStar}.
\end{align}
The zeroth-order moment of Eq.~\eqref{eq:S4_twpcStar} -- the splitting probability -- is in turn simply given by
\begin{equation}
 \phiStar_{j|i}\equiv \twpcStar_{j|i}(0)=\frac{
 \JOuter_{j|i}(l_{j|i},0)
 /
 \gOuter_{j|i}(0,0)
 }{\sum_{k\in\mathcal{N}_i}
 \JOuter_{k|i}(0,0)
 /
 \gOuter_{k|i}(0,0)
 }.
 \label{eq:S4_phiStar}
\end{equation}
In the following we decompose \eqref{eq:S4_twpcStar} \emph{exactly} into the splitting probability, transition-path-time and dwell-time statistics.

\subsection{Transition-path-time statistics from Green's function along a single leg with absorbing boundary conditions}

Before taking the limit in Eq.~\eqref{eq:ptrans_Gtr_limit} we Laplace
transform the Green's function $\tilde G_{j|i}^{\rm
  tr}(x,s|y)=\int_0^\infty \e^{-st}G_{j|i}^{\rm tr}(x,t|y)\dd t$
which, using the solutions Eq.~\eqref{eq:S4_sol_hom}, can be written
in the form
\cite{keil64,meln12,hart19}
\begin{equation}
 \tilde G_{j|i}^{\rm tr}(x,s|y)=
 \begin{cases}
  \dfrac{\gInner_{j|i}(y,s)\gOuter_{j|i}(x,s)}{w_{j|i}(y,s)}&\text{if $x\ge y$,}\\[1mm]
  \dfrac{\gInner_{j|i}(x,s)\gOuter_{j|i}(y,s)}{w_{j|i}(y,s)}&\text{if $x\le y,$}
 \end{cases}
\end{equation}
where we defined the Wronskian satisfying \cite{keil64,hart19}
\begin{align}
 w_{j|i}(y,s)&=\gInner_{j|i}(y,s)\JOuter_{j|i}(y,s)-\JInner_{j|i}(y,s)\gOuter_{j|i}(y,s)\nonumber\\
 &= w_{j|i}(x,s)\e^{\beta U_{j|i}(x)-\beta U_{j|i}(y)}.
 \label{eq:S4_Wronskian}
\end{align}
At the boundaries the Wronskian becomes $w_{j|i}(l_{j|i},s)=\gInner_{j|i}(l_{j|i},s)\JOuter_{j|i}(l_{j|i},s)$ and $w_{j|i}(0,s)=-\JInner_{j|i}(0,s)\gOuter_{j|i}(0,s)$ due to $\gOuter_{j|i}(l_{j|i},s)=\gInner_{j|i}(0,s)=0$.
Using 
 \begin{equation}
   \tilde J_{j|i}^{\rm tr}(x,s|y)=\dLF_{j|i}\tilde G_{j|i}^{\rm tr}(x,s|y)=\frac{\gInner_{j|i}(y,s)\JOuter_{j|i}(x,s)}{w_{j|i}(y,s)}
 \end{equation}
 for $x>y$, the Laplace image of the probability density of the transition-path time
becomes
\begin{widetext}
 \begin{multline}
 \twpctrans_{j|i}(s)
 =\lim_{y\to0} \frac{\tilde J_{j|i}^{\rm tr}(l_{j|i},s|y)}{\tilde J_{j|i}^{\rm tr}(l_{j|i},0|y)}
 =\lim_{y\to0}\frac{\gInner_{j|i}(y,s)}{\gInner_{j|i}(y,0)}\times\frac{w_{j|i}(y,0)}{w_{j|i}(y,s)}\times
 \frac{\JOuter_{j|i}(l_{j|i},s)}{\JOuter_{j|i}(l_{j|i},0)}
   \\[1mm]
 =
 \lim_{y\to0}\frac{\JOuter_{j|i}(y,0)\big[
 w_{j|i}(y,s)
+\JInner_{j|i}(y,s)\gOuter_{j|i}(y,s)\big]
}{\JOuter_{j|i}(y,s)\big[
w_{j|i}(y,0)+\JInner_{j|i}(y,0)\gOuter_{j|i}(y,0)\big]
}\times
\smash{
 \overbrace{
 \frac{w_{j|i}(0,0)}{w_{j|i}(0,s)}
 }^{\text{\normalsize${=}\frac{\JInner_{j|i}(0,0)\gOuter_{j|i}(0,0)}{\JInner_{j|i}(0,s)\gOuter_{j|i}(0,s)}$}\hidewidth}
 }
 \times
 \frac{\JOuter_{j|i}(l_{j|i},s)}{\JOuter_{j|i}(l_{j|i},0)}
 %
 %
 %
 %
 %
,
\label{eq:S4_trans_greens_pre}
\end{multline}
\end{widetext}
where we performed the following algebraic steps.  From the first to
the second line of Eq.~\eqref{eq:S4_trans_greens_pre} we rewrote the
first fraction (which is formally undetermined ``$0/0$''
in the limit $y\to 0$) first, by using Eq.~\eqref{eq:S4_Wronskian} with $x=0$,
inserting the resulting $\gInner_{j|i}(y,s)$, and using $\lim_{y\to0}w_{j |i}(y,s)=w_{j |i}(0,s)$. Since $\JOuter_{j|i}(y,s)$
does not have a singularity in the limit $\lim_{y\to 0}\JOuter_{j|i}(y,s)$,
the singularity ``$0/0$''
is solely encoded in the bracketed term  ``$[\cdots]\to0$'', and cancels in the limit $y\to 0$
in both numerator and denominator.  Employing l'Hospital's rule (on the bracketed terms ``$[\cdots]$'')
we now determine their first derivative with respect to $y$ at $y=0$
\begin{multline}
 D_{j|i}(0)\frac{\partial}{\partial y} \Big[w_{j|i}(y,s)
+\JInner_{j|i}(y,s)\gOuter_{j|i}(y,s)\Big]_{y=0}\\
=\beta D_{j|i}(0) \Big[-\overbrace{w_{j|i}(0,s)}^{=-\JInner_{j|i}(0,s)\gOuter_{j|i}(0,s)\hidewidth}\beta U_{j|i}'(0)-s\overbrace{\gInner_{j|i}(0,s)}^{=0}\gOuter_{j|i}(0,s)\\+\JInner_{j|i}(0,s)\frac{\del}{\del y}\gOuter_{j|i}(y,s)\Big]_{y=0}
=-\JInner_{j|i}(0,s)\JOuter_{j|i}(0,s),
\label{eq:S4_trans_greens_aux}
\end{multline}
where $U_{j|i}'(y)=\del_yU_{j|i}(y)$, and we have deduced $\del_yw_{j|i}(y,s)=-U_{j|i}'(y)w_{j|i}(y,s)$ from the left side of Eq.~\eqref{eq:S4_Wronskian}.
Furthermore, we used
$\del_y \JInner_{j|i}(y,s)=-s \gInner_{j|i}(y,s)$ following from Eq.~\eqref{eq:S4_sol_hom}, and finally employed $\JOuter_{j|i}(y,s)=- D_{j|i}(y)\beta U_{j|i}'(y)\gOuter_{j|i}(y,s)-D_{j|i}(y)\del_y\gOuter_{j|i}(y,s)$. Inserting Eq.~\eqref{eq:S4_trans_greens_aux} into Eq.~\eqref{eq:S4_trans_greens_pre}
and applying l'Hospital's rule finally yields the Laplace transform of the probability density of the
transition time
\begin{multline}
  \twpctrans_{j|i}(s)=\frac{\JOuter_{j|i}(0,0)}{\JOuter_{j|i}(0,s)}
  \frac{\JInner_{j|i}(0,s)\JOuter_{j|i}(0,s)}{\JInner_{j|i}(0,0)\JOuter_{j|i}(0,0)}
  \\
  \times 
\frac{\JInner_{j|i}(0,0)\gOuter_{j|i}(0,0)}{\JInner_{j|i}(0,s)\gOuter_{j|i}(0,s)}
  \times
   \frac{\JOuter_{j|i}(l_{j|i},s)}{\JOuter_{j|i}(l_{j|i},0)}
   \\=\frac{\gOuter_{j|i}(0,0)\JOuter_{j|i}(l_{j|i},s)}{\gOuter_{j|i}(0,s)\JOuter_{j|i}(l_{j|i},0)}.
  \label{eq:S4_trans_greens_final}
\end{multline}
Since a function and its Laplace image are mapped one-to-one, 
Eq.~\eqref{eq:S4_trans_greens_final} fully characterizes the statistics of transition-path time.

\subsection{Forward/backward symmetry of transition-path-time statistics}\label{subsec:forward/backward}
The statistics of the corresponding backward transition can be determined in an analogous manner as Eq.~\eqref{eq:S4_trans_greens_final}.
Identifying $\gOuter_{i|j}(y,s)=\gInner_{j|i}(l_{j|i}-y,s)$ and $\JOuter_{i|j}(y,s)=-\JInner_{j|i}(l_{j|i}-y,s)$ the backward transition-path time statistics $\twpctrans_{j|i}(s)$
become
\begin{multline}
  \twpctrans_{i|j}(s)=\frac{\gInner_{j|i}(l_{j|i},0)\JInner_{j|i}(0,s)}{\gInner_{j|i}(l_{j|i},s)\JInner_{j|i}(0,0)}=
  \\
  \overbrace{\frac{w_{j|i}(l_{j|i},0)w_{j|i}(0,s)}{w_{j|i}(l_{j|i},s)w_{j|i}(0,0)}}^{\stackrel{\hidewidth\text{Eq.~}\eqref{eq:S4_Wronskian}}{=}1}
  \frac{\gOuter_{j|i}(0,0)\JOuter_{j|i}(l_{j|i},s)}{\gOuter_{j|i}(0,s)\JOuter_{j|i}(l_{j|i},0)}=\twpctrans_{j|i}(s),
\end{multline}
where the first step follows from Eq.~\eqref{eq:S4_trans_greens_final} and
in the second step we used Eq.~\eqref{eq:S4_Wronskian}; in the last step
we identified $\twpctrans_{j|i}(s)$ in  Eq.~\eqref{eq:S4_trans_greens_final}, which completes the proof of $\twpctrans_{i|j}(s)=\twpctrans_{j|i}(s)$.
In other words, we have hereby  proven that  the duration of both transitions
$i\to j$ and $j\to i$ is identically distributed.

A similar derivation can be found in Ref.~\cite{bere06} for
underdamped  systems in which the momentum is assumed to be
equilibrated.

\subsection{Dwell-time statistics obey a state-symmetry}
Let $t$ denote the time of exiting from state $i$ towards state $j$ for the first time
and $\delta t$ the corresponding transition-path time, which are distributed according  to the probability densities $\wpcStar_{j|i}(t)$ and $\wpctrans_{j|i}(\delta t)$, respectively. The Laplace transform of $\wpcStar_{j|i}(t)$ and $\wpctrans_{j|i}(\delta t)$ will be denoted by
$\twpcStar_{j|i}(s)$ and $\twpctrans_{j|i}(s)$. The transition-path-time statistics $\wpctrans$ (or $\twpctrans$) do \emph{not} depend on the time at which a transition path
begins and, hence, are independent of the time interval before $\tau=t-\tau$
that is called the dwell-time period.
Therefore, we can obtain the statistics of the dwell time via
de-convolution
which in Laplace space becomes a simple division
\begin{multline}
 \twpcdwell_{j|i}(s)=\frac{\twpcStar_{j|i}(s)}{\phiStar_{j|i}\twpctrans_{j|i}(s)}\\=
 \frac{\frac{\JOuter_{j|i}(l_{j|i},s)}{\gOuter_{j|i}(0,s)}}{\sum_{k\in\mathcal{N}_i}\frac{\JOuter_{k|i}(0,s)}{\gOuter_{k|i}(0,s)}}
 \frac{\sum_{k\in\mathcal{N}_i}\frac{\JOuter_{k|i}(0,0)}{\gOuter_{k|i}(0,0)}}{\frac{\JOuter_{j|i}(l_{j|i},0)}{\gOuter_{j|i}(0,0)}}
 \frac{\gOuter_{j|i}(0,s)\JOuter_{j|i}(l_{j|i},0)}{\gOuter_{j|i}(0,0)\JOuter_{j|i}(l_{j|i},s)}
\\
=
\frac{\sum_{k\in\mathcal{N}_i}
\JOuter_{k|i}(0,0)
/
\gOuter_{k|i}(0,0)
}{
\sum_{k\in\mathcal{N}_i}
\JOuter_{k|i}(0,s)
/
\gOuter_{k|i}(0,s)
}
=\twpcdwell_{i}(s),
\label{eq:S4_dwell}
\end{multline}
where in the second line we inserted $\twpcStar_{j|i}(s)$ from Eq.~\eqref{eq:S4_twpcStar},
$\phiStar_{j|i}$ from Eq.~\eqref{eq:S4_phiStar} and
$\twpctrans_{j|i}(s)$ from Eq.~\eqref{eq:S4_trans_greens_final}; in
the last line of Eq.~\eqref{eq:S4_dwell} we canceled equal terms in
the numerator and denominator, respectively.
Strikingly, we find that the result does \emph{not} depend on the final state $j$,
which is why the dwell-time statistics obeys a state-symmetry, meaning that it only depends on the initial state $i$.
Therefore, we can write $ \twpcdwell_{j|i}(s)$ as $\twpcdwell_{i}(s)$ in the last step of Eq.~\eqref{eq:S4_dwell}. Since the product $\twpcStar_{j|i}(s)=\phiStar_{j|i}\twpctrans_{j|i}(s)\twpcdwell_{j|i}(s)$ in Laplace space becomes a convolution in the time domain, we have hereby completed the proof of
Eq.~\eqref{eq:local_decomposition}.

\subsection{Concluding remarks on the proofs}

To summarize this section we have shown in
Appendix~\ref{subsec:informal} that each change of state $i\to j$ in a
network is taken with (splitting) probability $\phiStar_{j|i}$ and has
a corresponding distribution of residence time $t$ in turn being a sum of conditionally independent dwell time $\tau$ and transition-path time $\delta t=t-\tau$ (for a given transition between the pair of states $i\to j$). We have proven two symmetries. First,
we proved in Appendix~\ref{subsec:forward/backward} that the statistics of transition-path time obeys a forward/backward symmetry $\wpctrans_{j|i}(\delta t)=\wpctrans_{i|j}(\delta t)$. Second, the statistics of dwell time
was proven in Eq.~\eqref{eq:S4_dwell} to depend solely on the initial
state $i$ -- that is, the dwell-time statistics does not depend on the state  $j$ to which the trajectory finally transits.

{\color{mynewcolor}
\section{Decomposition of paths}
}\label{sec:path_decomposition}
\subsection{Generalized Renewal theorem}\label{subsec:Renewal}

The classical renewal theorem \cite{sieg51} connects the first passage time density to the
propagator,  $P(a,T|i_0)$,
which is the probability to find the network in state $i_T=a$
at time $T$ given that it was initially in state $i_0$.
It can be understood as a decomposition of paths:
any system that starts at  $i_0$ and arrives at  $i_T=a$ at time $T$ 
must have reached $a$ for the first time at time $t\le T$  ($i_t=a$)
and then either stayed there or returned again after time $T-t$.
In mathematical terms this corresponds to  \cite{sieg51}
\begin{equation}
 P(a,T|i_0)=\int_{0}^T\dd t   \underbrace{P(a,T-t|a)}_{\hidewidth\text{return probability}\hidewidth}\overbrace{\wps_{a|i_0}(t)}^{\hidewidth\text{first passage $i_0\to a$}\hidewidth},
 \label{eq:S1_renewal_classic}
\end{equation}
where $\wps_{a|i_0}(t)\dd t$ denotes the probability that
the process starting from $i_0$ reaches the position $i_{t'}=a$ for the first time within the interval $t\le t'\le t+\dd t$. We refer to $\wps_{a|i_0}(t)$ as the unconditioned first passage time density to  the target state $a$ given that the system initially started from $i_0$. We call a  first passage problem  ``unconditioned'' if there is just one target state $a$ as in Eq.~\eqref{eq:S1_renewal_classic}.
The renewal theorem  \eqref{eq:S1_renewal_classic} that connects the propagation of a system, characterized by $P$,
to unconditioned first passage functionals embodied in the
probability density $\wps_{a|i_0}(t)$ has been routinely used to study first passage phenomena \cite{redn01,hart18}.

\begin{figure*}
 \centering
\includegraphics{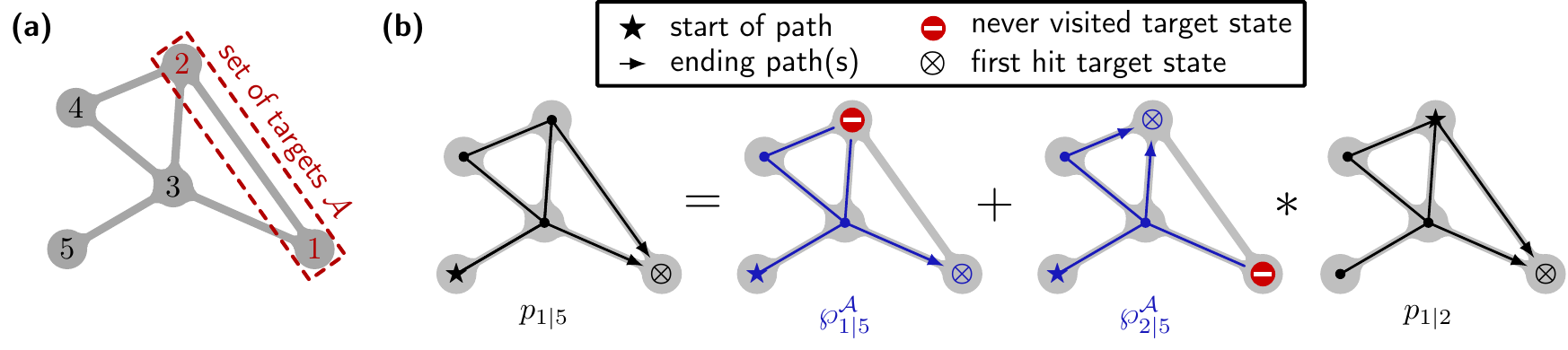}
 \caption{Illustration of generalized renewal theorem. (a)~Network with five states $\Omega=\{1,2,3,4,5\}$.
 The dashed box encloses one set of target state $\mathcal{A}$, which is here chosen to be $\mathcal{A}=\{1,2\}$. The network corresponds to the subnetwork from
 Fig.~\ref{fig:illustrantion_all}e
 in the main text (see also
 Fig.~\ref{fig:blinking}
 therein).  (b)~Path decomposition according to generalized renewal theorem from Eq.~\eqref{eq:S1_renewal_cond_pre} or equivalently Eq.~\eqref{eq:S1_renewal_cond} for the special case of two target states $\mathcal{A}=\{1,2\}$ with $a=1$. Each arrow tip corresponds to a path ending at the particular state without having visited that state before; each filled dot corresponds is a state that may have been visited.}
 \label{fig:renewal_gen_SI}
\end{figure*}

To study conditional first passage problems \cite{redn01} that involve more then a single target state $a$, we need to generalize the renewal theorem \eqref{eq:S1_renewal_classic}
in the following way. Let us consider a set of target states $\mathcal{A}$
corresponding to a subset of all network states $\Omega=\{1,\ldots,N\}$, that is, $\mathcal{A}\subset\Omega$.
A conditional first passage problem asks for the first time until the system reaches
the target state
$a\in\mathcal{A}$
given that it has not yet visited any of the other target states from $ \mathcal{A}\backslash\{a\}$. 
The problem is characterized by the joint density, $\wpc_{a|i_0}(t)$,
to enter the set of target $\mathcal{A}$ for the first time at time $t$ and hitting the specific target $a\in\mathcal{A}$ upon
starting from $i_0$, with normalization $\sum_{a\in \mathcal{A}}\int_0^\infty\wpc_{a|i_0}(t)\dd t=1$.
Note that the full set of neighboring states as targets, $\mathcal{A}=\mathcal{N}_{i_0}$, corresponds to $\wpc_{a|i_0}=\wp^{\rm loc}_{a|i_0}$ for all $a\in \mathcal{N}_{i_0}$.
In the spirit of the classical renewal theorem \eqref{eq:S1_renewal_classic}
we find that the conditional first passage density  to any subset $\mathcal{A}$,
$\wpc$, is related to the simpler unconditioned first passage time densities
according to
\begin{align}
 \wps_{a|i_0}(t)&=\wpc_{a|i_0}(t)+\sum_{a'\in \mathcal{A}\backslash\{a\}}\int_0^t \wps_{a|a'}(t'-t)\wpc_{a'|i_0}(t')\dd t'\nonumber\\
 &\equiv \wpc_{a|i_0}(t)+\sum_{a'\in \mathcal{A}\backslash\{a\}}\wps_{a|a'}*\wpc_{a'|i_0}(t),
 \label{eq:S1_renewal_cond_pre}
\end{align}
which is a generalization of the renewal theorem to conditioned first passage problems; in the last step we introduced ``$*$''
as the one-sided convolution operation.
An illustration of the generalized renewal theorem for a network with five states $\Omega=\{1,2,3,4,5\}$ and two target states $\mathcal{A}=\{1,2\}$
with initial condition $i_0=3$ is shown in Fig.~\ref{fig:renewal_gen_SI}.
In the simplest case, when the subset $\mathcal{A}$ contains a single element $\mathcal{A}=\{a\}$, we consistently obtain $\wps_{a|i_0}(t)=\wp^{\{a\}}_{a|i_0}(t)$, i.e., the conditioned first passage problem becomes a unconditioned one.
Laplace transforming the renewal theorem \eqref{eq:S1_renewal_cond_pre}, where the Laplace transform of some generic function $f(t)$ is defined as $\tilde f(s)=\int_0^\infty \e^{-st}f(t)\dd t$ (tacitly assuming that all functions are of exponential order),
we obtain
\begin{equation}
 \twps_{a|i_0}(s)=\twpc_{a|i_0}(s)+\sum_{a'\in \mathcal{A}\backslash\{a\}} \twps_{a|a'}(s)\twpc_{a'|i_0}(s),
\label{eq:S1_renewal_cond}
\end{equation}
where the convolution in the last term of Eq.~\eqref{eq:S1_renewal_cond_pre}
becomes a product after the Laplace transform.
It is worth mentioning that Eq.~\eqref{eq:S1_renewal_cond} via Eq.~\eqref{eq:error_propagation} links diffusion to (unconditioned) first passage statistics of currents as studied in \cite{ging17} (see also Refs.~\cite{garr17,ptas18,proe19a}).

In 
Supplementary Section 1.A \cite{Note2}
we show how the generalized renewal theorem Eq.~\eqref{eq:S1_renewal_cond} can be used to deduce explicit conditional moments of first passage time, which correspond to a multi-target
search problem, in terms of simpler unconditioned ``single-target'' quantities.
As we explain below (see Appendix~\ref{subsec:from_starlike})
one can in fact construct any network problem by
solving for networks with a specific and simpler star-like
topology. This sequential strategy, which we explain in the following, allows for a systematic study of general networks.
%

%
%
%
%
%

\subsection{Renewal theorem on star-like graphs}\label{subsec:renewal_star}

\begin{figure}
 \centering
\includegraphics[width=\columnwidth]{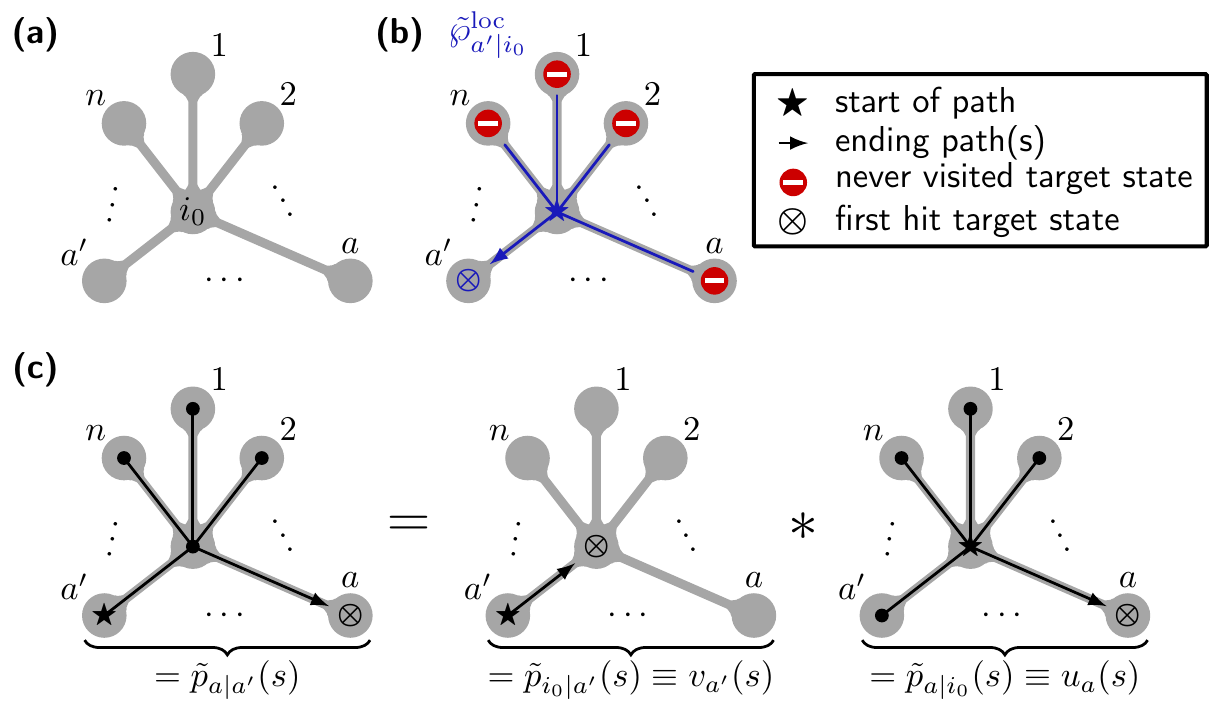}
 \caption{Star-like graph. (a)~Star-like graph with $N$ states from which $n=N-1$ are ``outer states'' $a,a'=1,\ldots,n$ and one state is called the ``inner state'' $N$. 
 (b)~Conditional first passage paths belonging to $\twpcStar_{a'|i_0}(s)$.
 (c)~Illustration of $\twps_{a|a'}(s)=\twps_{a|i_0}(s)\twps_{i_0|a'}(s)$, which holds for $a'\neq a$ and
 effectively means that each path starting from $a'$
 must pass through the center $i_0$ to reach the other end $a$ ($a\neq a'$). The matrix $\mD(s)$ in Eq.~\eqref{eq:matrixform} accounts and corrects for $\twps_{a|a}(s)=1\neq \twps_{a|i_0}(s)\twps_{i_0|a}(s)$. }
 \label{fig:renewal_star_SI}
\end{figure}

Let us for now focus on graphs with a star-like topology,
where all  $n=N-1$ ``outer nodes'' are target states, that is, $\mathcal{A}=\{1,2,\ldots, n\}$,
and the starting node is the  ``inner state''$i_0=N$ as depicted in Fig.~\ref{fig:renewal_star_SI}.
In the case of a star-like topology (i.e., $\mathcal{A}=\mathcal{N}_{i_0}$ and $\wpc_{a|i_0}=\wpcStar_{a|i_0}$) the renewal theorem Eq.~\eqref{eq:S1_renewal_cond} simplifies, meaning that it can be inverted more easily.

In a first, crucial step we realize that each path on a star-like graph, which starts from one end of the star $a$ to another end $a'\neq a$, must pass through the center $i_0$. That is,
the unconditioned first passage time from $a$ to $a'$ is the sum of first passage time from $a'$ to $i_0$ and
the first passage time from $i_0$ to $a$,
which effectively implies $\twps_{a|a'}(s)=\twps_{a|i_0}(s)\twps_{i_0|a'}(s)$ (see Fig.~\ref{fig:renewal_star_SI}c).
Using $\twps_{a|a'}(s)=\twps_{a|i_0}(s)\twps_{i_0|a'}(s)$ for $a'\neq a$,
the renewal theorem \eqref{eq:S1_renewal_cond}
in matrix form becomes
\begin{equation}
u_a(s)=\sum_{a'=1}^n \Big[\mD(s)+\bu(s)\bv(s)^\T\Big]_{aa'}\twpcStar_{a'|i_0}(s),
\label{eq:matrixform}
\end{equation}
where $\bu(s)$ and $\bv(s)$ are vectors with elements $u_a(s)\equiv \twps_{a|i_0}(s)$ and $v_a(s)\equiv \twps_{i_0|a}(s)$, respectively,
and  $\mD(s)$ denotes a diagonal matrix with elements $\mDc_{ii}(s)=1-u_i(s)v_i(s)$,
which corrects for $\twps_{a|a}(s)=1\neq\twps_{a|i_0}(s)\twps_{i_0|a}(s)$. 
Using the Sherman-Morrison-Woodbury formula we are able to invert
the matrix $\mD+\bu\bv^\T$ to get
\begin{align}
 \twpcStar_{a|i_0}(s)
 &=\frac{\frac{u_a(s)}{1-u_a(s)v_a(s)}}{1-n+\sum_{a'}[1-u_{a'}(s)v_{a'}(s)]^{-1}}\nonumber\\
& =\frac{\frac{su_a(s)}{1-u_a(s)v_a(s)}}{(1-n)s+\sum_{a'}\frac{s}{1-u_{a'}(s)v_{a'}(s)}}
 \label{eq:twpc}
\end{align}
which is the \emph{central result of this subsection} that allows us to
obtain conditional many-target first passage time distributions
from simpler unconditioned single-target first passage time densities.
The local splitting probability which formally reads $\phiStar_{a|i_0}=\twpcStar_{a|i_0}(0)$,
can be obtained by taking the limit $s\to 0$.

In the SM \cite{Note2} (see Supplementary Section 1.B)
 we show how Eq.~\eqref{eq:twpc} can be used to express the splitting probability
$\phiStar_{a|i_0}$, the conditional mean first passage $\avgcStar{t}_{a|i_0}$ and the second moment of exit time $\avgexit{t^2}_{i_0}=\sum_a\phiStar_{a|i_0}\avgcStar{t^2}_{a|i_0}$, merely in terms of simpler first and second moments of unconditioned first passage time, $\avgs{t}_{j|i}=-\del_s\twps_{j|i}(s)|_{s=0}$, $\avgs{t^2}_{j|i}=\del_s^2\twps_{j|i}(s)|_{s=0}$ with only a single target state. 
The results for $\phiStar_{a|i_0}$, $\avgcStar{t}_{a|i_0}$
and $\avgexit{t^2}_{i_0}$ in terms of unconditioned moments of first passage times are displayed in 
SM \cite{Note2} (see Supplementary Section 1.B)
 and can be used to derive the main practical result (see Sec.~\ref{subsec:main_practical}) after some quite tedious calculations, which are carried out as follows.


First, we derive all first and second moments of the unconditioned first passage time alongside with the first two moments of transition time (see SM \cite{Note2},
Supplementary Section 2.D).
Second, we insert them into the expressions for
$\phiStar_{a|i_0}$, $\avgcStar{t}_{a|i_0}$
and $\avgexit{t^2}_{i_0}$ listed in 
the SM \cite{Note2}
 [see 
  Eqs.~(S14), (S15), and (S19)
therein],
which is carried out in
the SM \cite{Note2} (Supplementary Section 3).
   This fully proves the results in Sec.~\ref{subsec:main_practical}.

%


%
%
%
%


\subsection{Networks with general
  topology from star-like subgraphs} \label{subsec:from_starlike}

  \begin{figure*}
\includegraphics{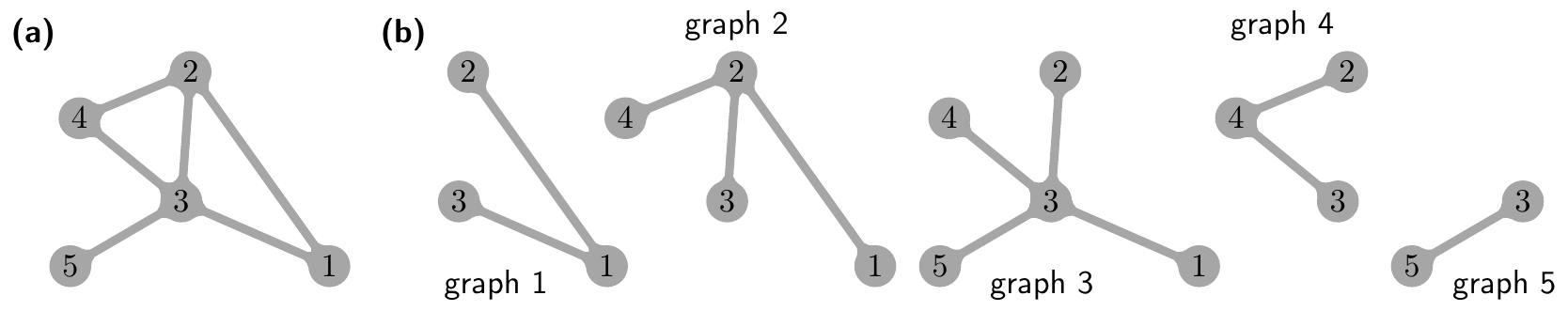}
  \caption{Network decomposition into star-like sub-graphs. (a)~Full graph with five network states. (b)~All five star-like sub-graphs.
  The set of neighboring states contains $\mathcal{N}_1=\{2,3\}$, $\mathcal{N}_2=\{1,3,4\}$,
  $\mathcal{N}_3=\{1,2,4,5\}$, $\mathcal{N}_4=\{2,3\}$ and $\mathcal{N}_5=\{3\}$.}
  \label{fig:star_decomposition_SI}
\end{figure*}

  The simplest network topology is a star-like topology, for which  we are able to conveniently
express moments of conditional first passage times in terms of simple unconditioned first passage moments
as explained in
Supplementary Section I.B in the SM \cite{Note2}.
 In the following we show that according to Ref.~\cite{land77} (see also Refs.~\cite{mont65,klaf87,haus87}) each network  can be decomposed exactly into a full set of subnetworks
with a star-like topology. Thereby, each star-like sub-graph characterizes
the local kinetics on a graph in the vicinity of a network state. Hence we will use all star-like sub-graphs as building blocks
to build and describe a general network.

Suppose that we are dealing with a large scale network with a set of $N$ states,
such that for each state $i\in \Omega=\{1,\ldots,N\}$
there exist a non-empty set of neighboring states $\mathcal{N}_i\subset\Omega$
with  $i\notin\mathcal{N}_i$. A fully connected network corresponds to $\mathcal{N}_i=\{1,\ldots,i-1,i+1,\ldots,N\}$.
For the five-state network in Fig.~\ref{fig:star_decomposition_SI} all sets of neighboring states are
$\mathcal{N}_1=\{2,3\}$, $\mathcal{N}_2=\{1,3,4\}$, $\mathcal{N}_3=\{1,2,4,5\}$, $\mathcal{N}_4=\{2,3\}$ and $\mathcal{N}_5=\{3\}$. 
The probability density that starting from state $i$
a nearest neighboring state $j\in\mathcal{N}_i$ will be reached for the first time
at time $t$ is distributed according to the probability density $\wpcStar_{j|i}(t)$ with a Laplace transform
$\twpcStar_{j|i}(s)$, where $\phiStar_{j|i}=\twpcStar_{j|i}(0)=\int_0^\infty\wpcStar_{j|i}(t)\dd t$ is the (splitting) probability that starting from $i$ the next visited state   will be  $j$.
We define the matrix $\ploc$ as
\begin{equation}
 \ploc_{ij}=
 \begin{cases}
           \twpcStar_{i|j}(s)&\text{if $i\in\mathcal{N}_j$ 
           },\\
           0&\text{otherwise}.
  \end{cases}
  \label{eq:S2_Pmatrix_def}
\end{equation}
Note that $\ploc$ is a hollow matrix, since $\ploc_{ii}=0$.
Taylor expanding \eqref{eq:S2_Pmatrix_def}
we obtain
\begin{equation}
 \ploc=\philoc-s\tloc+\mathrm{O}(s)^2.
 \label{eq:S2_taylor}
\end{equation}
where $\philoc_{ij}=\phiStar_{i|j}$ and $\tloc_{ij}=\phiStar_{i|j}\avgcStar{t}_{i|j}$ for $i\in\mathcal{N}_j$ and $j=1,\ldots,N$.
We emphasize that working in Laplace space allows us to conveniently add independent random variables. Namely, for any  two independent random variables $t_1$ and $t_2$, distributed
according to the densities $f_1$ and $f_2$ with Laplace transforms $\tilde f_i(s)=\int_0^\infty f_i(t)\e^{-s t}\dd t\equiv\avg{\e^{-s t_i}}$, we have that $\tilde f_{1+2}(s)=\avg{\e^{-s (t_1+t_2)}}=\avg{\e^{-s t_1}}\avg{\e^{-s t_2}}=\tilde f_1(s)\tilde f_2(s)$. That is, in Laplace space the sum of random variables is reflected by product of the Laplace transforms of the corresponding probability densities
[see also last terms in the generalized renewal theorem Eqs.~\eqref{eq:S1_renewal_cond_pre} and \eqref{eq:S1_renewal_cond}].
Conversely, a plain product in Laplace space becomes a convolution in the time domain,
$f_{1+2}(t)=f_1*f_2(t)$.

Having established the local kinetics we can now determine the first passage time to a set of
target states $\mathcal{A}$ starting from state $i_0\notin\mathcal{A}$ ($i_0\in\Omega\backslash\mathcal{A}$) for a general network as follows.
To select a target state and remaining states we first define the projection
matrix onto target state $\mathcal{A}$ and the rest, i.e., $\mathcal{A}^{\rm c}\equiv\Omega\backslash\mathcal{A}$,
which are given by
\begin{equation}
\PA\equiv \sum_{i\in\mathcal{A}}\ket{i}\bra{i}\quad \text{and}\quad\PAc\equiv \sum_{i\notin\mathcal{A}}\ket{i}\bra{i}=\id-\PA,
\label{eq:S2_PA_def}
\end{equation}
respectively, where  $\bra{i}=\ket{i}^\T$ is a unit column vector with all elements
zero except the $i$th component and $\id$ is the identity matrix.
The matrices $\PA$ and $\PAc$ are the indicator functions of $\mathcal{A}$ and $\mathcal{A}^{\rm c}$, respectively.
For example, for all target states $\alpha\in\mathcal{A}$
we find $\PA\ket{\alpha}=\ket{\alpha}$ and $\PAc\ket{\alpha}=0$, whereas for all remaining states $\beta\in\mathcal{A}^{\rm c}$
we have $\PA\ket{\beta}=0$ and $\PAc\ket{\beta}=\ket{\beta}$.
Starting from $i_0\in \mathcal{A}^{\rm c}$
the Laplace transform of the probability density to hit the target state $a\in \mathcal{A}$
``after the first step'' (without having visited any of the remaining states)
is given by $\bra{a}\boldsymbol{\mathcal{Q}}^{(1)}(s)\ket{i_0}=\bra{a}\ploc\ket{i_0}$;
similarly, if we select all elements that perform exactly one jump into a state $j\in\mathcal{A}^{\rm c}$ and then enter $a$ in the second jump we obtain $\bra{a}\ploc\ket{j}\bra{j}\ploc\ket{i_0}$, which after summing over all
intermediate non-target states with Eq.~\eqref{eq:S2_PA_def}, yields $\bra{a}\boldsymbol{\mathcal{Q}}^{(2)}(s)\ket{i_0}=\bra{a}\ploc\PAc\ploc\ket{i_0}$.
More generally, the Laplace transform of the probability density to hit
target $a$ for the first time exactly after $k$-th transitions while transiting $k-1$ times
between non-target states is given
by
 \begin{equation}
\bra{a}\boldsymbol{\mathcal{Q}}^{(k)}\ket{i_0}=\bra{a}\ploc[\PAc \ploc]^{k-1}\ket{i_0}.
\label{eq:S2_Qk}
\end{equation}
Summing now over all possible numbers of intermediate transitions we obtain a geometric sum that yields  \cite{land77}
\begin{equation}
 \twpc_{a|i_0}(s)\equiv\sum_{k=1}^\infty\bra{a}\boldsymbol{\mathcal{Q}}^{(k)}\ket{i_0}=\bra{a}\ploc[\id-\PAc\ploc]^{-1}\ket{i_0},
  \label{eq:S2_montroll}
\end{equation}
which is the main result of this subsection.
This result allows us to express conditional first passage times towards any set of targets $\mathcal{A}$ in terms of the local first passage time densities. 
The inverse Laplace transform ($s\to t$) of Eq.~\eqref{eq:S2_montroll}
yields the joint probability density $\wpc_{a|i_0}(t)$ that the continuous
trajectory starting from node $i_0\in \Omega\backslash\mathcal{A}=\mathcal{A}_{\rm c}$ arrives at time $t$ for the first time
in state $a\in\mathcal{A}$
without having visited any other state within $\mathcal{A}$.
The probability is normalized according to
$\sum_{a\in\mathcal{A}}\int_0^\infty\wpc_{a|i_0}(t)\dd t=1$.
The case in which $\mathcal{A}$ contains all neighbors of $i_0$, that is $\mathcal{A}=\mathcal{N}_{i_0}$, one immediately obtains $\PAc\ploc\ket{i_0}=0$, which
simplifies Eq.~\eqref{eq:S2_montroll} to $\twpc_{a|i_0}=\bra{a}\ploc\ket{i_0}=\twpcStar_{a|i_0}(s)$ for all $a\in \mathcal{A}=\mathcal{N}_{i_0}$. The independence between transition time and dwell time, Eq.~\eqref{eq:local_decomposition}, allows to express non-zero
elements of the
hollow matrix $\ploc$ in form of the product $\ploc_{ji}\equiv
\twpcStar_{j|i}(s)=\phiStar_{j|i}\twpctrans_{j|i}(s)\twpcdwell_{i}(s)$ for
any pair $i,j\neq
i$.

From
Eq.~\eqref{eq:S2_montroll} follows the splitting probability, i.e. the
probability to reach $a$ from $\textcolor{black}{i_0}$ before reaching any other state
within $\mathcal{A}$ which
reads
\begin{multline}
 \phi_{a|\textcolor{black}{i_0}}^\mathcal{A}\equiv\int_0^\infty \wpc_{a|i_0}(t)\dd t=\twpc_{a|i_0}(s)|_{s=0}
 \\
 =\bra{a}\philoc(\id -\PAc\philoc)^{-1}\ket{\textcolor{black}{i_0}},
 \label{eq:splitting_res}
\end{multline}
where we used $\philoc_{ij}=\phiStar_{i|j}$ from Eq.~\eqref{eq:S2_taylor}. 
Inserting the matrix
$\boldsymbol{\mathcal{T}}_{ij}=\phiStar_{i|j}\avgcStar{t}_{i|j}$
from Eq.~\eqref{eq:S2_taylor}
the  mean first passage time from $\textcolor{black}{i_0}$ to $a$, conditioned not to
visit any state $j\in\mathcal{A}\backslash\{a\}$, in turn reads
\begin{equation}
 \avg{t}^\mathcal{A}_{a|\textcolor{black}{i_0}}
   =\bra{a}(\id -\philoc\PAc)^{-1}\tloc(\id -\PAc\philoc)^{-1}\ket{\textcolor{black}{i_0}}/\phi_{a|\textcolor{black}{i_0}}^\mathcal{A}.
 \label{eq:cond_mean}
\end{equation}
A few remarks are in order before we prove Eq.~\eqref{eq:cond_mean}.
Eqs.~\eqref{eq:Idef}-\eqref{eq:binomial} render Eqs.~\eqref{eq:splitting_res} and \eqref{eq:cond_mean} fully
explicit.
 As an illustration in the following Appendix~\ref{subsec:fingerprints_ill} 
  we apply Eqs.~\eqref{eq:splitting_res} and 
Eq.~\eqref{eq:cond_mean}  to the synthetic network used in  Fig.~\ref{fig:blinking}, where
a diffusive barrier along
link $1-3$  mimics the effect of an entropic bottle-neck \cite{zwan92} (details about the model are given in Appendix~\ref{subsec:synthetic_network}).

The conditional mean first passage time  \eqref{eq:cond_mean} follows from
\begin{multline}
\phic_{a|i_0} \avgc{t}_{a|i_0}
\equiv
\int_0^\infty \wpc_{a|i_0}(t) t\dd t
=
-
 \frac{\del }{\del s}\twpc_{a|i_0}(s)|_{s=0}\\
=\bra{a}\tloc(\id-\PAc\philoc)^{-1}\ket{i_0}\\
+ \bra{a}\philoc(\id-\PAc\philoc)^{-1}\PAc\tloc(\id-\PAc\philoc)^{-1}\ket{i_0}\\
=\bra{a}[\id+\philoc\PAc(\id-\philoc\PAc)^{-1}]\tloc(\id-\PAc\philoc)^{-1}\ket{i_0}\\
=\bra{a}(\id-\philoc\PAc)^{-1}\tloc(\id-\PAc\philoc)^{-1}\ket{i_0}
\label{eq:S1_mean}
\end{multline}
where we have used the product rule of differentiation ``$\del (fg)=(\del f) g+f\del g$'' and the formula $\frac{\dd}{\dd s}\mat{A}^{-1}=-\mat{A}^{-1}\frac{\dd \mat{A}}{\dd s}\mat{A}^{-1}$ in the second line,
and $\PAc(\id-\philoc\PAc)^{-1}=(\id-\PAc\philoc)^{-1}\PAc$ in the second last line,
which finally leads to
Eq.~\eqref{eq:cond_mean}.
Note that the conditional mean first passage time, $\avgc{t}_{a|i_0}\equiv [\int_0^\infty \wpc_{a|i_0}(t) t\dd t]/[\int_0^\infty \wpc_{a|i_0}(t)\dd t]$, is obtained by dividing, Eq.~\eqref{eq:S1_mean} by the splitting probability Eq.~\eqref{eq:splitting_res}. This completes the proof of Eq.~\eqref{eq:cond_mean}.

Higher moments can formally be obtained along the same lines via
Eq.~\eqref{eq:S2_montroll}, such that the $k$th moment satisfies
\begin{equation}
 \phi_{a|i_0}^\mathcal{A}\avgc{t^k}_{a|i_0}\equiv\int_0^\infty \wpc_{a|i_0}(t)t^k\dd t =(-1)^k\frac{\del^k}{\del s^k}\twpc_{a|i_0}(s)|_{s=0}.
 \label{eq:S1_moment_gen_func}
\end{equation}
Using Eq.~\eqref{eq:S2_montroll} we can effectively  deduce any moment of the first passage time within the network 
  from $\ploc$.

If the network can be described by memory-less jump dynamics \cite{seif12} as, for instance, in the celebrated Gillespie algorithm  \cite{gill77,gill07}, the transitions between network states  are characterized by constant transition rates $w_{i\to j}$ from state $i$ to state $j$.
In this case the time until the state changes is exponentially distributed
with the rate of leaving state $i$, $r_i=\sum_{j\in\mathcal{N}_i}w_{i\to j}$,
yielding the \emph{same} exit time distribution $\wpcexit_i(t)=\sum_k\ploct_{ki}^{\rm M}=r_i\e^{-r_i t}$
irrespective of the final state $j$, with probability $\phiStar_{j|i}=w_{i\to j}/r_i$, i.e.,  $\wpcStar_{j|i}(t)/\phiStar_{j|i}=\wpcexit_i(t)$.
A Laplace transform $t\to s$ of such  memory-less kinetics would yield
$\ploc_{j|i}^{\rm M}=\phiStar_{j|i}r_i(s+r_i)^{-1}$ along with the $k$th moment $\avgcStar{t^k}_{j|i}=\avgdwell{\tau^k}_{i}=k!r_i^{-k}$.

\subsection{Fingerprints of memory}\label{subsec:fingerprints_ill}

Unique fingerprints of memory in state-to-state kinetics emerge already under minimal assumptions. Consider the kinetics from state $i_0=3$ to the pair of target states $\mathcal{A}=\{1,2\}$ in the network depicted in Fig.~\ref{fig:renewal_gen_SI}a (see also trajectory in Figs.~\ref{fig:blinking}a and \ref{fig:cg_traj2}),
which corresponds to
\begin{equation}
 \PA=
 \begin{pmatrix}
  1&0&0&0&0\\
  0&1&0&0&0\\
  0&0&0&0&0\\
  0&0&0&0&0\\
  0&0&0&0&0\\
 \end{pmatrix},
 \quad
 \PAc=
 \begin{pmatrix}
  0&0&0&0&0\\
  0&0&0&0&0\\
  0&0&1&0&0\\
  0&0&0&1&0\\
  0&0&0&0&1\\
 \end{pmatrix}.
  \label{eq:S8_network_matrices2}
 \end{equation}
In order to infer the  waiting time distribution $\wpcStar$ for the
exit from all states, respectively, we simulated $4\times 10^5$ exits from each state  (see Fig.~\ref{fig:S8_data0003_hist_states} in Appendix~\ref{subsec:synthetic_network}). 
The normalized waiting time distribution for the exit from state 3 is
genuinely heterogeneous \cite{greb18a}, i.e. it shows strong
variations between the respective legs (see Fig.~\ref{fig:blinking}c and see Fig.~\ref{fig:S8_data0003_hist_states} for a more
detailed analysis of the complete network).
The splitting probability and the local mean waiting time  are given by
\begin{align}
\begin{aligned}
 \philoc
 &\approx
  \begin{pmatrix}
  0  &0.33&0.25&0  &0\\
  0.5&0     &0.25&0.5&0\\
  0.5&0.33&0   &0.5&1\\
  0  &0.33&0.25&0  &0\\
  0  &0     &0.25&0  &0\\
 \end{pmatrix},  
\\
 \tloc&\approx
 \begin{pmatrix}
  0      &0.44&4.02&0        &0\\
  5.64&0       &1.58&0.66&0\\
  10.52&0.44&0      &0.66&1.33\\
  0      &0.44&1.58&0       &0\\
  0      &0       &1.58&0       &0\\
 \end{pmatrix}.
 \end{aligned}
 \label{eq:S8_network_matrices}
\end{align}
One can confirm that the system satisfies detailed balance, since
$\ln ( \philoc_{ji}/  \philoc_{ij})=\ln|\mathcal{N}_j|-\ln|\mathcal{N}_i|$ holds, where $|\mathcal{N}_i|$ is the number of states adjacent to $i$.

We now inspect the probability
to reach the target state $1$ ($2$) within $\mathcal{A}$ before
reaching state $2$ ($1$). Note that a trajectory may reach state $1$
through the link $1-3$ or via state $4$. 
Such conditioned transition kinetics quantify non-local effects and
are particularly important for marginal observations, i.e. when we do not monitor
all states but instead only a subset (in this case states 1, 2 and 3)
while the remaining states are left as part of the ``heat bath''. This
scenario is very relevant from an experimental point of view, since we
can typically monitor only a limited number of states.

After inserting Eqs.~\eqref{eq:S8_network_matrices2}  and \eqref{eq:S8_network_matrices}
into Eqs.~\eqref{eq:splitting_res} and \eqref{eq:cond_mean}, respectively, we find
$\phi^\mathcal{A}_{1|3}=0.4$ and $\phi^\mathcal{A}_{2|3}=0.6$ while $\avg{t}^\mathcal{A}_{1|3}\approx 8.26$, 
$\avg{t}^\mathcal{A}_{2|3}\approx 6.80$. 
A Markov process would presume isotropic mean waiting times
$\mathcal{T}_{ij}^{\rm M}=\phi_{i|j}\avgexit{t}_j$, 
which yields $\avg{t}^{\cA,{\rm M}}_{1|3}\approx 5.92<\avg{t}^{\cA,{\rm M}}_{2|3}\approx 9.14$, whereas $\avg{t}^\mathcal{A}_{1|3}\approx 8.26>\avg{t}^\mathcal{A}_{2|3}\approx 6.80$.

\section{Thermodynamic consistency of coarse-graining and discontinuous force fields}\label{sec:SI_thermodynamic_consistency}
In this Appendix we connect the thermodynamic consistency of the coarse-graining
to the preservation of cycles \cite{pugl10}, which implies that the 
entropy production rate \cite{schn76,seif12}  is
preserved in the coarse-grained process as discussed in Sec.~\ref{subsec:entropy}. 
This underlines that the violation of detailed balance is entirely encoded in the splitting probabilities (see also \cite{wang07}). In addition we derive the special limit of local detailed balance in the presence of a time-scale separation. Finally, we consider discontinuous diffusion coefficients and/or force fields.

\subsection{Thermodynamic consistency follows from
the preservation of cycle affinities}
\label{subsec:SI_cycles}

\begin{figure}
 \centering
\includegraphics{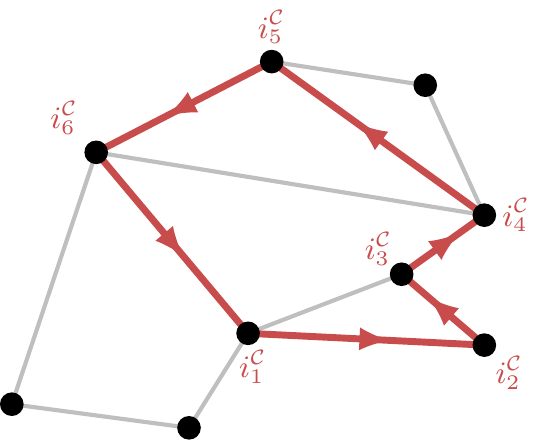}
\caption{One directed cycle $\mathcal{C}=i^\mathcal{C}_1\to i^\mathcal{C}_2\to\ldots\to i^\mathcal{C}_{\nu}\to i^\mathcal{C}_{\nu+1}=i^\mathcal{C}_1$ (see thick red arrow lines) is 
highlighted which encloses $\nu=|\mathcal{C}|=6$ nodes within a total set of $\Omega=9$ nodes.
}
\label{fig:SI_cycle}
\end{figure}

Whether or not the system relaxes to an equilibrium  distribution is
entirely encoded in the microscopic force field $\bF$. If the force field is conservative, that is $\bF(\bx)=-\nabla U(\bx)$, the resulting stationary state corresponds to thermodynamic equilibrium. 
Detailed balance is said to be broken if the time independent force field has a non-zero rotation
or equivalently, if at least one directed cycle $\mathcal{C}$ (see thick arrows in Fig.~\ref{fig:SI_cycle}) 
exists, for which the integral
\begin{equation}
  \mathcal{A}[\mathcal{C}]\equiv\oint_\mathcal{C}\bF(\bx)\dd x=\sum_{\alpha=1}^\nu\int_{i^\mathcal{C}_\alpha}^{i^\mathcal{C}_{\alpha+1}}
  \bF(\bx)\dd x
   \label{eq:S_affinity_cycle1}
\end{equation}
called affinity \cite{schn76}, is non-zero. If we insert the first main result, Eq.~\eqref{eq:local_detailed_balance_like},
into
\eqref{eq:S_affinity_cycle1} we obtain
\begin{equation}
  \mathcal{A}[\mathcal{C}]=\sum_{\alpha=1}^\nu \ln\Bigg[\frac{\phiStar_{i^\mathcal{C}_{\alpha+1}|i^\mathcal{C}_\alpha}}{\phiStar_{i^\mathcal{C}_{\alpha}|i^\mathcal{C}_{\alpha+1}}}\Bigg], 
   \label{eq:S_affinity_cycle2}
\end{equation}
where  we used the fact that the terms involving the function ``$g$''
form a vanishing telescope sum.
Equation \eqref{eq:S_affinity_cycle2} relates the affinity of \emph{all} cycles $\mathcal{C}$
exactly to the splitting probabilities $\{\phiStar_{j|i}\}$.
Therefore, detailed balance is satisfied as soon as 
\emph{all} closed cycles in a network satisfy $\mathcal{A}[\mathcal{C}]=0$. This proves that the splitting probabilities alone encode the breaking or validity of detailed balance.

Due to the preservation of cycle affinities [see Eq.~\eqref{eq:S_affinity_cycle1} and Eq.~\eqref{eq:S_affinity_cycle2}], the steady state  entropy production is entirely encoded in the splitting probabilities $\{\phiStar_{j|i}\}$.
This can be understood as an alternative proof of the preservation of the entropy production shown in Sec.~\ref{subsec:entropy} (see also Ref.~\cite{pugl10}).

\subsection{The peculiar local equilibration}
\label{subsec:S_local_detailed_balance}

%

Let us now address the limit of a time-scale separation that leads to a local equilibration prior to any change of state.
In the limit of high free energy barriers (i.e., $B_{j|i}\to \infty$ in Fig.~\ref{fig:loc_equilibration_methods}) the first two auxiliary integrals \eqref{eq:Idef}
simplify to
\begin{equation}
 \begin{aligned}
  I_{j|i}^{(1)}&=\int_{0}^{l_{j|i}}\dd y_1\frac{\e^{\beta U_{j|i}(y_1)}}{D_{j|i}(y_1)}\simeq \int_{x^*_{j|i}-\epsilon}^{x^*_{j|i}+\epsilon}\dd y_1\frac{\e^{\beta U_{j|i}(y_1)}}{D_{j|i}(y_1)}, \\
  I_{j|i}^{(2)}&=\int_{0}^{l_{j|i}}\dd y_1\int_{0}^{y_1}\dd y_2\frac{\e^{\beta U_{j|i}(y_1)-\beta U_{j|i}(y_2)}}{D_{j|i}(y_1)}
  \\&
  \simeq\int_{x^*_{j|i}-\epsilon}^{x^*_{j|i}+\epsilon}\dd y_1\int_{0}^{y_1}\dd y_2\frac{\e^{\beta U_{j|i}(y_1)-\beta U_{j|i}(y_2)}}{D_{j|i}(y_1)}
  \\&
   \simeq I_{j|i}^{(1)}\int_{0}^{x^*_{j|i}}\dd y_2\e^{-\beta U_{j|i}(y_2)},
 \end{aligned}
 \label{eq:Iaux_LDB_approx}
\end{equation}
where we assume $\epsilon\ll l_{j|i}$.
The (saddle-point) 
approximations in the first and third line of Eq.~\eqref{eq:Iaux_LDB_approx}
follow from $\e^{\beta U_{j|i}}(y_1)$ being largest in  the vicinity of $y_1\simeq x^*_{j|i}$
(see Fig.~\ref{fig:loc_equilibration_methods}).
The last approximation follows from $y_2\le y_1\simeq x^*_{j|i}$ 
and hence $\int_0^{y_{1}}\dd y_2 \e^{\beta U_{j|i}(y_2)}\simeq \int_0^{x^*_{j|i}}\dd y_2 \e^{-\beta U_{j|i}(y_2)}$.
Inserting Eq.~\eqref{eq:Iaux_LDB_approx}
into
the splitting probability in
Eq.~\eqref{eq:phi_explicit},
and 
the mean exit time in
Eq.~(S50) in the SM \cite{Note2}
yields the asymptotic rate of jumping from state $i$ to state $j$
\begin{align}
 w_{i\to j}&=\phiStar_{j|i}/\avgexit{t}_i\simeq\frac{1}{I_{j|i}^{(1)}}\sum_{k\in\mathcal{N}_i}\int_{0}^{x^*_{k|i}}\e^{-\beta U_{k|i}(y_2)}\dd y_2,\nonumber\\
 &\equiv\frac{\e^{\beta \mathcal{U}_{i}-\beta \mathcal{ F}_{i}}}{I_{j|i}^{(1)}},
 \label{eq:rate_LDB}
\end{align}
where in the last step we have defined the free energy of state $i$, $\mathcal{F}_i=-k_{\rm B}T\ln\mathcal{Z}_i$,
to be given by the partition function $\mathcal{Z}_i=\e^{-\beta \mathcal{F}_i}\equiv\sum_{k\in\mathcal{N}_i}\int_0^{x^*_{k|i}}\e^{-\beta\mathcal{U}_i-\beta U_{j|i}(x)}\dd x$. Note that $\mathcal{U}_i$
denotes the energy at node $i$ since $U_{j|i}(0)=0$.
Inserting the rates Eq.~\eqref{eq:rate_LDB} along with Eq.~\eqref{eq:LDB_like_last_term2} into the left hand side of Eq.~\eqref{eq:methods_local_detailed_balance} yields the right hand side of Eq.~\eqref{eq:methods_local_detailed_balance}. This completes the proof of the local detailed balance relation  in the limit of a time-scale separation.


Let us now briefly comment on transition-path times in the limit of a time-scale separation.
Since, high free energy barriers between any pair of state will eventually render all higher order integrals $\smash{I_{i|j}^{(k)}}$ negligibly small  if $k\ge3$,
the transition-path time $\avgtrans{\delta t}_{j|i}= I_{j|i}^{(3)}/I_{j|i}^{(1)}$
is negligibly short compared to the mean exit time from state $i$. {\color{mynewcolor}More precisely, it has been found for a parabolic barrier that the transition-path time scales $\propto |U_{j|i}''(x^*_{j|i})|^{-1}\ln B_{j|i}$ \cite{chun09,neup12,maka15,kim15,zijl20}, i.e., 
decreases with $B_{j|i}$ due to $|U_{j|i}''(x^*_{j|i})|^{-1}\propto 1/B_{j|i}$  \cite{kim15}, while the exit time grows much faster \cite{kram40,haen90,gard04}, i.e.  $\propto \e^{\beta B_{j|i}}$. 
One can show that  a rectangular shaped potential with a barrier height $B_{j|i}$ in fact yields a constant finite transition-path time $\avgtrans{\delta t}_{j|i}= I_{j|i}^{(3)}/I_{j|i}^{(1)}$ in the limit $B_{j|i}\to \infty$ while at the same time the exit time diverges $\propto\e^{\beta B_{j|i}}\to\infty$. The shape of the barrier may therefore decide 
whether or not the transition-path time is affected by the barrier height \cite{kim15}.}

\subsection{Generalization to discontinuous local potentials and discontinuous diffusion landscapes}\label{subsec:SI_discontinuous_landscapes}
We will first explain how one deals with discontinuous local potentials in general. Next, we account for possible discontinuities in the diffusion landscape by removing them through a  linear stretch of coordinates. Therefore, discontinuous diffusion landscapes can always be accounted for by mapping the coordinate system onto a continuous diffusion landscapes, but with possible discontinuities in the local potential.

\begin{figure}%
\centering%
\includegraphics{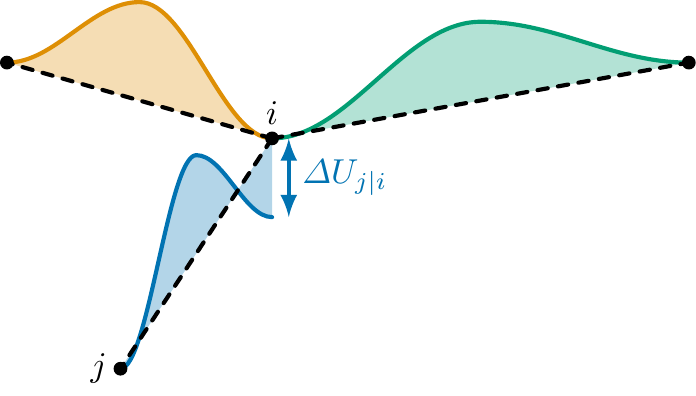}
 \caption{Discontinuous potential. \normalfont Local potential between state $i$ 
 three neighbor-states. Along the leg from state $i$ to state  $j$  the potential has a discontinuity of strength $\varDelta U_{j|i}$.}
 \label{fig:S_discontinuous_potential}
\end{figure}

\paragraph{Discontinuous local potential.} Let us begin with a discontinuous ``diverging force kick'' at the node $i$ towards state
$j$ which effectively means
$F_{j|i}(x)=F_{j|i}^{\rm cont}(x)+\varDelta U_{j|i}\delta (x)$, where $F_{j|i}^{\rm cont}(x)$ is some continuous force field,
$\delta (x)$ denotes the Dirac delta-function, and $\varDelta U_{j|i}$ is the strength of the discontinuity.
The ``force kick'' yields the potential $U_{j|i}(x)=-\varDelta U_{j|i}-\int_0^xF_{j|i}^{\rm cont}(x')\dd x'$. The local potential has  a discontinuity once $U_{j|i}(0)=-\varDelta U_{j|i}\neq0$. A single discontinuity between states $i$ and $j$ is schematically depicted in Fig.~\ref{fig:S_discontinuous_potential} (see blue line). The transition-path time is not affected by such ``kicks'' since the transition path spans the time interval \emph{after} the last passage  of state $i$ until the first entrance into state $j$, which can be confirmed by the following argument. To formally avoid a discontinuity we replace the discontinuity $\varDelta U_{j|i}\delta (x)$ by a smoothened force $\varDelta U_{j|i}/\epsilon$ within the interval $0\le x\le \epsilon$ and afterwards take the limit $\epsilon\to0$. The auxiliary integrals according to  
Eq.~\eqref{eq:Idef}
 become $\lim_{\epsilon\to0}I^{(2k-1)}_{j|i}=[I^{(2k-1)}_{j|i}|_{\varDelta U_{j|i}=0}]\times \e^{-\beta\varDelta U_{j|i}}$ and $
\lim_{\epsilon\to0}I^{(2k)}_{j|i}=I^{(2k)}_{j|i}|_{\varDelta U_{j|i}=0}
$ (for $k=1,2\ldots$).
Since all the odd-valued $k$ auxiliary integrals are affected by the discontinuity in precisely the same manner
``$I^{(2k-1)}_{j|i}\propto\e^{-\beta\varDelta U_{j|i}}$'', we find that first two moments of transition-path time,
 Eq.~\eqref{eq:trans_methods},
are \emph{not} affected by the discontinuity.

%

Importantly, a kick of strength $\varDelta  U_{j|i}$ affects the splitting probability $\phiStar_{j|i}$ of choosing a transition due to $\phiStar_{j|i}\propto 1/I_{j|i}^{(1)}\propto \e^{\beta\varDelta  U_{j|i}}$ [cf.
Eqs.~\eqref{eq:Idef} and \eqref{eq:phi_explicit}].
 Since the dwell time is affected by both, the splitting probability and the transition-path time [cf.
Eqs.~\eqref{eq:Idef} and \eqref{eq:dwell_methods}],
 a force-kick of strength
$\varDelta U_{j|i}$ \emph{does affect} the dwell-time statistics.

Therefore, as an interim summary we find that force-kicks arising from a discontinuous local potential (see Fig.~\ref{fig:S_discontinuous_potential}) affect both the splitting probability and the dwell-time statistics, whereas the transition-path time is \emph{not} affected.

\paragraph{Discontinuous diffusion landscape.}
Discontinuous diffusion landscapes, i.e. $D_{j|i}(x)$ satisfying  $D_{j|i}(0)\neq D_{k|i}$ for some $k\neq i$, are dealt with in the following manner. First,
we locally re-scale the coordinate system such that the
discontinuity disappears. Specifically,
 we locally stretch the coordinates between nodes
$i$ and $j$, $l_{j|i}$, homogeneously  by a factor $\alpha_{j|i}$ ($l_{j|i}\to \alpha_{j|i} l_{j|i}$) to obtain a re-scaled diffusion landscape,
$\alpha_{j|i}^2 D_{j|i}(x/\alpha_{j|i})$, and a correspondingly re-scaled local potential, $ U_{j|i}(x/\alpha_{j|i})-\beta^{-1}\ln \alpha_{j|i}$, where $\beta^{-1}=k_{\rm B} T$ is the thermal energy (see also Sec. II.E in Ref.~\cite{hinc10}). By choosing $\alpha_{j|i}$ such that the diffusion landscape becomes continuous, we obtain a mapping from a discontinuous diffusion landscape onto a continuous one. Hence, discontinuous diffusion
landscapes can be removed entirely via a linear change of local
coordinates. Such a re-scaling  gives rise to a discontinuous
potential -- a problem we have already solved above. In this sense all of the results presented here apply to dynamics on a graph with both, discontinuous diffusion  and discontinuous local potentials equally well. 
Notably, the results derived
in the Supplementary Section 3 in the SM \cite{Note2}
i.e.
Eqs.~\eqref{eq:phi_explicit}-\eqref{eq:dwell_methods}
in Sec.~\ref{subsec:main_practical}, can be used unaltered in the case of discontinuities in the potential and diffusion landscapes.

\section{Model parameters and additional confirmation of the main practical result}\label{sec:SI_examples}
In Appendix~\ref{subsec:synthetic_network} we provide details about the ``synthetic'' model discussed in Fig.~\ref{fig:blinking} and Sec.~\ref{subsec:from_starlike}.
We then define the catch-bond model in Appendix~\ref{subsec:SI_catch}, which is 
discussed in Sec.~\ref{subsec:super_Markov} (see Fig.~\ref{fig:illustrantion_all}d, Fig.~\ref{fig:allresults}b,c). The model from Fig.~\ref{fig:allresults}d-f is provided in Appendix~\ref{subsec:ATPase}, where we also derive the upper bound depicted in Fig.~\ref{fig:allresults}f. The lower bound in Fig.~\ref{fig:allresults}f is proven in Appendix~\ref{subsec:TUR_proof} and shown to saturate in Appendix~\ref{subsec:ATPase_nobarrier}.

In addition, we further corroborate all of our main findings.
In particular, we verify  symmetry (i) in Eq.~\eqref{eq:local_decomposition} in Fig.~\ref{fig:S_catch_hist2}c.
We test the reflection symmetry of the transition-path time -- symmetry (ii) in Eq.~\eqref{eq:local_decomposition} -- in Fig.~\ref{fig:S_ATPase_FB_trans} as well as Tab.~\ref{tab:ATPase_forward/backward_symmetry}.
In Tab.~\ref{tab:mu_list}, \ref{tab:moments_atp}, and Fig.~\ref{fig:S_catch_more_statistics} we corroborate our main practical result shown in Sec.~\ref{subsec:main_practical}.

\subsection{Dynamics in the synthetic network from Fig.~\ref{fig:blinking} and Sec.~\ref{subsec:from_starlike}}
\label{subsec:synthetic_network}
We briefly state \emph{all} model parameters and then provide details about the analysis. Moreover, we use the model to corroborate  the main practical result (see Sec.~\ref{subsec:main_practical}).

\begin{figure*}
 \includegraphics[width=0.85\textwidth]{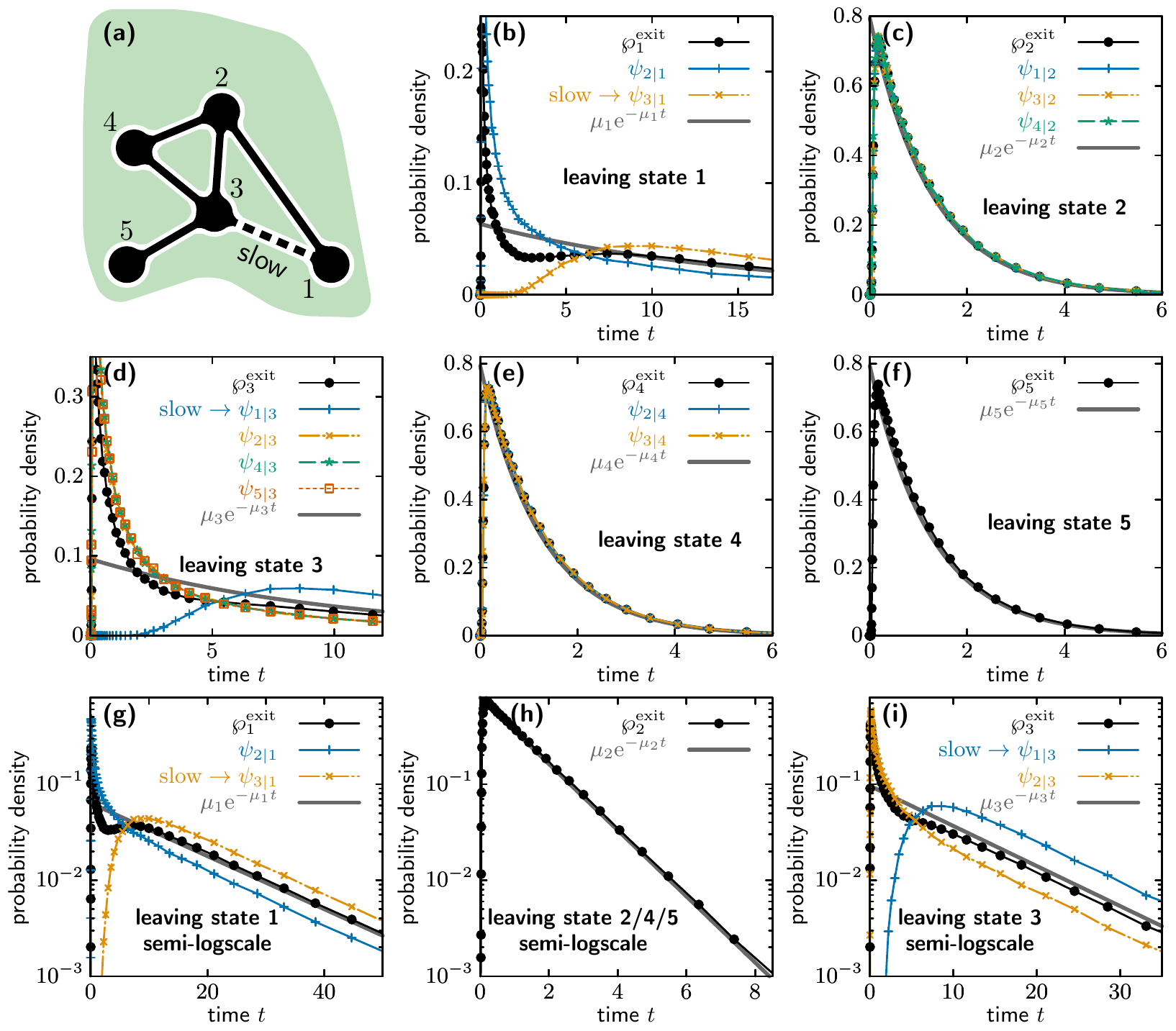}
 \caption{State-to-state kinetics in the synthetic network.  (a)~Schematics of the network with a ``slow'' transition between states 1 and 3 (dashed line). All remaining connections (solid lines) are chosen to be equally fast with $3\,k_{\rm B}T$ barriers separating the network states. (b-f)~Probability density of local residence time, $\wpcStar_{j|i}$,
 starting from (b) $i=1$, (c) $i=2$, (d) $i=3$, (e) $i=4$ and (f) $i=5$, which are deduced from 400000\,repeated exit events, respectively. The gray solid line denotes the estimated long time asymptotics with the values from Tab.~\ref{tab:mu_list}. The densities of leaving state 2, 4 and 5 identical for synthetic model.
 (g-i)~probability densities on a log-scale; redundant densities are omitted since 
 $\wpcexit_2=\wpcexit_4=\wpcexit_5$ holds, and all histograms plotted in (c) and (e) correspond to the same density, respectively, as well as $\wpcStar_{2|3}=\wpcStar_{4|3}=\wpcStar_{5|3}$.}
 \label{fig:S8_data0003_hist_states}
\end{figure*}

\paragraph{Definition of the dynamics.}
The synthetic network in  Fig.~\ref{fig:blinking}a (see also Fig.~\ref{fig:S8_data0003_hist_states}a) is chosen to have one ``slow'' link
between states 1 and 3 being separated by a connection of length $l_{1|3}=l_{3|1}=12.5518$
with embedded diffusion coefficient $D_{3|1}=D_{1|3}=1+4\times16(x/l_{3|1}-1)^2(x/l_{3|1})^2$
and local force $F_{1|3}(x)=32\,k_{\rm
  B}T((2x/l_{1|3}-1)(1-x/l_{1|3})x/l_{1|3}^2$ 
(corresponding to a local potential
$\beta U_{1|3}=\beta U_{3|1}=16(x/l_{3|1}-1)^2(x/l_{3|1})^2$ with  a $1\,k_{\rm B}T$
barrier). Note that the local force is illustrated in Fig.~\ref{fig:network_microstate_redundancy}a and the corresponding local potential is taken from Eq.~\eqref{eq:local_potential}.
To assure the mildest of conditions all remaining states
are chosen to be separated by the same distance $l_{j|i}=1$, diffusion
landscape $D_{j|i}=1$, and
have the same
force field $F_{j|i}(x)=96\,k_{\rm
  B}T\times(2x/l_{j|i}-1)(1-x/l_{j|i})x/l_{j|i}^2$ (i.e.
local potential
$\beta U_{j|i}(x)=48(1-x)^2x^2$, which corresponds to a 3\,$k_{\rm B}T$ barrier separating
each pair of states). 
This network (globally) satisfies detailed balance, since $U_{j|i}(l_{j|i})=\mathcal{U}_j-\mathcal{U}_i$ for all $i,j$ with $\mathcal{U}_i=\mathcal{U}_j=0$ (for the definition of detailed balance see  paragraph including Eq.~\eqref{eq:local_potential} or Appendix~\ref{subsec:SI_cycles}).

\paragraph{Simulation of microstate dynamics.}

We generate individual trajectories using the stochastic Milstein scheme
from Appendix~\ref{subsubsec:Milstein} with time-increment $\varDelta t=10^{-4}$.
A short segment of the trajectory is shown in Fig.~\ref{fig:blinking}a.
In total we simulated 400\,000\,exits from each state and evaluated the probability density of the waiting time between all pair of states.
The results are summarized Fig.~\ref{fig:S8_data0003_hist_states}(b-f), while panels (g-i) display the \emph{same} probability densities on a semi-logarithmic scale. The probability densities depicted in Fig.~\ref{fig:blinking}c 
in the main text are
 taken from Fig.~\ref{fig:S8_data0003_hist_states}d, while the inset in Fig.~\ref{fig:blinking}c represents Fig.~\ref{fig:S8_data0003_hist_states}i.
 The gray lines in Fig.~\ref{fig:S8_data0003_hist_states} corresponding
 to the solid black line in Fig.~\ref{fig:blinking}c denote the long-time asymptotics of the waiting time distribution, which are determined as explained in the following paragraph.



\paragraph{Analysis of the long-time asymptotics.} The long time asymptotics of waiting time density in state $i$ becomes a \emph{single} exponential decay $\psi_{j|i}\propto\e^{\mu_i^\infty t}$ with the same exponent $\mu_i^\infty$ for  all exits to states $j$ adjacent to $i$. One can show that this implies the long time asymptotics to be determined by $\mu_i^\infty=\lim_{n\to \infty}n\avgexit{t^{n-1}}/\avgexit{t^{n}}$.
Note that when the waiting time distribution and long-time asymptotics coincide \cite{kram40,haen90,gard04}
one can instead simply use ($n=1$), i.e., 
 $ \mu_i^\infty\approx 1/\avgexit{t}$.
For examples violating the latter assumption, it turned out taking $n=2$ provides a fairly good estimate, $\mu_i=2\avgexit{t}/\avgexit{t^{2}}\approx \mu_i^\infty$ (see also Ref.~\cite{hart18}).
We deduced the mean exit time and $\mu_i$ both from the theory (Sec.~\ref{subsec:main_practical}) and the simulation. These estimates are shown 
in Tab.~\ref{tab:mu_list}.
\begin{table}
 \caption{Mean versus asymptotics. Each ``experimental'' value is deduced from $N_{\rm sim}=4\times 10^{5}$ simulated exits from each state generated by the stochastic Milstein scheme with $\varDelta t=10^{-4}$. The ``theory'' values are obtained from a numerically evaluation of the results in Sec.~\ref{subsec:main_practical} [see also
  Eqs.~(S47) and (S50)
 in the SM \cite{Note2}].
 Each experimental value has a relative statistical error of about $1/\sqrt{N_{\rm sim}}\approx 0.0016$.}
 \label{tab:mu_list}
\begin{tabular}{c|c|c|c|c}
\hline\hline
 &\multicolumn{2}{c}{ mean exit rate $1/\avgexit{t}_i$}& \multicolumn{2}{|c}{rate
 $\mu_i=2\avgexit{t}_i/\avgexit{t^2}_i$}\\\hline
 state $i$&experiment&theory&experiment &theory\\\hline
 1&0.0617&\textbf{0.0618}&0.0634&\textbf{0.0636}\\
  2&0.7496&\textbf{0.7523}&0.7909&\textbf{0.7934}\\
  3 &0.1140&\textbf{0.1143}&0.0960&\textbf{0.0961}\\
  4 &0.7509&\textbf{0.7523}&0.7929&\textbf{0.7934}\\
  5 &0.7523&\textbf{0.7523}&0.7944&\textbf{0.7934}\\
\hline\hline
\end{tabular}
\end{table}
The thick gray lines in Fig.~\ref{fig:S8_data0003_hist_states}
were deduced from the theory values in Tab.~\ref{tab:mu_list}.
This corroborates the results in Sec.~\ref{subsec:main_practical} and shows that the long-time asymptotics can fairly accurately be determined from the first two moments of the exit time.


\paragraph{Slow transitions  amplify the long time asymptotics.}
Whenever transitions are slow we observe in Fig.~\ref{fig:S8_data0003_hist_states} that
the long-time asymptotics of the local probability density lie above the normalized gray line, which can be explained as follows. When  transitions are long
the probability density $\psi_{j|i}(t)$ becomes negligibly small on time-scales shorter than the transition time $t\lesssim \avgtrans{\delta t}_{j|i}$. Since  $\psi_{j|i}(t)$ must be normalized  $\int_0^\infty\psi_{j|i}(t)\dd t=1$ one inevitably requires more weight of the probability density at long times. Note that all lines plotted in Fig.~\ref{fig:S8_data0003_hist_states}b-i are probability densities which are normalized to unity. In other words, the blue solid line in Fig.~\ref{fig:S8_data0003_hist_states}i is above the gray thick line at long times since it is below the thick gray line at short times.



\subsection{Catch-bond analysis}\label{subsec:SI_catch}

In this subsection we provide details about the catch-bond analysis shown in Fig.~\ref{fig:illustrantion_all}d and
Fig.~\ref{fig:allresults}b,c.
Dissecting the life time of a bond into the dwell and transition time we also corroborate symmetry (i)  entering the  main result in Eq.~\eqref{eq:local_decomposition} (see Fig.~\ref{fig:S_catch_hist2}c).

\begin{figure}
 \centering
 \includegraphics{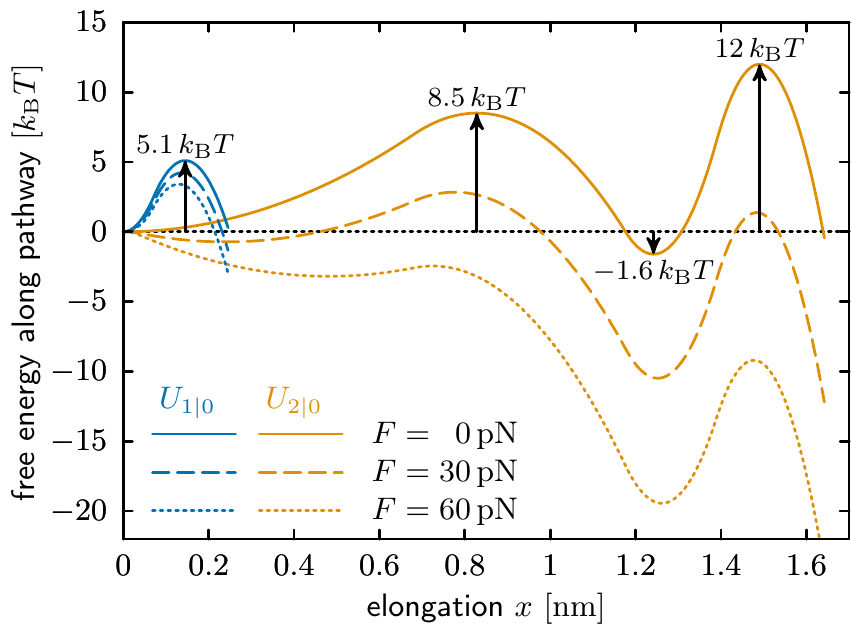}
 \caption{Catch-bond free energy landscape.  The local potential is set to $U_{j|0}(x)=U_{j|0}^{(0)}(x)+ U^{\rm load}(F,x)$. The precise form of
$U_{1|0}$ and $U_{2|0}$ is given in
Eq~\eqref{eq:potential_catch}. {\color{mynewcolor}For a biophysical setting underpinned by the potential see Fig.~5a in Ref.~\cite{thom08a} (see also here Fig.~\ref{fig:illustrantion_all}d)}.
}
 \label{fig:catch_potential}
\end{figure}

\paragraph{The model.}
We employ a so-called switch catch-bond model \cite{bart02}
with parameters chosen to reproduce experimental results on bacterial
adhesion bonds \cite{thom06,thom08a} (see also
Ref.~\cite{buck14} for related experiments). 
The local potential, Eq.~\eqref{eq:local_potential}, along the $j$th pathway ($j=1,2$)
is decomposed into 
$U_{j|0}(x)=U_{j|0}^{(0)}(x)+U^{\rm load}(F,x)$, where
$U_{j|0}^{(0)}(x)\equiv U^{\rm load}(0,x)$ is
the (free) energy profile at zero pulling-force and $U^{\rm
  load}(F,x)$ accounts for a nonzero pulling-force $F$.
The potential along pathway 1,  $U_{1|0}(x)$, and along pathway 2, $U_{2|0}(x)$, is depicted in Fig.~\ref{fig:catch_potential},
where solid lines
represent potential values at zero pulling-force $F=0$, dashed lines show
the tilted potential under a moderate force $F=30\,$pN, and the
dotted line corresponds to a pulling force $F=60\,$pN.
Blue lines depict the potential along the fast pathway 1, and orange lines
the potential along the slow pathway 2. The
potential is formally defined as follows.
Defining the scaled dimensionless distance $\tilde x\equiv  x/(4.14\,\text{nm})$, the potentials are given by 
\begin{widetext}
 \begin{equation}
\begin{aligned}
\beta U^{\rm pull}(x,F)&=\frac{F}{1\,\text{pN}}\times
\begin{cases}
 50 \tilde x^2 & \text{if $\tilde x\le 0.01$,}\\
 \tilde x-0.005&\text{if $\tilde x>0.01$,}
\end{cases}\\
 \beta U_{1|0}^{(0)}(x)
 &=
 \begin{cases}
  8500 \tilde x^2&\text{if $\tilde x\le 0.01714$},\\
  5.1-8160(\tilde x-0.035)^2&\text{if $0.01714<\tilde x\le 0.06=l_{1|0}/(4.14\,$nm)},
 \end{cases}\\
  \beta U_{2|0}^{(0)}(x)
 &=
 \begin{cases}
  258.228 \tilde x^2&\text{if $\tilde x\le 0.16458$},\\
  8.5-1200(\tilde x-0.2)^2&\text{if $0.16458<\tilde x\le 0.28417$,}\\
  -1.6+6378.95(\tilde x-0.3)^2&\text{if $0.28417<\tilde x\le 0.33553$,}\\
12-9264.39(\tilde x-0.36)^2&\text{if $0.33553<\tilde x\le 0.397=l_{2|0}/(4.14\,$nm)},
 \end{cases}
\end{aligned}
\label{eq:potential_catch}
 \end{equation}
\end{widetext}
where $l_{1|0}=0.06\times4.14\,\text{nm}=0.248\,\text{nm}$ and $l_{2|0}=0.397\times4.14\,\text{nm}=1.64\,\text{nm}$. The dimensionless unit-length $\tilde x=x/(4.14\,\text{nm})$ is used to connect thermal energy and force according to $k_{\rm B}T/(1\,\text{pN})=4.14\,\text{nm}$.
The diffusion coefficient is set to be constant $D_{j|0}(x)=(4.14)^2\,\text{nm}^2\text{s}^{-1}=17.1\,\text{nm}^2\text{s}^{-1}$ along both pathways $j=1$ and $j=2$.

\begin{figure*}
\includegraphics{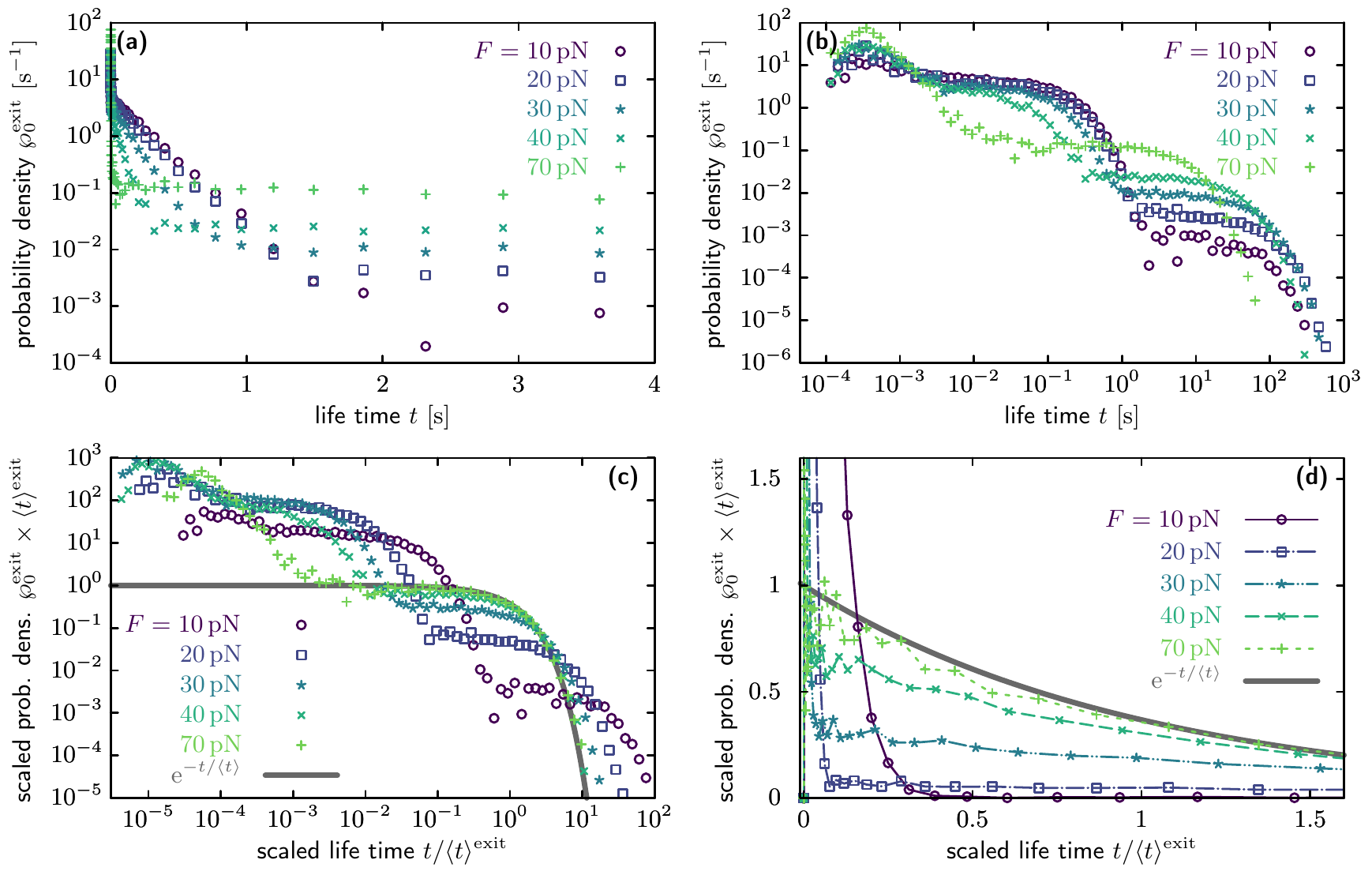}
 \caption{Probability density of bond life-time. \normalfont
 (a)~Probability densities on a semi-logarithmic scale.
 (b)~Probability densities on a log-log scale. (c)~Scaled probability density and scaled time on a log-log scale; the mean life-time corresponds to $t/\avgexit{t}=1$. (d)~Scaled probability densities on a linear scale. Each probability density is estimated from 10\,000 ruptures. All symbols are deduced from histograms with equidistant binning in logarithmic time.  
 }
 \label{fig:S_catch_hist}
\end{figure*}

\paragraph{Simulation results in Fig.~\ref{fig:illustrantion_all}d.}
We propagate the microstate with the stochastic Runge-Kutta scheme given in Appendix~\ref{subsubsec:Runge_Kutta} with time increment $\varDelta t=10^{-6}\,\mathrm{s}$.
Setting the initial distance to $x_0=0$ we
simulate $10^4$ ruptures
for each force $F=0,5\,{\rm pN},\ldots,85\,{\rm pN}$, respectively.
Some selected probability densities  with a logarithmic (increasing) binning are shown in Fig.~\ref{fig:S_catch_hist}. Fig.~\ref{fig:S_catch_hist}a  depicts the probability densities on a semi-logarithmic scale only over a short period of time (4 seconds). In Fig.~\ref{fig:S_catch_hist} depicts the full time range on a double-logarithmic scale,
which after normalization of time $t\to t/\avgexit{t}$,
Fig.~\ref{fig:S_catch_hist}c, allows us to conveniently depict the shape of all distributions simultaneously on a linear scale as in Fig.~\ref{fig:S_catch_hist}d. That is,
all scaled densities in
in Fig.~\ref{fig:S_catch_hist}d
have the \emph{same} scaled  mean at  $t/\avgexit{t}=1$. We adopted the density belonging to rectangeles  ($F=20\,{\rm pN}$) in Fig.~\ref{fig:S_catch_hist}
in the blue shaded plot
in Fig.~\ref{fig:illustrantion_all}d.
In Tab.~\ref{tab:asymmetry_catch} we list the mean rupture times along the pathway 1 and 2, respectively, whereby length of  orange and black bars in
 Fig.~\ref{fig:illustrantion_all}d 
 along the $j$-th pathway represent the values $\avg{t}_{j|0}$ at
 $F=20\,{\rm pN}$ from the table.

\begin{table}
 \caption{Asymmetry of the mean rupture time. The mean rupture time $\avg{t}_{1|0}$ and $\avg{t}_{2|0}$ along pathway 1 and 2, respectively.
 We highlight the results with a strong asymmetry $\avg{t}_{2|0}/\avg{t}_{1|0}>100$. 
 }
 \label{tab:asymmetry_catch}
 \begin{tabular}{c|c|c|c}\hline\hline
 pulling force $F$ &$\phi_{1|0}$& $\avg{t}_{1|0}$& $\avg{t}_{2|0}$\\\hline
 10\,pN&0.98&2.81\,s&62\,s\\
 20\,pN&0.81&6.32\,s&77\,s\\
 30\,pN&0.44&2.51\,s&58\,s\\
 40\,pN&0.17&0.43\,s&33.2\,s\\
 50\,pN&0.07&\textbf{0.015}\,s&\textbf{19.4\,s}\\
 60\,pN&0.05&\textbf{0.004\,s}&\textbf{11.35\,s}\\\hline\hline
 \end{tabular}
\end{table}

 In contrast to the experiment \cite{thom06} we assumed here that \emph{all} trajectories instead of $99.2$\,\%  start from $x=0$. We note that the fit of the experimental data carried out in Ref.~\cite{thom06} found the initial binding to take place with $99.2$\,\% in what was called state 1, which corresponds here to the distance $x=0$. Correspondingly, about $0.8$\,\% of experimental ruptures carried out in Ref.~\cite{thom06} were estimated to start in the first intermediate minima along the slow path 2 (potential is depicted in
 Fig.~\ref{fig:catch_potential}
 in the main text). 

\begin{figure}%
 \centering%
   \includegraphics[scale=1]{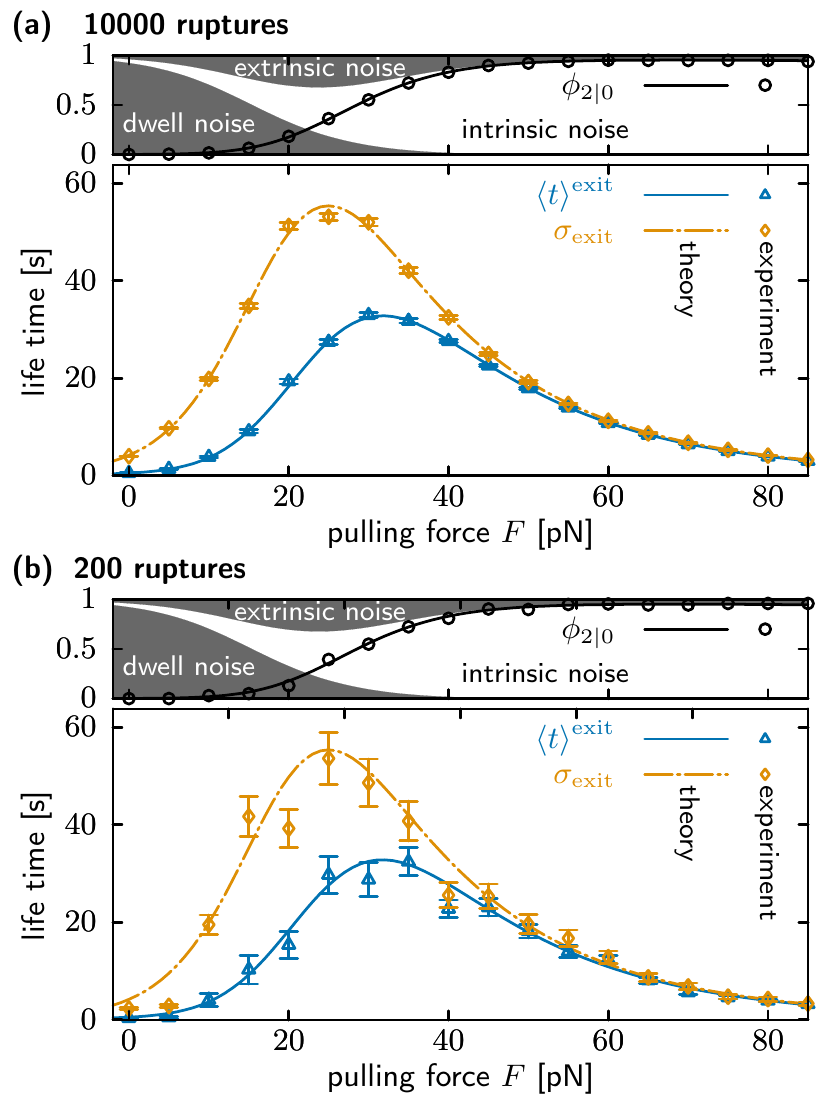}%
 \caption{Catch-bond analysis with improved statistics.  The results for (a) $10^4$ ruptures (b) 200 ruptures. The error bars denote the standard deviation and the lines the theoretical results.}%
 \label{fig:S_catch_more_statistics}%
 \end{figure}

\paragraph{Simulation of Fig.~\ref{fig:allresults}b,c}
Using \emph{all} $10^4$\,rupture events
we deduce in Fig.~\ref{fig:S_catch_more_statistics}a (see symbols)
the splitting probability
$\phi_{2|0}=1-\phi_{1|0}$, 
the mean life time, $\avgexit{t}$, and the its standard deviation $\sigma_{\rm exit}=\sqrt{\avgexit{t^2}-(\avgexit{t})^2}$.
The error bars denote the root mean square error. 
The theoretical lines in Fig.~\ref{fig:S_catch_more_statistics}a were obtained by
numerical integration of Eq.~\eqref{eq:Idef} along both pathways (1 and 2)
and consecutive use of
Eqs.~\eqref{eq:phi_explicit}-\eqref{eq:binomial}. This example, nicely corroborates the validity of our results. In Fig.~\ref{fig:allresults}b,c the number of ruptures is chosen to be similar as in typical experiments \cite{thom06,buck14} (here 500 rupture events). 

\paragraph{Interplay of fast and slow transitions at finite statistics.}
The interplay between fast and slow transitions can lead to severe misinterpretation of experimental error estimates (here at low pulling force). To see this we presume that we had only 200 ruptures measured (see Fig.~\ref{fig:S_catch_more_statistics}b). Clearly, errors are expected to become larger, whereas at low pulling forces we mistakenly estimate the errors to be far too small. To undestand this we need to inspect the full probability densities depicted in Fig.~\ref{fig:S_catch_hist} (see circles, $F=10\,{\rm pN}$). The probability density is negligibly small at values $t\le \avgexit{t}/2$, i.e., the mean is mainly dominated by extremely rare and extremely long transitions. This becomes more severe at smaller forces. For example, at $F=5\,{\rm pN}$ 
we do not encounter a single rupture along the slow path in the first 200 ruptures, which is why
we experimentally would not be able to see them.
This is the reason that 
the theory lines in \ref{fig:S_catch_more_statistics}b are ten standard deviations away from the theory line at $F=5\,{\rm pN}$. 
In other words, in reality 200 ruptures \emph{alone} at $F=5\,{\rm pN}$  lead to the \emph{same} quality of statistics as an experiment with only one or two ruptures.

\begin{figure}
 \centering
  \includegraphics{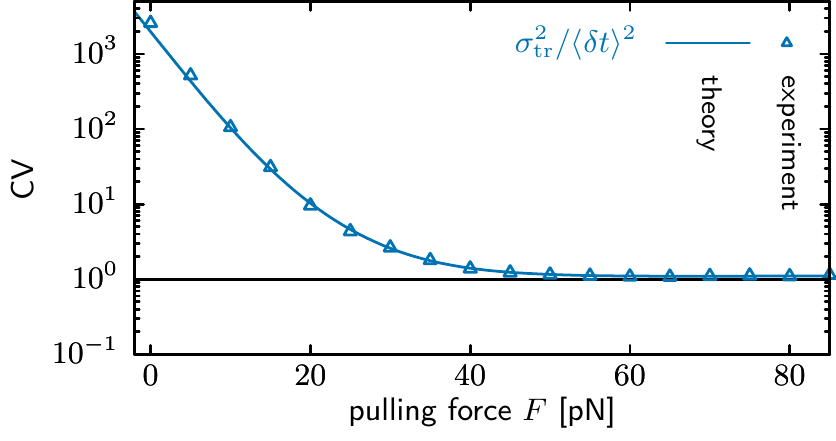}
 \caption{\color{mynewcolor}Coefficient of variation of transition-path time. Symbols are obtained from $10^4$ ruptures as in Fig.~\ref{fig:S_catch_more_statistics} and  the theory lines are deduced from Eq.~\eqref{eq:trans_methods}. We defined the mean transition time $\langle\delta t\rangle=\phi_{1|0}\langle\delta t\rangle_{1|0}+\phi_{2|0}\langle\delta t\rangle_{2|0}$ and variance $\sigma^2_{\rm tr}=\phi_{1|0}\langle\delta t^2\rangle_{1|0}+\phi_{2|0}\langle\delta t^2\rangle_{2|0}-\langle\delta t\rangle_0^2$.}
 \label{fig:S_CV}
\end{figure}

\begin{figure*}%
\includegraphics[width=\textwidth]{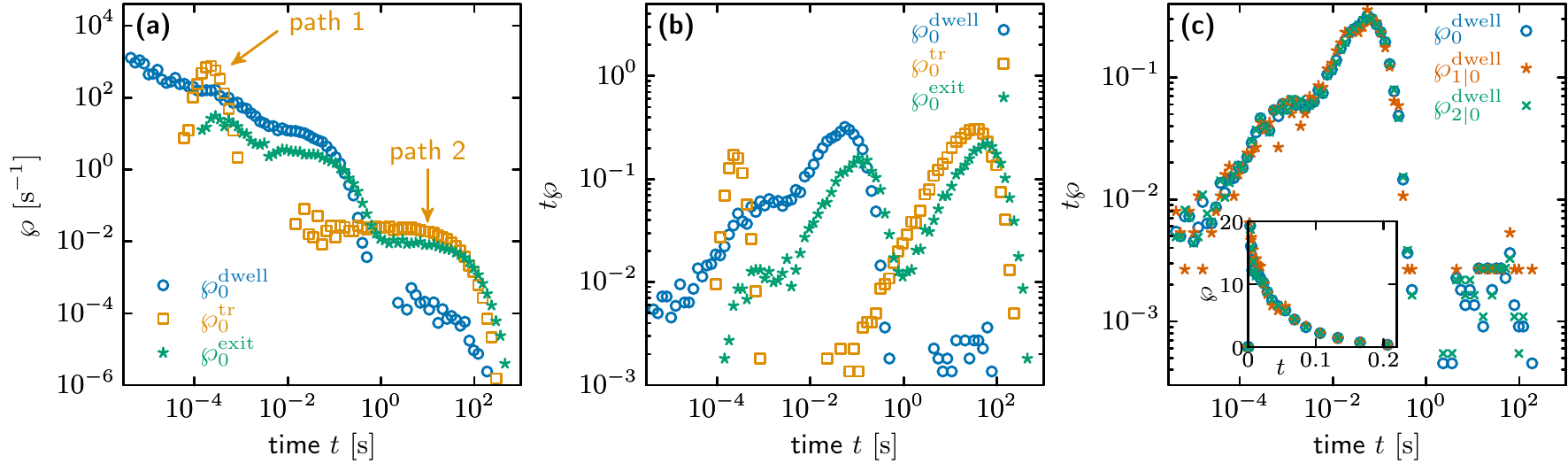}
 \caption{Decomposition of the probability density of  bond life-time at pulling force $F=30$\,pN into  the probabilities of transition and dwell time, respectively.   (a)~Probability densities of bond life-time $\wpcexit$, dwell-time $\wpcdwell$,  and transition-time $\wpctrans$ on a log-log scale.
 (b)~Result from (a) where the probability densities are transformed to logarithmic densities, $t\wp (t)$, which is normalized according to $\int_{-\infty}^\infty t\wp (t)\,\dd (\ln t)=\int_0^\infty\wp (t)\,\dd t=1$. (c)~Test of $\wpcdwell_0=\wpcdwell_{1|0}=\wpcdwell_{2|0}$. 
 }
 \label{fig:S_catch_hist2}
\end{figure*}

{\color{mynewcolor}\paragraph{Detection of parallel transition paths according to Ref.~\cite{sati20}.}\label{paragraph:sati20}
 If we are able transition path times $\delta t$ directly one can also
 evaluate the coefficient of variation $\sigma_{\rm tr}^2/\avg{\delta
   t}^2=(\avg{\delta t^2}-\avg{\delta t}^2)/\avg{\delta t}^2$. The
 result is shown in Fig.~\ref{fig:S_CV}. The coefficient clearly
 exceeds 1, which according to Ref.~\cite{sati20} correctly implies multidimensional (parallel) transition paths. Note in Sec.~\ref{subsec:super_Markov}  we detect parallel transition paths  ``merely'' from measuring the life time of the bond $t=\delta t+\tau$, which formally represents a first passage time.
 }

\paragraph{Dwell time is a property of the state.} Using the individual trajectories for the catch-bond system we
dissect the life-time of a bond into a transition- and dwell-period according to
Eq.~\eqref{eq:local_decomposition}
in the main text. For brevity, we merely show the result for $F=30$\,pN at which the two paths are taken with approximately equal probability.
In Fig.~\ref{fig:S_catch_hist2}a we depict the histogram of the life-time of the bond $\wpcexit_0(t)$ (see green stars), which is dissected into
the statistics of dwell time $\wpcdwell_0$ (see blue circles) and the transition-time statistics $\wpctrans_0\equiv \phi_{1|0}\wpctrans_{1|0}+\phi_{2|0}\wpctrans_{2|0}$ (see orange rectangles), respectively. The
arrow ``path 1'' in Fig.~\ref{fig:S_catch_hist2}a indicate $\phi_{1|0}\wpctrans_{1|0}$, whereas the arrow ``path 2''
indicates $\phi_{2|0}\wpctrans_{2|0}$. 
Note that 
 the life-time $\wpcexit_0$ is equal to the convolution of dwell- and transition-time distributions, $\wpcexit_0=\wpctrans_0*\wpcdwell_0$, which signifies their statistical independence.
Fig.~\ref{fig:S_catch_hist2}b depicts the probability density of logarithmic time, $\ln t$, which is $t\wp(t)$ since $\int_{-\infty}^\infty t\wp(t)\dd [\ln t]=1$.

We now use the data to verify symmetry (i) in our second main result,
Eq.~\eqref{eq:local_decomposition},
which states 
that the dwell-time statistics is
identically distributed along both pathways 1 and 2. To test this we
compared the histogram of dwell time along path 1, $\wpcdwell_{1|0}$,
and the histogram along path 2, $\wpcdwell_{2|0}$, with the estimated
probability density along both pathways $\wpcdwell_{0}$ in
Fig.~\ref{fig:S_catch_hist2}c, where the inset depicts the results on a linear scale. Fig.~\ref{fig:S_catch_hist2}c nicely illustrates 
their distribution to be equal $\wpcdwell_{0}=\wpcdwell_{1|0}=\wpcdwell_{2|0}$ (deviations are merely arising from finite statistics). This example illustrates that the dwell-time statistics does not depend on the pathway of the rupture (states 1 and 2) but only on the initial state 0, i.e. the dwell-time statistics solely depends on initial state (not on the final one). This example corroborates
Eq.~\eqref{eq:local_decomposition}
in the main text. Using fast three-color single-molecule Foster resonance energy transfer (FRET) it was possible to detect similar parallel transition paths in the binding of disordered proteins  \cite{kim20}.

\paragraph{Alternative experiment from Ref.~\cite{buck14}.}
Finally, we want to comment on the  effect of changing the length-scale. Suppose the length $x$ is stretched by a factor $\lambda$ such that $U_{j|i}(x)\to U_{j|i}(x/\lambda)$, i.e. $F_{j|i}(x)\to F_{j|i}(x/\lambda)/\lambda$, which implies that the
loading force $F$ becomes equivalent to the loading force $F/\lambda$ after rescaling. To address a related experiment \cite{buck14} displaying quite different time- and length-scales we need to scale the
length by a factor $\lambda$ ($\lambda\approx3$) such that the maximum  life-time is found at $F\approx 10$\,pN as reported in Ref.~\cite{buck14} instead of $F\approx 30$\,pN, which is shown here in Fig.~\ref{fig:S_catch_more_statistics} (see also Ref.~\cite{thom06}). Moreover, scaling the diffusion constant $D\to\alpha D$ corresponds to an accelerated  time, which re-scales the bond life-time $\propto\alpha^{-1}\lambda^{-2}$. To shift the maximum life time from $\simeq 30$\,s (see Fig.~\ref{fig:S_catch_more_statistics}) to 1.2\,s${}={}$30\,s$/$25 from the experiment in Ref.~\cite{buck14} we, in addition to the scaled location  of the maximum, scale the diffusion constant by $\alpha=25\times\lambda^{-2}\approx2.78$. With this scaled units we obtain the same plots as shown in Fig.~\ref{fig:S_catch_more_statistics} (see also
Fig.~\ref{fig:allresults}b,c
in the main text) but with the x-axis scaled by a factor of $1/3$ and the y-axis is scaled by a factor of $1/25$ to quantitatively account for different experiment reported in Ref.~\cite{buck14}.

Summarizing, in this subsection we further confirmed
Eqs.~\eqref{eq:phi_explicit}-\eqref{eq:binomial}
in the main text, by numerical experiments, which are shown in Figs.~\ref{fig:S_catch_more_statistics}
using more statistics (up to $10^4$\,rupture events). We tested the decomposition of the bond life-time into its
dwell- and transition-period according to
Eq.~\eqref{eq:local_decomposition}
in the main text and we corroborated the theoretical prediction that the dwell time indeed  depends only on the initial state but not the final state (see Fig.~\ref{fig:S_catch_hist2}c).

%
%
%
%

\subsection{ATPase with sine-wave potential}\label{subsec:ATPase}
\paragraph{Model and energetics.}
We assume the dynamics of an idealized ATPase\-  to be described  by the following model.
The ATPase\- rotates stochastically about one axis and experiences 
an angle dependent torque at rotation angle $\theta_t$ at time $t$.
The torque is assumed to have the following two contributions: 
\begin{itemize}
 \item[\textbf{(i)}]a rotational free energy potential
(see blue shaded lines Fig.~\ref{fig:allresults}d) that displays three well-defined rotational states (minima) that are separated by $120^\circ$. The free energy exerts a conservative torque  proportional to the slope of the blue line.  The potential is given by
$U^\text{rot}(\theta)=\frac{B}{2}[1-\cos(\theta/3)]$ with the implied conservative torque given by $-\partial_\theta U^\text{rot}(\theta)$.

\item[\textbf{(ii)}]
a non-equilibrium
 torque $M$  that 
 embodies a sum of a mechano-chemical force arising from the hydrolysis of an ATP molecule and a mechanical torque that is applied to the shaft.  More precisely, a tight coupling with $M=\varDelta \mu/120^\circ-M^{\rm mech}$ is assumed, where $\varDelta\mu=\mu_{\rm ATP}-\mu_{\rm ADP}-\mu_{\rm Pi }$
 is the chemical free energy released in the hydrolysis  reaction ${\rm ATP}\to{\rm ADP}+{\rm P}_i $ and $M^{\rm mech}$ reflects a mechanical torque \cite{toya11}.
\end{itemize}
In Fig.~\ref{fig:allresults}d-f we set $B=5\,k_{\rm B}T$ and assume the
diffusion coefficient to be constant and without loss of generality $D=1$. Moreover, we  use
scaled units $x_t=\theta_t/120^\circ$, that is, distances are measured in units of a third of a revolution.
Using the scaled coordinate 
the local potential, which accounts for both torque (i) and mechano-chemical force (ii), is given by
$U_\pm(x)=\frac{B}{2}[1-\cos(2\pi x)]\pm (M\times 120^\circ) x$ with $l_\pm=120^\circ/120^\circ=1$, where ``$+$'' accounts for the potential along the counterclockwise direction and ``$-$''corresponds to the potential along the opposite direction.
Detailed balance   
is established whenever the chemical free energy released per $120^\circ$ step is balanced by the mechanical torque (multiplied by $120^\circ$), i.e.  $M=0$ (see item (ii) above).

\begin{figure}
 \centering
    \includegraphics[width=\columnwidth]{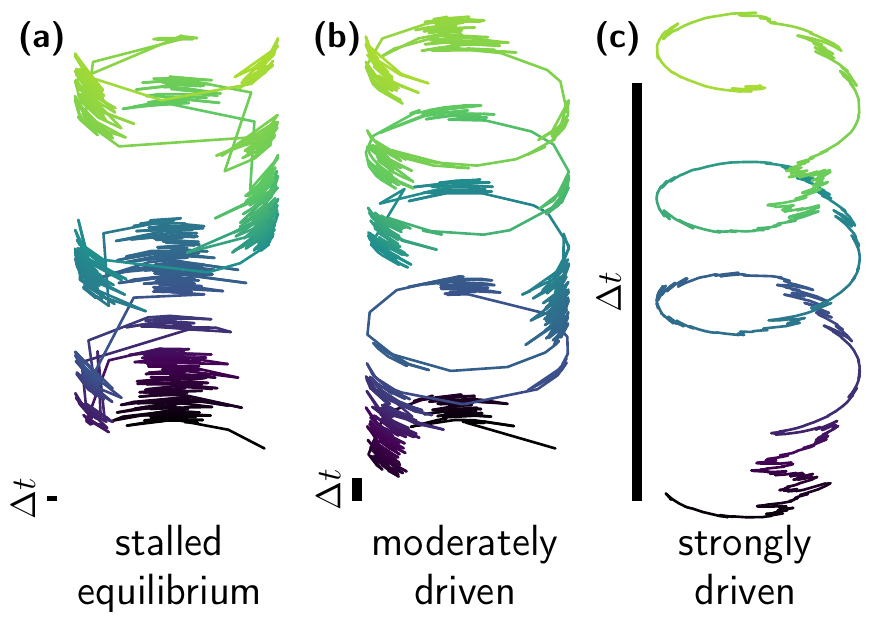}
 \caption{Effect of non-equilibrium driving on single trajectories. Realization of a stochastic  trajectory of the ATPase\-
 toy-model as function of time (running from dark to bright) at (a) equilibrium $M=0$, (b) at moderate non-equilibrium driving $M\times 120^\circ=B=5\,k_{\rm B}T$ and  (c) at strong driving $M\times 120^\circ=20\,k_{\rm B}T$.  The length of each trajectory is chosen to have the same average number of transitions. To compare the different time units we added the bars $\Delta t$, which all span the \emph{same} period of time.
 }
 \label{fig:S_rotator_3dtraj}
\end{figure}

For convenience, we restrict our analysis to a  periodic rotation which has a sine wave shape with barriers of height $B$ that separate two minima. Counting the minima in the counterclockwise direction yields the set of states $\Omega=\{1,2,3\}$, such that for each state $i\in \Omega$
the local potential formally reads $U_{i\pm 1|i}\equiv U_{\pm}$ with the convention ``$i-1=3$
if $i=1$'' and ``$i+1=1$ if $i=3$''. The sets of neighboring states are then $\mathcal{N}_1=\{2,3\}$, $\mathcal{N}_2=\{1,3\}$ and $\mathcal{N}_4=\{1,2\}$, that is, the three-state network is fully-connected.
We want to compare the minima-to-minima dynamics, which are generally non-Markovian,
to a Markov kinetics corresponding to an exponentially distributed waiting time
with the same expected time $\avgexit{t}$ of leaving each minimum. Only two numbers become relevant, $B/(k_{\rm B} T)$ and $M\times 120^\circ/(k_{\rm B}T)$
representing, respectively, the barrier-height separating two minima  $B$ and the non-equilibrium driving $M$ in units of the thermal energy $k_{\rm B}T$. The diffusion constant is set to $D_\pm=1$. To obtain the numerical results in Fig.~\ref{fig:allresults}e,f we fixed the barrier height to $B/(k_{\rm B} T)=5$
and use the stochastic Runge Kutta scheme (see Appendix~\ref{subsubsec:Runge_Kutta}) with time increment $\varDelta t=10^{-4}$ in dimensionless simulation units. 
We simulate all trajectories until we observe in total 500\,000 state-to-state changes (i.e., minima-to-minima transitions).
Note that a different value for $D\neq 1$ would \emph{not} affect Fig.~\ref{fig:allresults}e,f.
The fluctuating rotational state as a function of time is illustrated in Fig.~\ref{fig:S_rotator_3dtraj}  for various strengths of driving $M\times 120^\circ/(k_{\rm B}T)=0,5,20$, where each all black bars  represent  an equal duration $\varDelta t$,
indicating that a stronger driving leads to faster rotation.

\paragraph{Splitting probability.} The splitting probability
involves the auxiliary integrals
\begin{equation}
\begin{aligned}
  I^{(1)}_+&=\int_0^1 \e^{\beta \frac{B}{2}[1-\cos(2\pi x)]- \beta(M\times 120^\circ) x}\dd x
  ,\\
 I^{(1)}_-&=\int_0^1 \e^{\beta \frac{B}{2}[1-\cos(2\pi x)]+ \beta(M\times 120^\circ) x}\dd x,
\end{aligned}
\end{equation}
where we used Eq.~\eqref{eq:Idef}
 with $k=1$ and inserted
$U_{\pm}(x)=\frac{B}{2}[1-\cos(2\pi x)]\mp (M\times 120^\circ) x$ and $l_\pm=1$.
Using Eq.~\eqref{eq:phi_explicit}%
, one obtains after some algebra
\begin{equation}
\begin{aligned}
  \phi_+&=\frac{1/I^{(1)}_+}{1/I^{(1)}_++1/I^{(1)}_-}
  =\frac{\e^{f}}{\e^{f}+1}=1-\phi_-,
\end{aligned}
 \label{eq:S8_splitting_ATP}
\end{equation}
where we defined $f=\beta(M\times 120^\circ)$. Note that splitting probabilities satisfy Eq.~\eqref{eq:local_detailed_balance_like}, which here corresponds to $\ln(\phi_+/\phi_-)=f$. Some values are listed in Tab.~\ref{tab:moments_atp}.
\begin{table*}
 \caption{Comparing theory to simulation.  Non-equilibrium driving is quantified in terms of $f\equiv M\times 120^\circ /(k_{\rm B} T)$. 
 The theoretical values
 for the splitting probability follow from Eq.~\eqref{eq:S8_splitting_ATP}. By evaluating
 the auxiliary integrals
 in  Eq.~\eqref{eq:Idef} and using 
 Eqs.~(S47), and (S50) in the SM \cite{Note2}, 
 we obtain the theoretical values for mean first exit time $\avgexit{t}$ and the second moment of the exit time
 $\avgexit{t^2}$, and therefrom the standard deviation $\sigma_{\rm exit}=\sqrt{\avgexit{t^2}-(\avgexit{t})^2}$. Note that for $f=20$ the system is driven so strongly that no backward transition is observed in $500\,000$ trajectories, which is why we experimentally determine $\phi_-=0$.
 }
 \label{tab:moments_atp}
\centering
\begin{tabular}{c||c|c||c|c||c|c}\hline\hline
non-equil.&\multicolumn{2}{c||}{splitting prob. $\phi_-=1-\phi_+$}&\multicolumn{2}{c||}{mean exit time $\avgexit{t}$}&
\multicolumn{2}{c}{standard deviation $\sigma_{\rm exit}$}
\\\cline{2-7}
driving $f$
&theory&experiment
&theory&experiment
&theory&experiment\\\hline
 0 &$\boldsymbol{0.500000}$&  0.500910    &$\boldsymbol{5.4115}$&5.4158&$\boldsymbol{5.3561}$&5.3533\\
 2 &$\boldsymbol{0.119203}$&  0.119490    &$\boldsymbol{3.7064}$&3.7066&$\boldsymbol{3.6513}$&3.6469\\   
 5 &$\boldsymbol{0.006692}$&  0.006654    &$\boldsymbol{1.2488}$&1.2474&$\boldsymbol{1.1951}$&1.1940\\
 10&$\boldsymbol{0.000045}$&  0.000048    &$\boldsymbol{0.2910}$&0.2916&$\boldsymbol{0.2425}$&0.2428\\
 20&$\boldsymbol{2.06\cdot10^{-9}}$&  0     &$\boldsymbol{0.0703}$&0.0705&$\boldsymbol{0.0350}$&0.0350\\\hline\hline
\end{tabular}
\end{table*}

\paragraph{Symmetry of the waiting time distribution.} 
We have proven in Appendix~\ref{sec:greens} a forward/backward symmetry 
of the transition time ``$\wpctrans_+(\delta t)=\wpctrans_-(\delta t)$''  (i.e.,  symmetry (ii) in our second main result in Eq.~\eqref{eq:local_decomposition}).
To numerically corroborate this main finding we compare in Tab.~\ref{tab:ATPase_forward/backward_symmetry}
\begin{table}
 \caption{Test of forward/backward symmetry of mean transition-time. The transition time is evaluated from 500\,000 state-to-state changes. The statistical error in the mean transition time $\avg{t}_-$ denotes the estimated $\sim 95\,\%$ confidence interval.
 }
 \label{tab:ATPase_forward/backward_symmetry}
\centering
\begin{tabular}{c||c|c|c}\hline\hline
non-equil.&\multicolumn{2}{c|}{mean transition time}&$\#$ steps\\
\cline{2-3}
driving& $\avgtrans{\delta t}_+$&$\avgtrans{\delta t}_-$ &backwards\\\hline
 0 &$0.0656$&$0.0656\pm0.0001$&250457 \\
 2 &$0.0659$&$0.0660\pm0.0002$&59745\\   
 5 &$0.0676$&$0.0685\pm0.0009$&3327 \\
 10&$0.0712$&$0.0573\pm0.0071$&24\\\hline\hline
\end{tabular}
\end{table}
the mean transition time along the forward ``$+$'' and backward ``$-$'' direction  (as explained in Appendix~\ref{subsubsec:functionals}). In Fig.~\ref{fig:S_ATPase_FB_trans} we further compare the entire probability densities.
\begin{figure}
 \centering
  \includegraphics{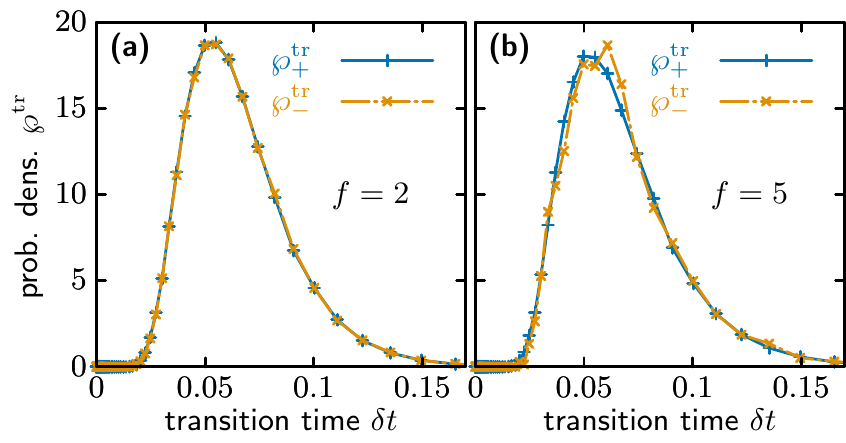}%
 \caption{Test of forward/backward symmetry of transition time. \normalfont Probability density of transition time $\wpctrans_\pm$ in forward ``$+$'' and backward ``$-$'' direction for weakly for (a) $f=2$ and (b) $f=5$. The number of trajectories entering $\wpctrans_-$ are (a) $n_-=59745$ and (b) $n_-=3327$ (see also Tab.~\ref{tab:ATPase_forward/backward_symmetry}). Deviations between blue and orange lines in (b) are due to finite statistics ($n_-=3327$); see also Tab.~\ref{tab:ATPase_forward/backward_symmetry} for the number of observed backward transitions.}
 \label{fig:S_ATPase_FB_trans}
\end{figure}
Due to the periodicity in each $120^\circ$ step and the forward/backward symmetry 
of the transition time ``$\wpctrans_+(\delta t)=\wpctrans_-(\delta t)$'', Eq.~\eqref{eq:local_decomposition} implies that the local waiting is given by $\wpcStar_{\pm}(t)=\phi_\pm \wpcexit(t)$, such that $\avg{t^k}_{\pm}=\avgexit{t^k}$. Therefore, the extrinsic transition-noise
vanishes, which in turn according the proof
  shown in
the last subsection in the SM \cite{Note2}
 implies the fluctuations to be sub-Markov: $\sigma_{\rm exit}^2\equiv\avgexit{t^2}-(\avgexit{t})^2\le(\avgexit{t})^2$, that is, $\avgexit{t^2}\le 2 (\avgexit{t})^2$.

A few comments are in order.
The symmetry of the local mean waiting time $\avg{t}_{+}=\avg{t}_{-}$, was, to the best of our knowledge, first discovered in Ref.~\cite{kolo05} for lattice models
of kinesin motors (see also Ref.~\cite{kolo07}). The extension to the entire distribution $\wpcStar_{\pm}(t)=\phi_\pm \wpcexit(t)$ was later found in studies for the stopping-time of the thermodynamic entropy production in active molecular processes \cite{neri17}.  The symmetry allows us to  simplify the discussion by merely focusing on the splitting probability $\phi_\pm$ and the exit time distributions $\wpcexit(t)$.

\paragraph{Exit-time statistics and implied number of transitions.}
The number of exits  after time $t$, $n_t$, where one exit corresponds to the event of leaving one minima and reaching any other minima for the first time,
is stochastic and influenced  solely by the exit time $\wpcexit(t)$. As explained  above the distribution of the waiting-time is the same  along both directions ``$+$'' and ``$-$'', i.e. $\psi_\pm=\wpcStar_{\pm}(t)/\phiStar_{\pm}=\wpcexit(t)$. At long times the central limit theorem for renewal processes \cite{fell71}
renders $n_t$ asymptotically normally distributed with mean
$\avg{n_t}\simeq t/\avgexit{t}$ and variance $\operatorname{var}(n_t)\equiv\avg{n_t^2}-\avg{n_t}^2\simeq t\sigma_{\rm exit}^2/(\avgexit{t})^3= t[\avgexit{t^2}-(\avgexit{t})^2]/(\avgexit{t})^3$, where ``$\simeq$'' denotes equality ``$=$'' in the limit $t\to\infty$ (see also Ref.~\cite{land77}).
Applying the
central limit theorem
for
the mean square angular deviation, $\avg{\delta \theta_t^2}\equiv\avg{[\theta_t-\avg{\theta_t}]^2}$,
we obtain
\begin{align}
 \frac{\avg{\delta \theta_t^2}}{(120^\circ)^2}&=
 4\phi_+\phi_-\avg{n_t}+(\phi_+-\phi_-)\operatorname{var}(n_t)
 \nonumber\\
 &\simeq 4\phi_+\phi_-\frac{t}{\avgexit{t}}+(\phi_+-\phi_-)\frac{t\sigma_{\rm exit}^2}{(\avgexit{t})^3},
 \label{eq:S8_error_propagation_proof}
\end{align}
where in the first step we related the number of state-to-state changes to the angular deviation, and in the second step we used the central limit theorem. Eq.~\eqref{eq:S8_error_propagation_proof} proves
Eq.~\eqref{eq:error_propagation}
in the main text.
As soon as $\wpcexit(t)$ becomes memory-less, that is, $\wpcexit(t)\propto\e^{-t/\avgexit{t}}$, $n_t$ becomes \emph{Poissonian} with mean $\avg{n_t}^{\rm M}=t/\avgexit{t}$ and variance $\operatorname{var}(n_t)^{\rm M}=t/\avgexit{t}$, where the superscript ``M'' signifies the restriction to memory-less Markov jumps.

\paragraph{Proof of the ``upper bound'' in Fig.~\ref{fig:allresults}f.}
Using
Eq.~\eqref{eq:S8_error_propagation_proof}
we obtain the ratio of the true
angular mean squared deviation, $\langle \delta\theta_t^2\rangle$, and the one deduced
from a Markov-jump model that corresponds to setting $\sigma_{\rm
  exit}^{\rm M}=\avgexit{t}$, i.e.
\begin{equation}
\frac{\langle \delta\theta_t^2\rangle}{\langle \delta\theta_t^2\rangle^{\rm M}}\simeq\frac{4\phi_+\phi_-(\avgexit{t})^2+(\phi_+-\phi_-)^2\sigma_{\rm exit}^2}{(\avgexit{t})^2}
\label{eq:diff_ratio}
\end{equation}
where equality holds as $t\to\infty$, and  the superscript ``M'' denotes the Markov-jump limit (see also Ref.~\cite{maes09}).
Vanishing extrinsic noise renders the kinetics sub-Markovian,
$\sigma_{\rm exit}^2\le (\avgexit{t})^2$ and immediately yields
$\langle \delta\theta_t^2\rangle/\langle \delta\theta_t^2\rangle^{\rm M}\le 1$. This implies the dotted line in
Fig.~\ref{fig:allresults}f  to be a general upper bound on angular diffusivity.

In this subsection we showed that the splitting probability for the ATPase modeled by a tilted  periodic potential is fully determined by the external driving $f$ and is given by Eq.~\eqref{eq:S8_splitting_ATP}, which notably holds for any $120^\circ$ periodic potential. We related the number of state-to-state transitions to the exit time
via the well-established central limit theorem for renewal processes \cite{fell71} (see also  Ref.~\cite{land77}). We illustrated the forward/backward symmetry of transition time in the mean (see Tab.~\ref{tab:ATPase_forward/backward_symmetry}) and the entire distribution of transition time (see Fig.~\ref{fig:S_ATPase_FB_trans}).

In the next section we provide details about the lower bound on the
diffusivity in Fig.~\ref{fig:allresults}f set by
the Thermodynamic
uncertainty relation (TUR), and in the subsection after that we further address biased
diffusion obtained in the limit of vanishing free energy barriers ($B\to0$).

\subsection{Thermodynamic uncertainty relation (TUR) in periodic systems}\label{subsec:TUR_proof}
In previous subsection (see Eq.~\eqref{eq:diff_ratio}), we derived an upper bound on the diffusivity when extrinsic noise vanishes.
Conversely,
a \emph{lower bound} on the diffusivity can be deduced from the
so-called thermodynamic uncertainty relation (TUR) \cite{bara15,horo20}.
In the limit $t\to\infty$ the TUR for unicyclic networks implies
\begin{equation}
 \frac{\langle \delta\theta_t^2\rangle}{\langle \theta_t\rangle^2}\times(\phi_+-\phi_-)\frac{t}{\avgexit{t}}\ln\frac{\phi_+}{\phi_-}\ge 2,
  \label{eq:ineq_TUR_pre}
\end{equation}
where $\langle \theta_t\rangle/120^\circ\to (\phi_+-\phi_-)t/\avgexit{t}$.
Inserting
Eqs.~\eqref{eq:S8_error_propagation_proof}
and \eqref{eq:ineq_TUR_pre}
into Eq.~\eqref{eq:diff_ratio} yields
\begin{align}
 \frac{\langle \delta\theta_t^2\rangle}{\langle \delta\theta_t^2\rangle^{\rm M}}
 &\ge
 \frac{2(\phi_+-\phi_-)}{\ln(\phi_+/\phi_-)}=\frac{2}{f}\frac{\e^{f}-1}{\e^{f}+1},
 \label{eq:ineq_TUR}
\end{align}
where in the last step we defined $f\equiv M\times 120^\circ/(k_{\rm
  B}T)$ and used $\e^{f}\equiv \phi_+/\phi_-$ which follows from Eq.~\eqref{eq:local_detailed_balance_like}. The right hand side of
the inequality \eqref{eq:ineq_TUR} is depicted in
Fig.~\ref{fig:allresults}f by the solid gray line
and coincides with the result for plain biased diffusion (i.e. with the
barrier set to zero, $B=0$; see below for more
details). This completes the proof that
the mean
squared angular deviation (angular diffusivity)
in \emph{all periodic one dimensional systems must} lie between
the dotted and solid gray lines in Fig.~\ref{fig:allresults}f.

\subsection{Plain biased diffusion saturates TUR}\label{subsec:ATPase_nobarrier}

Let us finally consider plain biased diffusion, which in the Model from Appendix~\ref{subsec:ATPase} corresponds to setting $B=0$. Adopting the reduced coordinates $x=\theta/120^\circ$  with $l_\pm=1$ the local potential simplifies to $\beta U_{\pm}=\mp\beta (M\times 120^\circ) x\equiv \mp f x$.
The splitting probability is still given by Eq.~\eqref{eq:S8_splitting_ATP}.
Using
 Eqs.~(S47), and (S50) in the SM \cite{Note2}
we obtain the
 mean and variance of exit time
 \begin{equation}
 \begin{aligned}
   \avgexit{t}&=\frac{\e^f-1}{f(\e^f+1)},\\
   \sigma_{\rm exit}^2&=\avgexit{t^2}-(\avgexit{t})^2=\frac{2(\e^{2f}-2f\e^{f}-1)}{f^3(\e^f+1)^2},
 \end{aligned}
  \label{eq:S8_exit_flat}
 \end{equation}
respectively, where we further inserted the local potential
$\beta U_{\pm}=\mp f x$ along with $D_\pm=1$
into the
first line of the corresponding auxiliary integrals in Eq.~\eqref{eq:Idef}.
Inserting  Eqs.~\eqref{eq:S8_splitting_ATP} and \eqref{eq:S8_exit_flat}
into Eq.~\eqref{eq:diff_ratio} yields
$\avg{\delta \theta_t^2}/\avg{\delta \theta_t^2}^{\rm M}\simeq 2f^{-1}(\e^f-1)(\e^f+1)^{-1}$
, which
 saturates the inequality
 Eq.~\eqref{eq:ineq_TUR}.

 \title{Supplemental Material: Emergent memory and kinetic hysteresis in strongly driven networks}
\author{David Hartich}
\email{david.hartich@mpibpc.mpg.de}
\affiliation{%
Mathematical bioPhysics Group, Max Planck Institute for Biophysical Chemistry, 37077 Göttingen, Germany}
 \author{Aljaž Godec}%
 \email{agodec@mpibpc.mpg.de}
\affiliation{%
Mathematical bioPhysics Group, Max Planck Institute for Biophysical Chemistry, 37077 Göttingen, Germany}



%
%


\onecolumngrid
\newpage

\noindent\textbf{\Large Supplemental Material:\\[2mm]Emergent memory and kinetic hysteresis in strongly driven networks}\\[8mm]
\hspace*{0.1\textwidth}\begin{minipage}{0.84\textwidth}
Derivation of the main practical result in three major technical steps.\\
\textbf{Step 1:} In the \ref{sec:gen_renewal} we show that show the singular renewal theorem be inverted. Thereby, we obtain moments of then conditional first passage time (local waiting time) in terms if higher moments of simpler unconditioned moments of first passage time.\\
\textbf{Step 2:}
\ref{sec:S_unconditioned_moments} derives the unconditioned first passage times from the backward-Focker-Planck equation. At the end of this  section we derive the first two moments of transition time. \\
\textbf{Step 3:} In \ref{sec:moments_wrapup} we insert the result from step 2 into the results from step 1 to obtain main practical result, which is shown in Sec.~II.D in the main text. \ref{subsec:sub-Markov_proof} further proves the main consequence of the main result that is proven in \ref{subsec:sub-Markov_proof} which is discussed in Sec.~V.A in the main text. 
\end{minipage}\vspace*{3mm}

\twocolumngrid


\appendix


\counterwithout{equation}{section}
\stepcounter{SIequation}
\setcounter{section}{0}
\stepcounter{SIfigure}
%



\renewcommand{\appendixname}{}
\renewcommand*{\thesection}{Supplementary Section \arabic{section}}
\renewcommand*{\thesubsection}{Supplementary Section \arabic{section}.\Alph{subsection}}
\renewcommand*{\thesubsubsection}{\normalfont \arabic{section}.\Alph{subsection}(\roman{subsubsection})}
\makeatletter
\renewcommand*{\p@section}{}
\renewcommand*{\p@subsection}{}
\renewcommand*{\p@subsubsection}{}
\makeatother



\renewcommand{\theequation}{S\arabic{equation}}

\renewcommand{\thefigure}{S\arabic{figure}}
\renewcommand{\thetable}{S\arabic{table}}



\section{{Solving generalized renewal theorem}}\label{sec:gen_renewal}
We first invert the generalized renewal theorem to represent moments of conditional first passage times in terms of simpler unconditioned ones. Thereby, we solve an underdetermined singular system of equations in the spirit of l'Hospital's rule. In the first part we discuss the renewal theorem in its most general form. 

\subsection{Deducing moments of first passage times from the renewal theorem {\normalfont (optional background)}}


\label{subsec:Renewal_inverse_general}
The Taylor expansions of the functions entering the renewal theorem
[Eq. (C3) in Appendix C in the main text]
 are connected to
 the moments of first passage times via
 \begin{widetext}
  \begin{equation}
 \begin{aligned}
  \twps_{j|i}(s)&=1-s\avgs{t}_{j|i}+\frac{1}{2}s^2\avgs{t^2}_{j|i}-\frac{1}{3!}s^3\avgs{t^3}_{j|i}\pm\ldots\qquad&\text{(for $i,j\in \Omega$ and $i\neq j$),}\\
  \twpc_{j|i}(s)&=\phic_{j|i}\bigg[1-s\avgc{t}_{j|i}+\frac{1}{2}s^2\avgc{t^2}_{j|i}-\frac{1}{3!}s^3\avgc{t^3}_{j|i}\pm\ldots\bigg]\qquad&\text{(for $j\in\mathcal{A}$ and $i\in\mathcal{A}^{\rm c}=\Omega\backslash\mathcal{A}$),}
 \end{aligned}
 \label{eq:S1_moments_def}
\end{equation}
where $\phic_{j|i}=\twpc_{j|i}(0)=\int_0^\infty \wpc_{j|i}(t)\dd t$ is the splitting probability,
which is a marginal of $\wpc_{j|i}(t)$ with normalization $\sum_{j\in\mathcal{A}}\phic_{j|i}=\sum_i\int\dd t \wpc_{j|i}(t)=1$. Moreover, $\avgs{t^k}_{j|i}$ is the $k$th (unconditional) moment of first passage time to reach a single target $j$, and $\avgc{t^k}_{j|i}$ is the $k$th moment of the conditional first passage time towards state $j$ given that none of the other states $\mathcal{A}\backslash\{j\}$ have been reached before.
A na\"ive Taylor expansion of Eq.~(C3) in Appendix C in the main text
using Eq.~\eqref{eq:S1_moments_def} yields
 \begin{equation}\addtocounter{equation}{1}\tag{\theequation}
\begin{aligned}
1&=\sum_{a'\in\mathcal{A}}\phic_{a'|i_0},&&\text{($0^\text{th}$ derivative)}\\
\avgs{t}_{a|i_0}&=\sum_{a'\in\mathcal{A}}\phic_{a'|i_0}\avgc{t}_{a'|i_0}+\sum_{a'\in\mathcal{A}\backslash\{a\}}\phic_{a'|i_0}\avgs{t}_{a|a'},&&\text{($1^\text{st}$ derivative)}\\
\avgs{t^2}_{a|i_0}&=\sum_{a'\in\mathcal{A}}\phic_{a'|i_0}\avgc{t^2}_{a'|i_0}+2\sum_{a'\in\mathcal{A}\backslash\{a\}}\phic_{a'|i_0}\avgs{t}_{a|a'}\avgc{t}_{a'|i_0}+\sum_{a'\in\mathcal{A}\backslash\{a\}}\phic_{a'|i_0}\avgs{t^2}_{a|a'},&&\text{($2^\text{nd}$ derivative)}
\\
&\;\,\vdots\\
\avgs{t^k}_{a|i_0}&=\sum_{a'\in\mathcal{A}}\phic_{a'|i_0}\avgc{t^k}_{a'|i_0}+\sum_{a'\in\mathcal{A}\backslash\{a\}}\sum_{l=1}^{k}\binom{k}{l}\phic_{a'|i_0}\avgs{t^l}_{a|a'}\avgc{t^{k-l}}_{a'|i_0},&&\text{($k^\text{th}$ derivative)}
\end{aligned}
\label{eq:S1_moments_setofequations}
\end{equation}
\end{widetext}
where $\binom{k}{l}=\frac{k!}{l!(k-l)!}$. Notably, Eq.~\eqref{eq:S1_moments_setofequations} leads to an underdetermined system of equations, that is,
the conditioned moments $\avgc{t}_{a|i_0}$ and splitting probabilities $\phic_{a|i_0}$ cannot directly be obtained from unconditioned first passage moments highlighted by the superscript ``single''. 
Below we show how one can nevertheless deduce the splitting probabilities and conditioned moments of first passage time.
Once the splitting probabilities are known, the $k$-th moment of the exit time, $\avgexit{t^k}_{i_0}=\sum_a\phi_{a|i_0}\avgc{t^k}_{a|i_0}$, can, somewhat surprisingly,  be determined
from  Eq.~\eqref{eq:S1_moments_setofequations} as
\begin{equation}\addtocounter{equation}{1}\tag{\theequation}
\label{eq:S1_exit_moments}
\begin{aligned}
\avgexit{t}_{i_0}&=\avgs{t}_{a|i_0}-\sum_{\hidewidth a'\in\mathcal{A}\backslash\{a\}\hidewidth}\phic_{a'|i_0}\avgs{t}_{a|a'},\\
  \avgexit{t^2}_{i_0}&=\avgs{t^2}_{a|i_0}-2\sum_{\hidewidth a'\in\mathcal{A}\backslash\{a\}\hidewidth}\phic_{a'|i_0}\avgs{t}_{a|a'}\avgc{t}_{a|i_0}
  \\&
  \qquad\qquad-\sum_{a'\in\mathcal{A}\backslash\{a\}}\phic_{a'|i_0}\avgs{t^2}_{a|a'},
  \\
&\;\,\vdots\\
\avgexit{t^k}_{i_0}&=\avgs{t^k}_{a|i_0}\\
&\phantom{{}={}}
-\!\!\!\sum_{ a'\in\mathcal{A}\backslash\{a\}}\sum_{l=1}^{k}\binom{k}{l}\phic_{a'|i_0}\avgs{t^l}_{a|a'}\avgc{t^{k-l}}_{a'|i_0},\hidewidth
\end{aligned}
\end{equation}
which holds for any $a\in \mathcal{A}$.
It is remarkable that we are able to obtain  the $k$-th moment of the exit time from ``just'' the first $k$ unconditioned
moments of first passage time and the first $k-1$ conditional moments,
that is, we apparently get an additional moment of exit time ``for free''.

To access the splitting probability as well as conditioned moments by means of the generalized renewal theorem
[Eq. (C3) in Appendix C in the main text]
we need an alternative strategy since Eq.~\eqref{eq:S1_moments_setofequations}
is an underdetermined system of equations. In other words, we need to invert a singular equation. First, we define the subset of target states $\mathcal{A}=\{a_1,a_2\ldots,a_n\}\subset\{1,2,\ldots,N\}$ with $n$ elements ($n<N$).
Second, we rewrite the renewal theorem
[Eq. (C3) in Appendix C in the main text]
in form of a matrix product ($i,j=1,\ldots,n$)
\begin{equation}
 \by(s)=\mM(s)\bx(s),\;\;\text{with}\;\; M_{ij}(s)=\begin{cases}\twps_{a_i|a_{j}}(s)&\text{if $i\neq j$,}\\1&\text{else},\end{cases}
 \label{eq:S1_renewal_matrices}
\end{equation}
and $y_i(s)=\twps_{a_i|i_0}(s)$, $x_{j}(s)=\twpc_{a_j|i_0}(s)$, where $\bx(s)$ encodes the conditioned moments of first passage times according to Eq.~\eqref{eq:S1_moments_def}; Eq.~\eqref{eq:S1_renewal_matrices} is explicitly shown in Fig.~\ref{fig:S1_renewal_matrices}.
\begin{figure*}
 \centering
 \begin{equation*}
 \overbrace{
 \begin{pmatrix}
  \twps_{a_1|i_0}(s)\\
  \twps_{a_2|i_0}(s)\\
  \vdots\\
  \twps_{a_{n-1}|i_0}(s)\\
  \twps_{a_n|i_0}(s)
 \end{pmatrix}}^{\displaystyle\by(s)}=
  \overbrace{
  \begin{pmatrix}
  1&\twps_{a_1|a_2}(s)&\hdots&\twps_{a_{1}|a_{n-1}}(s)&\twps_{a_{1}|a_{n}}(s)\\
  \twps_{a_2|a_1}(s)&1&\hdots&\twps_{a_{2}|a_{n-1}}(s)&\twps_{a_{2}|a_{n}}(s)\\
  \vdots&\vdots&\ddots&\vdots&\vdots\\
  \twps_{a_{n-1}|a_1}(s)&\twps_{a_{n-1}|a_{2}}(s)&\hdots&1&\twps_{a_{n-1}|a_{n}}(s)\\
  \twps_{a_{n}|a_1}(s)&\twps_{a_{n}|a_{2}}(s)&\hdots&\twps_{a_{n}|a_{n-1}}(s)&1
 \end{pmatrix}}^{\displaystyle\mM(s)}
  \overbrace{
  \begin{pmatrix}
  \twpc_{a_1|i_0}(s)\\
  \twpc_{a_2|i_0}(s)\\
  \vdots\\
  \twpc_{a_{n-1}|i_0}(s)\\
  \twpc_{a_n|i_0}(s)
 \end{pmatrix}
 }^{\displaystyle\bx(s)}
 \end{equation*}
 \caption{\normalfont Explicitly written equation \eqref{eq:S1_renewal_matrices} with $\mathcal{A}=\{a_1,a_2,\ldots,a_n\}\subset\{1,\ldots,N\}=\Omega$.}
 \label{fig:S1_renewal_matrices}
\end{figure*}
\setcounter{figure}{0}
\begin{figure*}
 \begin{equation*}
 \mM_\gamma(s)=
  \begin{pmatrix}
  1&\twps_{a_1|a_2}(s)&\hdots&\twps_{a_{1}|a_{\gamma-1}}(s)&\textcolor{blue!70!black!70}{\smash{\overbrace{\twps_{a_1|i_0}(s)}^{\displaystyle \by(s)}}}&\twps_{a_{1}|a_{\gamma+1}}(s)&\hdots&\twps_{a_{1}|a_{n-1}}(s)&\twps_{a_{1}|a_{n}}(s)\\
  \twps_{a_2|a_1}(s)&1&\hdots&\twps_{a_{2}|a_{\gamma-1}}(s)&\textcolor{blue!70!black!70}{\twps_{a_{2}|i_0}(s)}&\twps_{a_{2}|a_{\gamma+1}}(s)&\hdots&\twps_{a_{2}|a_{n-1}}(s)&\twps_{a_{2}|a_{n}}(s)\\[2mm]
  \vdots&\vdots&\ddots&\vdots&\vdots&\vdots&&\vdots&\vdots
  \\[2mm]
   \twps_{a_{\gamma-1}|a_{1}}(s)&\twps_{a_{\gamma-1}|a_{2}}(s)&&1&\textcolor{blue!70!black!70}{\twps_{a_{\gamma-1}|i_0}(s)}&\twps_{a_{\gamma-1}|a_{\gamma+1}}(s)&&\twps_{a_{\gamma-1}|a_{n-1}}(s)&\twps_{a_{\gamma-1}|a_{n}}(s)
   \\
   \twps_{a_{\gamma}|a_{1}}(s)&\twps_{a_{\gamma}|a_{2}}(s)&&\twps_{a_{\gamma}|a_{\gamma-1}}(s)&\textcolor{blue!70!black!70}{\twps_{a_{\gamma}|i_0}(s)}&\twps_{a_{\gamma}|a_{\gamma+1}}(s)&&\twps_{a_{\gamma}|a_{n-1}}(s)&\twps_{a_{\gamma}|a_{n}}(s)
   \\
   \twps_{a_{\gamma+1}|a_{1}}(s)&\twps_{a_{\gamma+1}|a_{2}}(s)&&\twps_{a_{\gamma+1}|a_{\gamma-1}}(s)&\textcolor{blue!70!black!70}{\twps_{a_{\gamma+1}|i_0}(s)}&1&&\twps_{a_{\gamma+1}|a_{n-1}}(s)&\twps_{a_{\gamma+1}|a_{n}}(s)
   \\[2mm]
    \vdots&\vdots&&\vdots&\vdots&\vdots&\ddots&\vdots&\vdots\\[2mm]
   \twps_{a_{n-1}|a_1}(s)&\twps_{a_{n-1}|a_2}(s)&&\twps_{a_{n-1}|a_{\gamma-1}}(s)&\textcolor{blue!70!black!70}{\twps_{a_{n-1}|i_0}(s)}&\twps_{a_{n-1}|a_{\gamma+1}}(s)&\hdots&1&\twps_{a_{n-1}|a_{n}}(s)\\
   \twps_{a_{n}|a_1}(s)&\twps_{a_{n}|a_2}(s)&\hdots&\twps_{a_{n}|a_{\gamma-1}}(s)&\textcolor{blue!70!black!70}{\twps_{a_{n}|i_0}(s)}&\twps_{a_{n}|a_{\gamma+1}}(s)&\hdots&\twps_{a_{n}|a_{n-1}}(s)&1
 \end{pmatrix}
 \end{equation*}
\caption{\normalfont Matrix $\mM_\gamma(s)$; the colored symbols represent the vector $\by(s)$ from Fig.~\ref{fig:S1_renewal_matrices} and the remaining elements are equal to the elements of $\mM(s)$.}
\label{fig:S1:CramersRule}
\end{figure*}

According to Cramer's rule the $\gamma$th component of $\bx(s)$ is given by
\begin{multline}
 x_\gamma(s)=\frac{\det \mM_\gamma(s)}{\det \mM(s)},\\\text{with}\quad
[\mM_\gamma(s)]_{ij}=
\begin{cases}
\twps_{a_i|a_j}(s)&\text{if $i\neq j$ and $j\neq \gamma$,}\\
1&\text{if $i= j$ and $j\neq \gamma$,}\\
\twps_{a_i|i_0}(s)&\text{if $j=\gamma$},
\end{cases}
 \label{eq:S1:CramersRule}
\end{multline}
that is, we replace the $\gamma$th column of $\mM$ by $\by(s)$ to obtain $\mM_\gamma(s)$. Eq.~\eqref{eq:S1:CramersRule} solves
 Eq.~\eqref{eq:S1_renewal_matrices}. The matrix $\mM_\gamma(s)$ is illustrated in Fig.~\ref{fig:S1:CramersRule}.
Since the
matrices $\mM(s)$ and $\mM_\gamma(s)$
both have all entries equal to 1 for $s=0$,
 the limit $s\to0$ in Eq.~\eqref{eq:S1:CramersRule} seems to be undetermined (i.e., yields ``zero divided by zero''). To avoid a division by zero we first identify
the dyadic product $\mM_\gamma(0)=\mM(0)=\bnu\bnu^\T$ with $\bnu^\T\equiv(1,\ldots,1)$
and then use the matrix determinant lemma to remove the singularity
\begin{align}
  x_\gamma(s)&=\frac{\det \mM_\gamma(s)}{\det \mM(s)}=\frac{\det [\frac{\mM_\gamma(s)-\bnu\bnu^\T}{s}+\frac{\bnu\bnu^\T}{s}]}{\det [\frac{\mM(s)-\bnu\bnu^\T}{s}+\frac{\bnu\bnu^\T}{s}]}\nonumber\\
&  =
\bigg[ \frac{s+\bnu^\T\big[\frac{\mM_\gamma(s)-\bnu\bnu^\T}{s}\big]^{-1}\bnu}{s+\bnu^\T\big[\frac{\mM(s)-\bnu\bnu^\T}{s}\big]^{-1}\bnu}\bigg]
 \frac{\det\big[\frac{\mM_\gamma(s)-\bnu\bnu^\T}{s}\big]}{\det\big[\frac{\mM(s)-\bnu\bnu^\T}{s}\big]}
\label{eq:S1_Cramer_solved}
\end{align}
where in the first step we used Cramer's rule in Eq.~\eqref{eq:S1:CramersRule}, the second step involves a division by $s^n$ of both numerator and denominator as well as an addition of zero; in the third step we employed the matrix determinant lemma.
According to Eqs.~\eqref{eq:S1_moments_def}, \eqref{eq:S1_renewal_matrices} and \eqref{eq:S1:CramersRule} we have
\begin{multline}
\bigg[\frac{\mM(s)-\bnu\bnu^\T}{s}\bigg]_{ij}=\\
\begin{cases}
\left[
\begin{aligned}
  -\avgs{t}_{a_i|a_j}+\frac{1}{2}\avgs{t^2}_{a_i|a_j}s\\-\frac{1}{3!}\avgs{t^3}_{a_i|a_j}s^2\pm\ldots
\end{aligned}
\right]
&\text{if $i\neq j$}\\
0 &\text{if $i=j$,}
\label{eq:S1_aux_matix1}
\end{cases}
\end{multline}
and
\begin{multline}
\bigg[\frac{\mM_\gamma(s)-\bnu\bnu^\T}{s}\bigg]_{ij}
=\\
\begin{cases}
\left[
\begin{aligned}
 -\avgs{t}_{a_i|a_j}+\frac{1}{2}\avgs{t^2}_{a_i|a_j}s\\-\frac{1}{3!}\avgs{t^3}_{a_i|a_j}s^2\pm\ldots
\end{aligned}\right]
  &\text{if $i\neq j$ and $j\neq \gamma$},
\\
0 &\text{if $i= j$ and $j\neq \gamma$},\\
\left[
\begin{aligned}
 -\avgs{t}_{a_i|i_0}+\frac{1}{2}\avgs{t^2}_{a_i|i_0}s\\-\frac{1}{3!}\avgs{t^3}_{a_i|i_0}s^2\pm\ldots
\end{aligned}\right]
&
\text{if $j= \gamma$},
\end{cases}
\label{eq:S1_aux_matix2}
\end{multline}
where $i,j=1,\ldots,n$ and
$i_0\notin\mathcal{A}=\{a_1,\ldots,a_n\}$. Notably, the zeroth moment
of the conditioned first passage time -- the splitting probability --
requires the first moment of unconditioned first passage times. Inspecting Eq.~\eqref{eq:S1_Cramer_solved}
we generally find that the $k$th moment of the conditional first passage time can be expressed in terms of the first $k+1$ unconditioned moments, which indicates that conditional first passage problems are notoriously more difficult to solve.

\subsection{Renewal theorem on star-like graphs}\label{subsec:Renewal_inverse_star}

In this subsection we consider a star-like graph as in
Appendix~C.2 in the main text
(see Fig.~10 in the main text).
Defining the forward and backward unconditioned \emph{single} target moments,
$f_a^{(k)}\equiv\avgs{t^k}_{a|i_0}$ and $b_a^{(k)}\equiv\avgs{t^k}_{i_0|a}$
we obtain
\begin{equation}
 \begin{aligned}
  u_a(s)&=\twps_{a|i_0}(s)=1-f_a^{(k)}s+\frac{f_a^{(2)}}{2}s^2-\frac{f_a^{(3)}}{6}s^3\pm\ldots,\\
   v_a(s)&=\twps_{i_0|a}(s)=1-b_a^{(k)}s+\frac{b_a^{(2)}}{2}s^2-\frac{b_a^{(3)}}{6}s^3\pm\ldots,
 \end{aligned}
\end{equation}
where we omit here and in the following subsection the index $i_0$.
Using the definitions
\begin{equation}
\begin{aligned}
 \gamma_a^{(1)}&=f_a^{(1)}+b_a^{(1)},\\
 \gamma_a^{(2)}&=f_a^{(2)}+b_a^{(2)}-2[f_a^{(1)}]^2,
 \label{eq:S1_aux}
\end{aligned}
\end{equation}
the numerator of the right hand side of
 Eq.~(C5) in Appendix C in the main text
satisfies
\begin{align}
 \frac{su_a(s)}{1-u_a(s)v_a(s)}
 &=\mathcal{A}^{(0)}_a+\mathcal{A}^{(1)}_as+\mathcal{A}^{(2)}_as^2+\mathrm{O}(s)^3 
 \nonumber\\
 &=\frac{1}{\gamma_a^{(1)}}+\frac{\gamma_a^{(2)}}{2[\gamma^{(1)}_a]^2}s+\mathcal{A}^{(2)}_as^2+\mathrm{O}(s)^3
 \label{eq:S1_A_aux}
\end{align}
and consequently using the product rule of differentiation the denominator
satisfies
\begin{align}
s+ \sum_{a=1}^n\frac{su_a(s)v_a(s)}{1-u_a(s)v_a(s)}&=
 \mathcal{B}^{(0)}+\mathcal{B}^{(1)}s+\mathcal{B}^{(2)}s^2+\mathrm{O}(s)^3,
\end{align}
where
\begin{equation}\addtocounter{equation}{1}\tag{\theequation}
\label{eq:S1_B_aux}
\begin{aligned}
  \mathcal{B}^{(0)}&=\sum_{j=1}^n\mathcal{A}^{(0)}_j
  =\sum_{j=1}^n\frac{1}{\gamma^{(1)}_j}
  \\
  \mathcal{B}^{(1)}&=1+\sum_{j=1}^n\Big[\mathcal{A}^{(1)}_j-\mathcal{A}^{(0)}_jb ^{(1)}_j\Big]\\
  &=1+\sum_{j=1}^n\Big[\frac{\gamma_j^{(2)}}{2[\gamma^{(1)}_j]^2}-\frac{b ^{(1)}_j}{\gamma^{(1)}_j}\Big]\\
  \mathcal{B}^{(2)}&=\sum_{j=1}^n
 \Big[\mathcal{A}^{(2)}_j-\mathcal{A}^{(1)}_jb ^{(1)}_j+\frac{1}{2}\mathcal{A}^{(0)}_jb ^{(2)}_j\Big]
 \\
& =
 \sum_{j=1}^n
 \Big[\mathcal{A}^{(2)}_j-\frac{\gamma_j^{(2)}b ^{(1)}_j}{2[\gamma^{(1)}_j]^2}+\frac{b ^{(2)}_j}{2\gamma^{(1)}_j}\Big].
\end{aligned}
\end{equation}
Since the Laplace transform of the local first passage time density
is given by
\begin{align}
 \twpcStar_{a|i_0}&=\frac{\mathcal{A}_a^{(0)}+\mathcal{A}_a^{(1)}s+\mathcal{A}_a^{(2)}s^2+\ldots}{\mathcal{B}^{(0)}+\mathcal{B}^{(1)}s+\mathcal{B}^{(2)}s^2+\ldots}
 \nonumber\\
 &=
 \phiStar_{a|i_0}\Big[1- \avgcStar{t}_{a|i_0}s+\frac12\avgcStar{t^2}_{a|i_0}s^2\pm\ldots\Big],
 \label{eq:S1_pcond_Taylor}
\end{align}
we directly obtain the splitting probability in the limit $s\to0$
yielding
\begin{equation}
 \phiStar_{a|i_0}=\frac{\mathcal{A}^{(0)}_a}{\mathcal{B}^{(0)}}=\frac{\frac{1}{\gamma_a^{(1)}}}{\sum_j\frac{1}{\gamma_j^{(1)}}},
 \label{eq:S1_splitting_from_unc}
\end{equation}
where in the last step we inserted the coefficients $\mathcal{A}^{(0)}_a$ and
$\mathcal{B}^{(0)}$ from equations \eqref{eq:S1_A_aux} and \eqref{eq:S1_B_aux}, respectively.
The first derivative of \eqref{eq:S1_pcond_Taylor} at $s=0$
yields the conditional mean first passage times
\begin{multline}
 \avgcStar{t}_{a|i_0}=\\\frac{\mathcal{B}^{(1)}}{\mathcal{B}^{(0)}}-\frac{\mathcal{A}^{(1)}_a}{\mathcal{A}^{(0)}_a}
 =\frac{1+\sum_{j=1}^n\Big[\frac{\gamma_j^{(2)}}{2[\gamma^{(1)}_j]^2}-\frac{b ^{(1)}_j}{\gamma^{(1)}_j}\Big]}{\sum_{i=1}^n\frac{1}{\gamma_j^{(1)}}}
 -\frac{\gamma_a^{(2)}}{2\gamma^{(1)}_a}
\\
 =
 \frac{1-\sum_{j=1}^n\frac{b ^{(1)}_j}{\gamma^{(1)}_j}}{\sum_{j=1}^n\frac{1}{\gamma_j^{(1)}}}
 +\frac{1}{2}\sum_{j=1}^n\phiStar_{j|i_0}\bigg[\frac{\gamma_j^{(2)}}{\gamma^{(1)}_j}-\frac{\gamma_a^{(2)}}{\gamma^{(1)}_a}\bigg],
\label{eq:S1_1st_moment_from_unc}
 \end{multline} 
where we inserted $\mathcal{A}^{(0)}_a$ and
$\mathcal{B}^{(0)}$ from Eqs.~\eqref{eq:S1_A_aux} and \eqref{eq:S1_B_aux}  in the first step, and finally
identified the splitting probability $\phiStar_{j|i_0}$ from Eq.~\eqref{eq:S1_splitting_from_unc} and used $\sum_j\phiStar_{j|i_0}=1$.
Using Eqs.~\eqref{eq:S1_splitting_from_unc} and  \eqref{eq:S1_1st_moment_from_unc}
the mean exit time from node $i$ becomes
\begin{align}
 \avgexit{t}_{i_0}&= \sum_{j=1}^n\phiStar_{j|i_0} \avgcStar{t}_{a|i_0}=
 \frac{\mathcal{B}^{(1)}-\sum_{j=1}^n\mathcal{A}_j^{(1)}}{\mathcal{B}^{(0)}}
 \nonumber\\
 &=
 \frac{1-\sum_{j=1}^n\frac{b ^{(1)}_j}{\gamma^{(1)}_j}}{\sum_{j=1}^n\frac{1}{\gamma_j^{(1)}}},
 \label{eq:S1_exit1_from_unc}
\end{align}
which is the first term in the result of Eq.~\eqref{eq:S1_1st_moment_from_unc}.
The second moment of the conditional first passage time
after differentiating Eq.~\eqref{eq:S1_pcond_Taylor} twice at $s=0$
yields
\begin{equation}
  \avgcStar{t^2}_{a|i_0}=
  2\frac{\mathcal{A}^{(2)}_a}{\mathcal{A}^{(0)}_a}-2\frac{\mathcal{B}^{(2)}}{\mathcal{B}^{(0)}}+2\frac{\mathcal{B}^{(1)}}{\mathcal{B}^{(0)}}\overbrace{\bigg[\frac{\mathcal{B}^{(1)}}{\mathcal{B}^{(0)}}-\frac{\mathcal{A}^{(1)}_a}{\mathcal{A}^{(0)}_a}\bigg]}^{=\avgcStar{t}_{a|i_0}}
  \label{eq:S1_2nd_moment_from_unc}
\end{equation}
and hence
\begin{widetext}
 \begin{align}
  \avgexit{t^2}_{i_0}&=
  \sum_{j=1}^n\phiStar_{j|i_0} \avgcStar{t^2}_{j|i_0}=
   2\sum_{j=1}^n\frac{\mathcal{A}^{(2)}_j}{\mathcal{B}^{(0)}}-2\frac{\mathcal{B}^{(2)}}{\mathcal{B}^{(0)}}+
  2\frac{\mathcal{B}^{(1)}}{\mathcal{B}^{(0)}}\avgexit{t}_{i_0}
  =\frac{ \sum_{j=1}^n
 \Big[\frac{\gamma_j^{(2)}b ^{(1)}_j}{[\gamma^{(1)}_j]^2}-\frac{b ^{(2)}_j}{\gamma^{(1)}_j}\Big]+2\mathcal{B}^{(1)}\avgexit{t}_{i_0}}{\mathcal{B}^{(0)}}
 \nonumber\\
 &=\frac{ \sum_{j=1}^n
 \Big[\frac{\gamma_j^{(2)}b ^{(1)}_j}{[\gamma^{(1)}_j]^2}-\frac{b ^{(2)}_j}{\gamma^{(1)}_j}\Big]+2\Big[1+\sum_{j=1}^n\Big(\frac{\gamma_j^{(2)}}{2[\gamma^{(1)}_j]^2}-\frac{b ^{(1)}_j}{\gamma^{(1)}_j}\Big)\Big]
 \dfrac{1-\sum_{j=1}^n\frac{b ^{(1)}_j}{\gamma^{(1)}_j}}{\sum_{j=1}^n\frac{1}{\gamma_j^{(1)}}}
 }{\sum_{j=1}^n\frac{1}{\gamma_j^{(1)}}}
  \label{eq:S1_exit2_from_unc}
\end{align}
\end{widetext}
where in the second last step of the first line we used
$\sum_j\phiStar_{j|i_0}=1$ and Eqs.~\eqref{eq:S1_splitting_from_unc}, \eqref{eq:S1_1st_moment_from_unc}, and \eqref{eq:S1_exit1_from_unc}; in the last step of the first line we used the third line of Eq.~\eqref{eq:S1_B_aux}, and the last step in the second line we inserted the first two lines of Eq.~\eqref{eq:S1_B_aux} as well as Eq.~\eqref{eq:S1_exit1_from_unc}.

\section{Unconditioned moments from backward Fokker-Planck equation on star-like graphs and moments of transition-path time along a single leg}\label{sec:S_unconditioned_moments}
Here we focus on star-like graphs spanned by the $i$-th node such that each state  is taken from the set of neighboring states $\mathcal{N}_i$. We determine unconditioned moments of the first passage time that are used in \ref{subsec:Renewal_inverse_star}.

\subsection{Backward Fokker-Planck equation}
In this subsection we determine the unconditioned first passage time to node $a\in\mathcal{N}_i$
starting from a point that lies between node $i$ and a connected neighboring node $j\in\mathcal{N}_i$.
Before determining the unconditioned moments of the first passage time it
proves convenient to translate the
forward Fokker-Planck equation
[Eq.~(A5) in Appendix~A in the main text]
into its adjoint, backward form that reads \cite{gard04}
\begin{multline}
 \ddel{t}\Ploc(x,a,t|y,j)
 \\=\e^{\beta U_{j|i}(y)}\ddel{y}\e^{-\beta U_{j|i}(y)}D_{j|i}(y)\ddel{y}\Ploc(x,a,t|y,j)\\
 \equiv\LB_{j|i}(y)\Ploc(x,a|y,j),
 \label{eq:S3_FPE_backward}
\end{multline}
and the boundary conditions at the $i$-th inner node from
[Eq.~(A5) in Appendix~A in the main text]
 become
\begin{equation}
\begin{aligned}
  &\Ploc(x,a,t|0,j)= \Ploc(x,a',t|0,k)\quad\forall a'\in\mathcal{N}_i,\\
&0= \sum_{a\in\mathcal{N}_i}D_{a|i}(y)\ddel{y} \Ploc(x,a,t|y,k)|_{y=0}.
\end{aligned}
 \label{eq:S3_backward_boundary}
\end{equation}
We first consider the unconditioned first passage problem to state $a$ by setting $\Ploc(l_{a|i},a,t|y,j)=\Ploc(x,j,t|l_{a|i},a)=0$ ($a$th node absorbing) whereas the remaining links  are made reflecting, i.e., $\del_y\Ploc(x,k,t|y,j)|_{y=l_{j|i}}=0$
for all $k\in\mathcal{N}_i$ and $j\in\mathcal{N}_i\backslash\{a\}$.

Note that the forward and backward Fokker-Planck equations both satisfy $\ddel{t}\Ploc(x,a,t|y,j)=\LB_{j|i}(y) \Ploc(x,a|y,j)=\LF_{a|i}(x) \Ploc(x,a|y,j)$.
The backward Fokker-Planck equation allows us to conveniently determine the moments of the unconditioned first passage time
using standard methods (see also Refs.~\cite{redn01,gard04} or Ref.~\cite{beni09} for a discussion on graphs).
We denote the survival probability by
\begin{equation}
 S_{a|i}(t|y,j)\equiv \sum_k\int_0^{l_{k|i}}\dd x\Ploc(x,k,t|y,j),
\end{equation}
which decays monotonically in time, $-\del _t S_{a|i}(t|y,j)\ge0$, due to the single absorbing end at $a$ and $x=l_{a|i}$.
From the backward Fokker-Planck equation \eqref{eq:S3_FPE_backward} 
follows the evolution equation for the survival probability
\begin{equation}
 \ddel{t} S_{a|i}(t|y,j) =\LB_{j|i}(y)S_{a|i}(t|y,j),
\end{equation}
with $S_{a|i}(t|l_{a|i},a)=0$ for all $t\ge0$ and $\del_y S_{a|i}(t|y,j)|_{y=l_j}=0$ if $j\neq a$
as well as $S_{a|i}(0|y,j)=1$ for all $j$ and $y<l_{j|i}$.
Since the first passage time density is the negative derivative of the
survival probability, $-\del_t S_a(t|y,j)$,
we can write the unconditioned $k$th moment as
\begin{align}
 T_{a|i}^{(k)}(y,j)&\equiv-\int_0^\infty t^k\del_tS_a(t|y,j)\dd t\nonumber\\
 &=k\int_0^\infty t^{k-1}S_a(t|y,j)\dd t,
 \label{eq:S3_moment_def}
\end{align}
where the last identity follows from partial differentiation.
Operating from the left by the backward operator $\LB_{j|i}(y)$ in Eq.~\eqref{eq:S3_FPE_backward}
yields the hierarchical connection between the moments
 \cite{gard04}
\begin{multline}
\LB_{j|i}(y)T_{a|i}^{(k)}(y,j)\\=e^{\beta U_{j|i}(y)}\ddel{y}\e^{-\beta U_{j|i}(y)}D_{j|i}(y)\ddel{y}T_{a|i}^{(k)}(y,j)\\=-kT_{a|i}^{(k-1)}(y,j),
\label{eq:S3_ODE_moments}
\end{multline}
with $T_{a|i}^{(0)}(y,j)=1$.
Henceforth we will use the short-hand notation $\ddel{y}T_{a|i}^{(k)}(y,j)\equiv T_{a|i}^{(k)}{}'(y,j)$
to write the boundary conditions as
\begin{equation}
\begin{aligned}
  T_{a|i}^{(k)}(0,a)&=T_{a|i}^{(k)}(0,j), & \sum_{\alpha\in\mathcal{N}_i} T_{\alpha|i}^{(k)}{}'(0,i)&=0,\\
T_{a|i}^{(k)}(l_a,a)&=0,& T_{a|i}^{(k)}{}'(l_{j|i},j)&=0,
\end{aligned}
 \label{eq:S3_boundary}
\end{equation}
for all $k>0$ and $j\neq a$.
The unconditioned moments are related to the renewal theorem
from \ref{subsec:Renewal_inverse_star}
via
\begin{equation}
T_{a|i}^{(k)}(0,a)=f_a^{(k)} ,\quad T_{a|i}^{(k)}(l_{j|i},j)=
\sum_{l=0}^k\frac{k!}{l!(k-l)!}f_a^{(l)}b_j^{(k-l)}.
\label{eq:S5_uncond_moments_link}
\end{equation}
For example, $T_{a|i}^{(1)}(l_{j|i},j)=f_a^{(1)}+b_j^{(1)}$ and $T_{a|i}^{(2)}(l_{j|i},j)=f_a^{(2)}+2f_a^{(1)}b_j^{(1)}+b_j^{(2)}$,
which follows from the decomposition of trajectories on star-like graphs as explicitly depicted in
Fig.~10c in the main text.
 Note that
$a'$ in
Fig.~10c in the main text
 here plays the role of  $j$    ($j\neq a$) and the inner node $i_0$ here becomes $i$.

\subsection{Hierarchy of moments}
In the following we translate Eq.~\eqref{eq:S3_ODE_moments} into a hierarchical integration formula
that allows us to deduce $T_{a|i}^{(k)}(y,j)$ from
$T_{a|i}^{(k-1)}(y,j)$ with $T_{a|i}^{(0)}(y,j)=1$.

First, for $j=a$ we use  Eq.~\eqref{eq:S3_ODE_moments}  
and set therein
$y=y_{2k-1}$ such that the integral $\int_0^y\dd y_{2k-1}(\cdots)$ over both sides yields
\begin{align}
 &\e^{-\beta U_{a|i}(y)}D_{a|i}(y)T_{a|i}^{(k)}{}'(y,a)- \e^{-\beta U_{a|i}(0)}D_{a|i}(0)T_{a|i}^{(k)}{}'(0,a)
 \nonumber\\
 &=-k\int_0^y\dd y_{2k-1}\e^{-\beta U_{a|i}(y_{2k-1})}T_{a|i}^{(k-1)}(y_{2k-1},a)
\end{align}
which in turn allows us to obtain
\begin{multline}
 T_{a|i}^{(k)}{}'(y,a)=\e^{\beta U_{a|i}(y)-\beta U_{a|i}(0)}\frac{D_{a|i}(0)}{D_{a|i}(y)} T_{a|i}^{(k)}{}'(0,a)\\-k\int_0^y\dd y_{2k-1}\frac{\e^{\beta U_{a|i}(y)-\beta U_{a|i}(y_{2k-1})}}{D_{a|i}(y)}T_{a|i}^{(k-1)}(y_{2k-1},a).
  \label{eq:S3_aux1}
\end{multline}
Similarly, for $j\neq a$
we get
\begin{multline}
  T_{a|i}^{(k)}{}'(y,j)=\\k\int_y^{l_{j|i}}\dd y_{2k-1}\frac{\e^{\beta U_{j|i}(y)-\beta U_{j|i}(y_{2k-1})}}{D_{j|i}(y)}T_{a|i}^{(k-1)}(y_{2k-1},j),
  \label{eq:S3_aux2}
\end{multline}
where we used $T_{a|i}^{(k)}{}'(l_{j|i},j)=0$ from Eq.~\eqref{eq:S3_boundary}.
According to the second condition from Eq.~\eqref{eq:S3_boundary} the continuity condition at the central node $i$ reads $T_{a|i}^{(k)}{}'(0,a)=-\sum_{j\neq a}T_{a|i}^{(k)}{}'(0,j)$, and $T_{a|i}^{(k)}{}'(0,j)$, which we insert alongside of Eq.~\eqref{eq:S3_aux2} into Eq.~\eqref{eq:S3_aux1}
to get
\begin{align}
& T_{a|i}^{(k)}{}'(y,a)=
 \nonumber\\
 &-k\int_0^y\dd y_{2k-1}\frac{\e^{\beta U_{a|i}(y)-\beta U_{a|i}(y_{2k-1})}}{D_{a|i}(y)}T_{a|i}^{(k-1)}{}(y_{2k-1},a)
 \nonumber\\
 &-k\sum_{j\neq a}\int_0^{l_{j|i}}\!\!\!\dd y_{2k-1}\frac{\e^{\beta U_{a|i}(y)-\beta U_{j|i}(y_{2k-1})}}{D_{a|i}(y)}T_{a|i}^{(k-1)}{}(y_{2k-1},j),
\end{align}
in turn leading to the first set of moments
\begin{widetext}
\begin{align}
  T_{a|i}^{(k)}(y,a)&=k\int_y^{l_{a|i}}\dd y_{2k}\int_0^{y_{2k}}\dd y_{2k-1}\frac{\e^{\beta U_{a|i}(y_{2k})-\beta U_{a|i}(y_{2k-1})}}{D_{a|i}(y_{2k})}T_{a|i}^{(k-1)}(y_{2k-1},a)\nonumber\\
&\quad  +
  k\sum_{j\neq a}\int_y^{l_{a|i}}\dd y_{2k}\int_0^{l_{j|i}}\dd y_{2k-1}\frac{\e^{\beta U_{a|i}(y_{2k})-\beta U_{j|i}(y_{2k-1})}}{D_{a|i}(y_{2k})}T_{a|i}^{(k-1)}(y_{2k-1},j).
  \label{eq:S3_hierarchy1}
\end{align}
The second set  of moments with $j\neq a$  follows from Eq.~\eqref{eq:S3_aux2} 
\begin{equation}
  T_{a|i}^{(k)}(y,j)= T_{a|i}^{(k)}(0,a)+k\int_0^{y}\dd y_{2k}\int_y^{l_{j|i}}\dd y_{2k-1}\frac{\e^{\beta U_{j|i}(y_{2k})-\beta U_{j|i}(y_{2k-1})}}{D_{j|i}(y_{2k})}T_{k-1,a}(y_{2k-1},j),
   \label{eq:S3_hierarchy2}
\end{equation}
\end{widetext}
where we used the first condition of Eq.~\eqref{eq:S3_boundary}, which reads $ T_{a|i}^{(k)}(0,a)= T_{a|i}^{(k)}(0,j)$. We note that Eqs.~\eqref{eq:S3_hierarchy1} and \eqref{eq:S3_hierarchy2} determine \emph{all} unconditioned moments of the first passage time for diffusion graphs with potentials.
In the following subsection we additionally provide explicit results for first two moments, i.e., for $k=1$ and $k=2$.

\subsection{Explicit unconditioned moments of the first passage time}
To render the rather convoluted calculation more tractable
we now simplify the notation by
removing within this subsection the subscript $i$
referring to the instantaneous, tagged node $i$. More precisely, in this subsection
we use
the shorthand notation $l_j\equiv l_{j|i}$,  $U_j\equiv U_{j|i}$, $D_j\equiv D_{j|i}$ and $\sum_{j\neq a}\equiv\sum_{j\in\mathcal{N}_i\backslash\{a\}}$. 

We calculate the first two unconditioned moments defined in
Eq.~\eqref{eq:S5_uncond_moments_link} (and
\ref{subsec:Renewal_inverse_star}) using Eq.~\eqref{eq:S3_hierarchy1} and Eq.~\eqref{eq:S3_hierarchy2}.
We use Eqs.~\eqref{eq:S5_uncond_moments_link} and \eqref{eq:S3_hierarchy1} to obtain the first moment of the  unconditioned ``forward'' first passage time
\begin{multline}
  f_a^{(1)}=T_{a|i}^{(1)}(0,a)=\int_0^{l_{a}}\dd y_{2}\int_0^{y_{2}}\dd y_{1}\frac{\e^{\beta U_a(y_{2})-\beta U_{a}(y_{1})}}{D_a(y_{2})}\\+
  \sum_{j\neq a}\int_0^{l_a}\dd y_{2}\int_0^{l_j}\dd y_{1}\frac{\e^{\beta U_a(y_2)-\beta U_j(y_2)}}{D_a(y_2)}
   \label{eq:S3_f1}
\end{multline}
and  then also  use Eq.~\eqref{eq:S3_hierarchy2} to derive the first moment of the  unconditioned ``backward'' first passage time
\begin{align}
 b_j^{(1)}&=T_{a|i}^{(1)}(l_j,j)-T_{a|i}^{(1)}(0,a)
 \nonumber\\
 &=\int_0^{l_j}\dd y_{2}\int_y^{l_j}\dd y_{1}\frac{\e^{\beta U_j(y_{2})-\beta U_j(y_{1})}}{D_j(y_2)},
 \label{eq:S3_b1}
\end{align}
see also Ref.~\cite{gard04} for an explicit solution of the latter result.
Similarly, using Eqs.~\eqref{eq:S3_hierarchy1} and \eqref{eq:S3_hierarchy2} the second moments
read
\begin{widetext}
\begin{multline}
f_a^{(2)}= T_{a|i}^{(2)}(0,a)
 =
2 \int_{0}^{l_a}\dd y_4\int_{0}^{y_4}\dd y_3 
 \int_{y_3}^{l_a}\dd y_2\int_{0}^{y_2}\dd y_1\frac{\e^{U_a(y_4)-U_a(y_3)+U_a(y_2)-U_a(y_1)}}{D_a(y_4)D_a(y_2)}
 \\
+2\sum_{j\neq a} \int_{0}^{l_a}\dd y_4\int_{0}^{y_4}\dd y_3 \int_{y_3}^{l_a}\dd y_2\int_0^{l_j}\dd y_1\frac{\e^{U_a(y_4)-U_a(y_3)+U_a(y_2)-U_{j}(y_1)}}{D_a(y_4)D_a(y_2)}
+2T_{a|i}^{(1)}(0,a)\sum_{j\neq a}\int_{0}^{l_a}\dd y_4\int_{0}^{l_{j}}\dd y_3\frac{\e^{U_a(y_4)-U_{j}(y_3)}}{D_a(y_4)}
\\
+
2\sum_{j\neq a}\int_{0}^{l_a}\dd y_4\int_{0}^{l_{j}}\dd y_3\int_{0}^{y_3}\dd y_2\int_{y_2}^{l_{j}}\dd y_1\frac{\e^{U_a(y_4)-U_{j}(y_3)+U_{j}(y_2)-U_{j}(y_1)}}{D_a(y_4)D_{j}(y_2)}
\label{eq:S3_f2}
\end{multline}
and
\begin{equation}
 b_j^{(2)}=2\int_0^{l_j}\dd y_4\int_{y_4}^{l_j}\dd y_3\int_0^{y_3}\dd y_2\int_{y_2}^{l_j}\dd y_1\frac{\e^{U_j(y_4)-U_j(y_3)+U_j(y_2)-U_j(y_1)}}{D_j(y_4)D_j(y_2)},
\label{eq:S3_b2}
\end{equation}
\end{widetext}
where we used $T_{2,a}(l_j,j)=f_a^{(2)}+2f_a^{(1)}b_j^{(1)}+ b_j^{(2)}$.
Note that each here ``backward'' first passage moment is effectively a ``simple'' 1-dimensional first passage problem.  One can, for example, use Ref.~\cite{gard04} for an alternative derivation of Eq.~\eqref{eq:S3_b2}.

Note that \emph{any} of the aforementioned integrals can be expressed in terms of the following auxiliary and elementary integrals
\begin{equation}
 \begin{aligned}
  I_{\alpha|i}^{(k)}&\equiv I_\alpha^{(k)}\equiv\int_0^{l_\alpha}\dd y_1\int_0^{y_1}\ldots \int_0^{y_{k-1}}\dd y_k g^{(k)}_{\alpha|i},\\
  R_{\alpha|i}^{(k)}&\equiv R_\alpha^{(k)}\equiv\int_0^{l_\alpha}\dd y_1\int_0^{y_1}\ldots \int_0^{y_{k-1}}\dd y_k h^{(k)}_{\alpha|i},
  \label{eq:S5_int_aux}
 \end{aligned}
\end{equation}
where auxiliary functions inside integrals 
are listed Tab.~\ref{tab:gh_def}. In what follows we merely used first six auxiliary integrals ($k=1,\ldots,6$).
%
%
\begin{table*}
 \caption{First six auxiliary functions at a glance. Since in this section we omitted for convenience the
 initial state $i$ in the subscript,  the general
 case is recovered by setting $g^{(k)}_{\alpha}=g^{(k)}_{\alpha|i}$ and $h^{(k)}_{\alpha}=h^{(k)}_{\alpha|i}$ along with $D_{\alpha}=D_{\alpha|i}$ and $U_{\alpha}=U_{\alpha|i}$, which is used in the main text and the remaining Appendices.}
 \label{tab:gh_def}
\begin{tabular}{r|c|c}\hline\hline
 $k$&$g^{(k)}_{\alpha|i}$&$h^{(k)}_{\alpha|i}$\\\hline &&\\[-4mm]
 $1$&$\dfrac{\e^{\beta U_{\alpha}(y_1)}}{D_\alpha(y_1)}$ & $\e^{-\beta U_{\alpha}(y_1)}$\\[2.5mm]
 $2$&$\dfrac{\e^{\beta U_{\alpha}(y_1)-\beta U_{\alpha}(y_2)}}{D_\alpha(y_1)}$&$\dfrac{\e^{-\beta U_{\alpha}(y_1)+\beta U_{\alpha}(y_2)}}{D_\alpha(y_2)}$ 
 \\[2.5mm]
 $3$&$\dfrac{\e^{\beta U_{\alpha}(y_1)-\beta U_{\alpha}(y_2)+\beta U_{\alpha}(y_3)}}{D_\alpha(y_1)D_\alpha(y_3)}$&$\dfrac{\e^{-\beta U_{\alpha}(y_1)+\beta U_{\alpha}(y_2)-\beta U_{\alpha}(y_3)}}{D_\alpha(y_2)}$
 \\[2.5mm]
 $4$&$\dfrac{\e^{\beta U_{\alpha}(y_1)-\beta U_{\alpha}(y_2)+\beta U_{\alpha}(y_3)-\beta U_{\alpha}(y_4)}}{D_\alpha(y_1)D_\alpha(y_3)}$&$\dfrac{\e^{-\beta U_{\alpha}(y_1)+\beta U_{\alpha}(y_2)-\beta U_{\alpha}(y_3)+\beta U_{\alpha}(y_4)}}{D_\alpha(y_2)D_\alpha(y_4)}$ 
 \\[2.5mm]
 $5$&$\dfrac{\e^{\beta U_{\alpha}(y_1)-\beta U_{\alpha}(y_2)+\beta U_{\alpha}(y_3)-\beta U_{\alpha}(y_4)+\beta U_{\alpha}(y_5)}}{D_\alpha(y_1)D_\alpha(y_3)D_\alpha(y_5)}$&$\dfrac{\e^{-\beta U_{\alpha}(y_1)+\beta U_{\alpha}(y_2)-\beta U_{\alpha}(y_3)+\beta U_{\alpha}(y_4)-\beta U_{\alpha}(y_5)}}{D_\alpha(y_2)D_\alpha(y_4)}$
  \\[2.5mm]
 $6$&$\dfrac{\e^{\beta U_{\alpha}(y_1)-\beta U_{\alpha}(y_2)+\beta U_{\alpha}(y_3)-\beta U_{\alpha}(y_4)+\beta U_{\alpha}(y_5)-\beta U_{\alpha}(y_6)}}{D_\alpha(y_1)D_\alpha(y_3)D_\alpha(y_5)}$&$\dfrac{\e^{-\beta U_{\alpha}(y_1)+\beta U_{\alpha}(y_2)-\beta U_{\alpha}(y_3)+\beta U_{\alpha}(y_4)-\beta U_{\alpha}(y_5)+\beta U_{\alpha}(y_6)}}{D_\alpha(y_2)D_\alpha(y_4)D_\alpha(y_6)}$\\
 $\vdots$&$\vdots$&$\vdots$\\\hline\hline
\end{tabular}
\end{table*}
Using these auxiliary integrals the first moments in Eqs.~\eqref{eq:S3_f1} and \eqref{eq:S3_b1}
along with the identity $\int_y^l=\int_0^l-\int_0^y$
yield
\begin{multline}
  f_a^{(1)}=I_a^{(2)}+\sum_{j\neq a}I_a^{(1)}R_j^{(1)}
 \quad\text{and}\quad b_a^{(1)}=I_a^{(1)}R_a^{(1)}-I_a^{(2)},\\
   \label{eq:S3_1st}
\end{multline}
which implies $f_a^{(1)}+b_a^{(1)}=I_a^{(1)}\sum_jR_j^{(1)}$.
Analogously, the second moments from Eqs.~\eqref{eq:S3_f2} and \eqref{eq:S3_b2}
become
\begin{align}
 \smash{\frac{f_a^{(2)}}{2}}={}&[I_a^{(2)}]^2-I_a^{(4)}+\big[I_a^{(2)}I_a^{(1)}-I_a^{(3)}\big]\sum_{j\neq a}R_j^{(1)}\nonumber\\[-1mm]
 &+f_a^{(1)}I_a^{(1)}\sum_{j\neq a}R_a^{(1)}
 +I_a^{(1)}\sum_{j\neq a}\big[R_j^{(2)}R_j^{(1)}-R_j^{(3)}\big],
 \nonumber\\[-1mm]
  \smash{\frac{b_a^{(2)}}{2}}={}&I_a^{(1)}\big[R_a^{(2)}R_a^{(1)}-R_a^{(3)}\big]+I_a^{(4)}-I_a^{(3)}R_a^{(1)}. 
  \label{eq:S3_2nd}
  \end{align}
For completeness we also list in Tab.~\ref{eq:S3_3rd} the third moments which allow for an alternative derivation of the main practical result, which is not pursued here.  We used them, however, to verify independently that the results in
Sec.~II.D in the main text
(see also \ref{subsec:second_moments})
 are correct. \vspace*{1mm}

\begin{table*}
\caption{Unconditioned third moments. \normalfont After some quite extended tedious but straight forward calculations we obtain the third moments. Third moments are listed for the sake of completeness. }
\label{eq:S3_3rd}
%
%
\begin{tabular}{c}
\hline\hline
\raisebox{-1.4mm}{
$\displaystyle
\begin{aligned}
 \frac{b_a^{(3)}}{6}
 =I_a^{(1)}[R_a^{(2)}]^2R_a^{(1)}-I_a^{(1)}R_a^{(2)}R_a^{(3)}-I_a^{(1)}R_a^{(4)}R_a^{(1)}+I_a^{(1)}R_a^{(5)}-I_a^{(3)}R_a^{(2)}R_a^{(1)}+I_a^{(3)}R_a^{(3)}+I_a^{(5)}R_a^{(1)}-I_a^{(6)}
\end{aligned}
$}
 \\[3.6mm]\hline
 \raisebox{-10.5mm}{
$\displaystyle
\begin{aligned}
  \frac{f_a^{(3)}}{6}&=I_a^{(6)}+[I_a^{(2)}]^3-2I_a^{(2)}I_a^{(4)}+\Big([I_a^{(2)}]^2I_a^{(1)}-I_a^{(2)}I_a^{(3)}+f_a^{(1)}[I_a^{(2)}I_a^{(1)}-I_a^{(3)}]+\frac{1}{2}f_a^{(2)}I_a^{(1)}
  +I_a^{(5)}-I_a^{(1)}I_a^{(4)}
  \Big)\smash{\sum_{j\neq a}R_j^{(1)}}\\
  &\quad+\Big(I_a^{(3)}-I_a^{(2)}I_a^{(1)}-f_a^{(1)}I_a^{(1)}
  \Big)\sum_{j\neq a}R_j^{(3)}
  +\Big(I_a^{(2)}I_a^{(1)}-I_a^{(3)}+f_a^{(1)}I_a^{(1)}\Big)\sum_{j\neq a}R_j^{(1)}R_j^{(2)}\\
  &\quad+I_a^{(1)}\sum_{j\neq a}\Big(R_j^{(5)}+[R_j^{(2)}]^2R_j^{(1)}-R_j^{(4)}R_j^{(1)}-R_j^{(2)}R_j^{(3)}\Big)
\end{aligned}
$}
\\\hline\hline
\end{tabular}

\end{table*}


\subsection{Transition-path time along one leg}\label{subsec:transition_results}
\vspace*{-1mm} The transition-path time was defined in 
Eq.~(6) in the main text.
 Here we determine the moments of transition-path time in two steps: we first consider the
definition of the transition-path time in 
Eq.~(6) in the main text
 at finite $y$ and only then take the limit $y\to0$.
To this end we now focus on the segment between a pair of nodes, the initial node $i$ and a target node $j$. Suppose that we start at a distance $y$ from node $i$ and ask for the $n$-th moment of the first passage time, $\delta T_{j|i}^{(n)}(y)$, after which the micro-state reaches for the first time the network state $j$ given that it did \emph{not} visit state $i$ before.
Using standard methods \cite{gard04} (see also explicitly Ref.~\cite{zhan07})
we obtain
\begin{widetext}
\begin{align}
\delta T_{j|i}^{(n)}(y)&=\frac{n}{I_{j|i}^{(1)}}\Bigg[\frac{\displaystyle\int_{y}^{l_{j|i}}\frac{\e^{\beta U_{j|i}(y')}}{D_{j|i}(y')}\dd y'}{\displaystyle\int_{0}^{y}\frac{\e^{\beta U_{j|i}(y')}}{D_{j|i}(y')}\dd y'}
\int_0^{y}\dd x'\e^{-\beta U_{j|i}(x')}\int_0^{x'}\dd x_1\frac{\e^{\beta U_{j|i}(x_1)}}{D_{j|i}(x_1)}\int_0^{x_1}\dd x_2 \frac{\e^{\beta U_{j|i}(x_2)}}{D_{j|i}(x_2)}\delta T_{j|i}^{(n-1)}(x')\nonumber\\[-1mm]
&\quad+
\int_{y}^{l_{j|i}}\dd x'\e^{-\beta U_{j|i}(x')}\int_0^{x'}\dd x_1\frac{\e^{\beta U_{j|i}(x_1)}}{D_{j|i}(x_1)}\int_{x'}^{l_{j|i}}\dd x_2 \frac{\e^{\beta U_{j|i}(x_2)}}{D_{j|i}(x_2)}\delta T_{j|i}^{(n-1)}(x')\Bigg]
\label{eq:S5_transition_hierarchy}
\end{align}
with $\delta T_{j|i}^{(0)}(x)\equiv 1$.
It is worth noting that Eq.~\eqref{eq:S5_transition_hierarchy}
for constant diffusion coefficient $D_{j|i}(x)=D$ is equivalent to Eq.~60 in Ref.~\cite{zhan07}, wherein one needs to replace $a\to 0$ and $b\to l_{j|i}$.
To make the calculations more efficient and to avoid redundant integrals we employ the shorthand notation from Eq.~\eqref{eq:S5_int_aux} along with the definitions
\begin{equation}
 I^{(k)}_{j|i}(y)\equiv I^{(k)}_{j|i}\big|_{l_{j|i}=y} \quad\text{and}\quad R^{(k)}_{j|i}(y)\equiv R^{(k)}_{j|i}\big|_{l_{j|i}=y},
\label{eq:S5_int_aux_x}
\end{equation}
such that $I^{(k)}_{j|i}=I^{(k)}_{j|i}(l_{j|i})$ and $R^{(k)}_{j|i}=R^{(k)}_{j|i}(l_{j|i})$.
After some algebra we obtain the first moment by means of the following calculation
\begin{align}
 \delta T_{j|i}^{(1)}(y)&=\frac{1}{I_{j|i}^{(1)}}\Bigg[\frac{\displaystyle\int_{y}^{l_{j|i}}\frac{\e^{\beta U_{j|i}(y')}}{D_{j|i}(y')}\dd y'}{\displaystyle\int_{0}^{y}\frac{\e^{\beta U_{j|i}(y')}}{D_{j|i}(y')}\dd y'}
\int_0^{y}\dd x'\e^{-\beta U_{j|i}(x')}\int_0^{x'}\dd x_1\frac{\e^{\beta U_{j|i}(x_1)}}{D_{j|i}(x_1)}\int_0^{x_1}\dd x_2 \frac{\e^{\beta U_{j|i}(x_2)}}{D_{j|i}(x_2)}
\nonumber\\
&\quad+
\int_{y}^{l_{j|i}}\dd x'\e^{-\beta U_{j|i}(x')}\int_0^{x'}\dd x_1\frac{\e^{\beta U_{j|i}(x_1)}}{D_{j|i}(x_1)}\int_{x'}^{l_{j|i}}\dd x_2 \frac{\e^{\beta U_{j|i}(x_2)}}{D_{j|i}(x_2)}
\Bigg]
\nonumber\\
&=
\frac{I^{(1)}_{j|i}-I^{(1)}_{j|i}(y)}{I^{(1)}_{j|i}I^{(1)}_{j|i}(y)}\bigg[R^{(2)}_{j|i}(y) I^{(1)}_{j|i}(y)-I^{(3)}_{j|i}(y)\bigg]+\frac{I^{(3)}_{j|i}-I^{(3)}_{j|i}(y)-R^{(2)}_{j|i}(y)[I^{(1)}_{j|i}-I^{(1)}_{j|i}(y)]}{I^{(1)}_{j|i}}
\nonumber\\
&=\frac{I^{(3)}_{j|i}}{I^{(1)}_{j|i}}-\frac{I^{(3)}_{j|i}(y)}{I^{(1)}_{j|i}(y)}
,
\label{eq:S5_trans1pre}
\end{align}
where in the first step we simply used \eqref{eq:S5_transition_hierarchy} with $\delta T_{j|i}^{(0)}(y)=1$, in the second last step we inserted the auxiliary integrals from Eqs.~\eqref{eq:S5_int_aux} and \eqref{eq:S5_int_aux_x}, and in the last step we have canceled redundant integrals. The limit $y\to 0$ gives the exact first moment of the transition-path time
\begin{equation}
 \avgcStar{\delta t}_{j|i}\equiv\lim_{y\to0}\delta T_{j|i}^{(1)}(y)=\frac{I^{(3)}_{j|i}}{I^{(1)}_{j|i}}.
 \label{eq:S5_trans1}
\end{equation}
Similarly, the second moment of the transition-path time reads, using Eq.~\eqref{eq:S5_transition_hierarchy} and some tedious but straightforward algebra,
\begin{align}
  \avgcStar{\delta t^2}_{j|i}&\equiv\lim_{y\to0}\delta T_{j|i}^{(2)}(y)=\frac{2}{I_{j|i}^{(1)}}\int_{0}^{l_{j|i}}\dd x'\e^{-\beta U_{j|i}(x')}\int_0^{x'}\dd x_1\frac{\e^{\beta U_{j|i}(x_1)}}{D_{j|i}(x_1)}\int_{x'}^{l_{j|i}}\dd x_2 \frac{\e^{\beta U_{j|i}(x_2)}}{D_{j|i}(x_2)}\bigg[\frac{I^{(3)}_{j|i}}{I^{(1)}_{j|i}}-\frac{I^{(3)}_{j|i}(x')}{I^{(1)}_{j|i}(x')}\bigg]
  \nonumber\\
  &=2\bigg[\frac{I^{(3)}_{j|i}}{I^{(1)}_{j|i}}\bigg]^2-2\frac{I^{(5)}_{j|i}}{I^{(1)}_{j|i}},
  \label{eq:S5_trans2}
\end{align}
\end{widetext}
where in the last step of the first line we inserted Eq.~\eqref{eq:S5_trans1pre}
and afterwards inserted the auxiliary integrals from Eqs.~\eqref{eq:S5_int_aux} and \eqref{eq:S5_int_aux_x} and removed redundant terms that cancel in the limit $y\to0$. Note that all integrals of the type ``$R^{(k)}_{j|i}$'' cancel. From Eqs.~\eqref{eq:S5_trans1} and \eqref{eq:S5_trans2}
we establish immediately that the transition-path time is generally  sub-Markovian, i.e.
$\avgcStar{\delta t^2}_{j|i}\le 2 [\avgcStar{\delta t}_{j|i}]^2$. This was also found in \cite{sati20}. This finding underlines the usefulness of
systematically decomposing nested integrals as defined in \eqref{eq:S5_int_aux}.
Eqs.~\eqref{eq:S5_trans1} and \eqref{eq:S5_trans2} 
prove
Eq.~(11) 
 in 
Sec.~II.D in the main text.
In the following
\ref{sec:moments_wrapup}
we systematically decompose nested integrals as in Eq.~\eqref{eq:S5_int_aux}
to derive the remainder of the main practical results
(see Sec.~II.D in the main text),
which entails
Eqs.~(10)
 and
 (12) in the main text.



\section{Explicit conditional moments of the first passage time, dwell time, and transition-path time}\label{sec:moments_wrapup}
In this Appendix each sum over $j$ (or $k$) runs over all neighboring states of the fixed state $i$, i.e., we use the short-hand notation $\sum_{j}\equiv\sum_{j\in\mathcal{N}_i}$ and $\sum_{k}\equiv\sum_{k\in\mathcal{N}_i}$, where $\mathcal{N}_i$ denotes  the set of states adjacent to the fixed initial state $i$.

\subsection{Splitting probability}

Inserting the first line of Eq.~\eqref{eq:S1_aux}  and Eq.~\eqref{eq:S3_1st} into  Eq.~\eqref{eq:S1_splitting_from_unc} results in the local splitting probability
\begin{equation}
 \phiStar_{a|i}=\frac{1/I^{(1)}_{a|i}}{\sum_{j}1/I^{(1)}_{j|i}},
 \label{eq:S6_splitting_fin}
\end{equation}
where $a\in\mathcal{N}_i$.

\subsection{First moment of exit time, conditional first passage time, and dwell time}
The first moment of the exit time is obtained by inserting the first line of Eq.~\eqref{eq:S1_aux} and Eq.~\eqref{eq:S3_1st} into Eq.~\eqref{eq:S1_exit1_from_unc}
yielding
\begin{equation}
 \avgexit{t}_{i}=\frac{1-\sum_{j}\frac{b ^{(1)}_j}{\gamma^{(1)}_j}}{\sum_{j}\frac{1}{\gamma_j^{(1)}}}=\frac{\sum_j\frac{I^{(2)}_{j|i}}{I^{(1)}_{j|i}}}{\sum_j\frac{1}{I^{(1)}_{j|i}}}=\sum_j \phiStar_{j|i}I^{(2)}_{j|i},
 \label{eq:S6_exit1_fin}
\end{equation}
where we first used Eq.~\eqref{eq:S1_exit1_from_unc} and then inserted  Eqs.~\eqref{eq:S1_aux} and \eqref{eq:S3_1st}; in the last step we used Eq.~\eqref{eq:S6_splitting_fin}.
The first moment of the dwell time $\avg{\tau}_i$ can now be deduced from the conditional independence of transition-path  and dwell time 
(see Appendix~B in the main text),
 implying $\avgexit{t}_i=\sum_j\phiStar_{j|i}[\avgtrans{\delta t}_{j|i}+\avgdwell{\tau}_i]$. Upon  using $\sum_j\phiStar_{j|i}=1$
we obtain
\begin{equation}
 \avgdwell{\tau}_i=\avgexit{t}_i-\sum_j\phiStar_{j|i}\avgtrans{\delta t}_{j|i}=\sum_{j}\phiStar_{j|i}\bigg[I^{(2)}_{j|i}-\frac{I^{(3)}_{j|i}}{I^{(1)}_{j|i}}\bigg],
  \label{eq:S6_dwell1_fin}
\end{equation}
where in the second step we inserted Eqs.~\eqref{eq:S5_trans1} and \eqref{eq:S6_exit1_fin}.
Using Eqs.~\eqref{eq:S5_trans1} and \eqref{eq:S6_dwell1_fin} the conditional mean first passage time in turn becomes
\begin{equation}
\avgcStar{t}_{a|i}=\avg{\tau}_i+\avgcStar{\delta t}_{a|i}=\frac{I^{(3)}_{a|i}}{I^{(1)}_{a|i}}+\sum_{j}\phiStar_{j|i}\bigg[I^{(2)}_{j|i}-\frac{I^{(3)}_{j|i}}{I^{(1)}_{j|i}}\bigg].
  \label{eq:S6_tc1_fin}
\end{equation}

\subsection{Second moment of exit time, conditional first passage time, and dwell time}\label{subsec:second_moments}
The exact expression for the second moment of the exit time
is obtained after some tedious algebra and reads
\begin{widetext}
\begin{align}
 \avgexit{t^2}_{i}&=
 \frac{ \sum_{j}
 \Big[\frac{\gamma_j^{(2)}b ^{(1)}_j}{[\gamma^{(1)}_j]^2}-\frac{b ^{(2)}_j}{\gamma^{(1)}_j}\Big]
  +2\Big[1+\sum_{j}\Big(\frac{\gamma_j^{(2)}}{2[\gamma^{(1)}_j]^2}-\frac{b ^{(1)}_j}{\gamma^{(1)}_j}\Big)\Big]
 \dfrac{1-\sum_{k}\frac{b ^{(1)}_k}{\gamma^{(1)}_k}}{\sum_{k}\frac{1}{\gamma_k^{(1)}}}
 }{\sum_{k}\frac{1}{\gamma_k^{(1)}}}
 \nonumber\\
&=   \frac{2\sum_j\Big[\frac{I_{j|i}^{(2)}I_{j|i}^{(3)}}{I_{j|i}^{(1)}}-\frac{I_{j|i}^{(4)}}{I_{j|i}^{(1)}}+\frac{I_{j|i}^{(2)}}{I_{j|i}^{(1)}}\Big(\frac{\sum_k [R^{(3)}_{k|i}-R^{(2)}_{k|i}R^{(1)}_{k|i}]}{\sum_kR^{(1)}_{k|i}}\Big)\Big]+2\sum_{j}\Big[\frac{I_{j|i}^{(2)}}{I_{j|i}^{(1)}}-\frac{I_{j|i}^{(3)}}{[I_{j|i}^{(1)}]^2}+\frac{1}{I_{j|i}^{(1)}}\Big(\frac{\sum_k [R^{(2)}_{k|i}R^{(1)}_{k|i}-R^{(3)}_{k|i}]}{\sum_kR^{(1)}_{k|i}}\Big)\Big]
 \dfrac{\sum_{k}\frac{I_{k|i}^{(2)}}{I_{k|i}^{(1)}}}{\sum_{k}\frac{1}{I_{k|i}^{(1)}}}
 }{\sum_{k}\frac{1}{I_{k|i}^{(1)}}}
  \nonumber\\
&=   \frac{2\sum_j\Big[\frac{I_{j|i}^{(2)}I_{j|i}^{(3)}}{I_{j|i}^{(1)}}-\frac{I_{j|i}^{(4)}}{I_{j|i}^{(1)}}\Big]+2\sum_{j}\Big[\frac{I_{j|i}^{(2)}}{I_{j|i}^{(1)}}-\frac{I_{j|i}^{(3)}}{[I_{j|i}^{(1)}]^2}\Big]
 \dfrac{\sum_{k}\frac{I_{k|i}^{(2)}}{I_{k|i}^{(1)}}}{\sum_{k}\frac{1}{I_{k|i}^{(1)}}}
 }{\sum_{k}\frac{1}{I_{k|i}^{(1)}}}=2\sum_j\phiStar_{j|i}\Big[I_{j|i}^{(2)}\avgcStar{t}_{j|i}-I_{j|i}^{(4)}\Big],
 \label{eq:S6_exit2_fin}
\end{align}
\end{widetext}
where in the first step we adopted Eq.~\eqref{eq:S1_exit2_from_unc},
in the second line we inserted Eqs.~\eqref{eq:S3_1st} and \eqref{eq:S3_2nd}; in the last line we canceled equal terms and finally inserted Eqs.~\eqref{eq:S6_splitting_fin} and \eqref{eq:S6_tc1_fin}. From Eq.~\eqref{eq:S6_exit2_fin}
follows that a vanishing extrinsic noise, which corresponds $\avgcStar{t}_{a|i}=\avgexit{t}_{i}$, immediately renders fluctuations sub-Markov (see after next subsection, \ref{subsec:sub-Markov_proof}, for more details).

Using the conditional independence of dwell and transition-path times, $\avgcStar{t^2}_{j|i}=\avgtrans{\delta t^2}_{j|i}+2 \avgtrans{\delta t}_{j|i}\avgdwell{\tau}_{i}+\avgdwell{\tau^2}_{i}$, we obtain
\begin{align}
 &\avgdwell{\tau^2}_{i}=\avgexit{t^2}_{i}-\sum_j\phiStar_{j|i}\Big[\avgtrans{\delta t^2}_{j|i}+2\avgtrans{\delta t}_{j|i}\avgdwell{\tau}_{i}\Big]
 \nonumber\\
 &
 =2(\avgdwell{\tau}_i)^2+\sum_j\phiStar_{j|i}\Big[2I^{(2)}_{j|i}\avgtrans{\delta t}_{j|i}-2I^{(4)}_{j|i}
 \nonumber\\
 &+2\avgtrans{\delta t}_{j|i}\avgdwell{\tau}_{i}\Big]-\sum_j\phiStar_{j|i}\Big[\avgtrans{\delta t^2}_{j|i}+2\avgtrans{\delta t}_{j|i}\avgdwell{\tau}_{i}\Big]
 \nonumber\\
 &=
 2(\avgdwell{\tau}_i)^2+\sum_j\phiStar_{j|i}\Big[2I^{(2)}_{j|i}\avgtrans{\delta t}_{j|i}-2I^{(4)}_{j|i}-\avgtrans{\delta t^2}_{j|i}\Big],
 \label{eq:S6_dwell2_fin}
\end{align}
where in the first step we used the definition $\avgexit{t^2}_i\equiv \sum_j\phiStar_{j|i}\avgcStar{t^2}_{j|i}$
and in the second step we inserted the transition-path time moments $\avgtrans{\delta t}_{j|i}$ and $\avgtrans{\delta t^2}_{j|i}$ from Eqs.~\eqref{eq:S5_trans1} and \eqref{eq:S5_trans2}, respectively, 
the splitting probability $\phiStar_{j|i}$ from Eq.~\eqref{eq:S6_splitting_fin}
as well as the first moment of the dwell time $\avgdwell{\tau}_{i}$
from Eq.~\eqref{eq:S6_dwell1_fin} into the second moment of the exit
time in Eq.~\eqref{eq:S6_exit2_fin}. In the final step of Eq.~\eqref{eq:S6_dwell1_fin} we combined the two sums. Eq.~\eqref{eq:S6_dwell2_fin} precisely proves the second line of
Eq.~(12) in the main text.

\subsection{Summary of the main practical result}\label{subsec:wrapup_summary}
Let us briefly summarize the proof carried out in this section,
which lead to the
main practical result, i.e., proofs of
Eqs.~(10)-(13) in the main text
given in
 Sec.~II.D in the main text.
Hereby,  Eq.~\eqref{eq:S6_splitting_fin}
proves
Eq.~(10)
in the main text. Furthermore,
Eqs.~\eqref{eq:S5_trans1} and \eqref{eq:S5_trans2} from the previous subsection
prove
Eq.~(11)
in  the main text and Eqs.~\eqref{eq:S6_dwell1_fin}
and \eqref{eq:S6_dwell2_fin} prove
Eq.~(12) in the main text.
Since
Eq.~(13)
in the main text follows from
Eq.~(3)
in the main text, which is proven in 
Appendix B in the main text,
we hereby have proven
Eqs.~(10)-(13)
in the main text, which represent the main practical result of our work. The auxiliary integrals $I^{(k)}_{j|i}$ are defined in Eq.~\eqref{eq:S5_int_aux} [see also
Eq.~(9)
in the main text] and contain the \emph{local} potential defined
in
Eq.~(5) in the main text,
which reads
$\beta U_{j|i}(x)\equiv-\beta\int_0^x\dd y F_{j|i}(y)=\beta\int_0^x\dd y F_{i|j}(l_{i|j}-y)$.

Note that due to the tediousness of the derivations 
in
Sec.~II.D in the main text,
we confirmed its correctness.
To this end we deduced the 
third moments of 
unconditioned first passage moments (see Table.~\ref{eq:S3_3rd}).
Inserting  Eq.~\eqref{eq:S1_exit_moments}
with $k=3$ into the first expression in the last line of Eq.~\eqref{eq:S1_moments_setofequations}
yields the third moment of unconditioned first passage time.
After a long calculation, which would go beyond the scope of this article,
we find that the left hand side of Eq.~\eqref{eq:S1_moments_setofequations}
(see Table.~\ref{eq:S3_3rd}) indeed agrees with the  right hand side, where we insert the results of
Sec.~II.D in the main text.
This can be seen as independent proof of the main practical result.

\subsection{Vanishing extrinsic noise renders fluctuations sub-Markov}\label{subsec:sub-Markov_proof}
Let us briefly prove that vanishing extrinsic noise implies sub-Markov fluctuations.
Vanishing extrinsic noise means that all transition from state $i$
to a neighboring state $a$ take on average equally long, that is,  $\avgtrans{\delta t}_{a|i}=\avgtrans{\delta t}_{j|i}$ for all $a,j\in \mathcal{N}_i$.

Let us from now on (in this subsection) assume that the extrinsic noise vanishes. 
According to Eq.~\eqref{eq:S6_exit1_fin} and Eq.~\eqref{eq:S6_tc1_fin} we obtain $\avgcStar{t}_{j|i}=\avgexit{t}_i$, which inserted into Eq.~\eqref{eq:S6_exit2_fin}
gives
\begin{align}
\avgexit{t^2}_i
&=2\sum_j\phiStar_{j|i}\Big[I^{(2)}_{j|i}\avgexit{t}_i-I^{(4)}_{j|i}\Big]
\nonumber\\
&=2[\avgexit{t}_i]^2-\sum_j\phiStar_{j|i}I^{(4)}_{j|i},
\end{align}
where in the last step we again used Eq.~\eqref{eq:S6_exit1_fin}. Since the auxiliary integrals are defined to be positive $I^{(4)}_{j|i}\ge0$,
we generally find $\avgexit{t^2}_i\le 2[\avgexit{t}_i]^2$, which proves that fluctuations become sub-Markov. We note that  $\avgexit{t^2}_i\ge 2[\avgexit{t}_i]^2$
is equivalent to $\sigma^2_{\rm exit}\equiv \avgexit{t^2}_i-[\avgexit{t}_i]^2\le [\avgexit{t}_i]^2$. This completes the proof of the main result of this paper, which states that vanishing extrinsic noise implies fluctuations to become sub-Markov.

We note that the following converse equivalent conclusion can  be drawn from the proven statement. Whenever the fluctuations are pronounced akin super-Markov, that is $\avgexit{t^2}_i\ge 2[\avgexit{t}_i]^2$, there must exist parallel transitions $i\to a$ and $i\to j$ that are unequally fast $\avgtrans{t}_{a|i}\neq\avgtrans{t}_{j|i}$.

\end{document}